\documentclass[11pt,a4page]{article}
\usepackage[pdftex]{graphics}
\usepackage{jheppub}
\usepackage{amsmath}
\usepackage{amssymb}
\usepackage{graphicx}
\usepackage{slashed}
\usepackage{booktabs}
\usepackage{dsfont}

\newcommand{\nn}{\nonumber}

\def\half{\frac{1}{2}}
\def\thalf{{\textstyle \frac{1}{2}}}
\def\imp{\Longrightarrow}

\def\goto{\rightarrow}

\def\Tr{{\rm Tr}}

\def\IC{\mathbb{C}}

\def\IR{\mathbb{R}}
\def\IZ{\mathbb{Z}}

\def\CE{{\cal E}}

\def\CH{{\cal H}}
\def\CI{{\cal I}}

\def\CL{{\cal L}}
\def\CM{{\cal M}}
\def\CN{{\cal N}}
\def\CO{{\cal O}}
\def\CP{{\cal P}}

\def\CR{{\cal R}}
\def\CS{{\cal S}}
\def\CT{{\cal T}}

\def\CV{{\cal V}}
\def\CW{{\cal W}}
\def\CX{{\cal X}}

\def\CZ{{\cal Z}}

\def\a{\alpha}
\def\b{\beta}

\def\d{\delta}
\def\e{\epsilon}

\def\z{\zeta}

\def\l{\lambda}

\def\t{\tau}

\def\O{\Omega}

\def \FN {\rm FN}
\def \SR {\rm SR}

\def\sD{\mathsf{D}}

\def\sH{\mathsf{H}}

\def\sL{\mathsf{L}}

\def\sO{\mathsf{O}}

\def\sR{\mathsf{R}}

\def\sT{\mathsf{T}}
\def\sU{\mathsf{U}}
\def\sV{\mathsf{V}}
\def\sW{\mathsf{W}}

\def\fl{\mathfrak{l}}
\def\sm{\mathfrak{m}}

\def\sr{\mathsf{r}}
\def\ss{\mathsf{s}}
\def\st{\mathsf{t}}

\def\sx{\mathsf{x}}

\title{Superconformal Index and \\ 3d-3d Correspondence for Mapping Cylinder/Torus}

\author[a]{Dongmin Gang,}
\author[a]{Eunkyung Koh,}
\author[b,c,d]{Sangmin Lee,}
\author[e,f]{Jaemo Park}

\affiliation[a]{School of Physics, Korea Institute for Advanced Study, Seoul 130-722, Korea}
\affiliation[b]{Center for Theoretical Physics, Seoul National University,
Seoul 151-747, Korea}
\affiliation[c]{Department of Physics and Astronomy, Seoul National University,
Seoul 151-747, Korea}
\affiliation[d]{College of Liberal Studies, Seoul National University,
Seoul 151-742, Korea}
\affiliation[e]{Department of Physics, POSTECH, Pohang 790-784, Korea}
\affiliation[f]{Postech Center for Theoretical Physics (PCTP), Postech, Pohang
  790-784, Korea}

\emailAdd{arima275@kias.re.kr}
\emailAdd{ekoh@kias.re.kr}
\emailAdd{sangmin@snu.ac.kr}
\emailAdd{jaemo@postech.ac.kr}

\abstract{ We probe the 3d-3d correspondence for mapping
cylinder/torus   using the superconformal index. We focus on the
case when the fiber is  a once-punctured torus ($\Sigma_{1,1}$). 
The corresponding 3d
field theories can be realized using  duality domain wall theories in 4d
$\CN=2^*$ theory. We show that the superconformal indices of the 3d
theories  are the $SL(2,\mathbb{C})$ Chern-Simons partition function
on the mapping cylinder/torus. For the mapping torus, we also consider  another
realization of the corresponding 3d theory associated with ideal
triangulation.  The equality between the indices from
the two descriptions for the mapping torus theory is reduced to a
simple basis change of the Hilbert space for the $SL(2,\mathbb{C})$ Chern-Simons theory on $\mathbb{R}\times \Sigma_{1,1}$. }

\begin{document}
\maketitle

\newpage
\section{Introduction and Summary} \label{sec : introduction}

There is an  interesting and fruitful approach of viewing lower dimensional superconformal field theories (SCFTs)
from the vantage point of the $(2,0)$ theory in six dimensions. Though we do not fully understand
the $(2,0)$ theory, this viewpoint leads to useful insight to understand  SCFTs in lower dimensions. Perhaps the most well-known example would be the celebrated AGT conjecture
\cite{AGT}.
Heuristically, the conjecture can be motivated by considering a twisted compactification of the $(2,0)$ theory on $S^4 \times \Sigma_{g,h}$
where  $\Sigma_{g,h}$ is a Riemann surface of genus  $g$ with $h$ punctures.
The compactification leads to interesting SCFTs with 8 supercharges in 4-dimensions \cite{GaiottoN=2}. A supersymmetric  partition function of the  $(2,0)$ theory on $S^4 \times \Sigma$ is expected to give the $S^4$-partition function of such 4d SCFTs. 
Pestun computed the $S^4$-partition function using localization techniques for theories 
whose Lagrangian is known \cite{Pestun}. Throughout this paper, we will mainly focus on the $A_1$ type of $(2,0)$ theory. In this case, the 4d theories, denoted by $T_\Sigma$, admit weakly coupled gauge theory descriptions with gauge group $SU(2)^{3g-3+h}$.
 On the other hand, the $(2,0)$ theory compactified on $S^4$ is expected to lead to a 2d conformal field theory.
It turns out that for the $A_1$ $(2,0)$ theory this is Liouville
theory (for a review of Liouville theory see \cite{Nakayama}). Hence, the $A_1$ $(2,0)$
theory on $S^4 \times \Sigma$ gives the partition
function (correlation function) of the Liouville theory on $\Sigma$.
Identifying the two partition functions obtained from two different regimes of the 
compactification, we obtain the AGT conjecture which relate $S^4$
partition function of $T_\Sigma$ theory with Liouville correlation
function  on $\Sigma$.

One might wonder if a similar relation may be found in 3-dimensions by 
compactifying the $(2,0)$ theory on some 3-manifold $M$. If so, it would lead to a plethora of  $\CN=2$ SCFTs in 3 dimensions.
In fact, in \cite{Dimofte:2011ju,Dimofte:2011py}, Dimofte, Gaiotto, Gukov (DGG) 
introduced an algorithm to construct 
the field theory $T_M$  associated with the 3-manifold $M$ using 
the ideal triangulation data of $M$.
By specifying the gluing rules of the field theory corresponding to those of the triangulation, one can construct a
huge class of 3d SCFTs.
One interesting feature is that the same manifold with two different
triangulations gives rise to two different descriptions of the same
SCFT. Some simple mirror pairs of 3d were shown to be described in
this way. Although it is difficult to  see from their construction,
the theory $T_M$ is  believed to be the 3d theory obtained by
compactifying  the $A_1$ $(2,0)$ theory on $M$.
Considering  the $(2,0)$ theory  on $ M \times S^2 \times_q S^1$ or $
M \times  S_b^3$,\footnote{$S^2 \times_q S^1$ denote $S^1$ bundle of two-sphere twisted with  holonomy for a  combination of $U(1)$ R-symmetry and space-time rotation symmetry. $S^3_b$ denote a squashed three-sphere (ellipsoid) \cite{Hama:2011ea}.  } we have the 3d-3d analogue of the AGT conjecture. 
If we first compacitify on $M$, we 
obtain the superconformal index ($S^2\times_q S^1$) or the sqaushed three-sphere partition function ($S_b^3$) for $T_M$.
On the other hand, if we compactify on  $  S^2 \times_q S^1$ or $
S_b^3$ first, the theory is expected to be a $SL(2,\IC)$ or $SL(2,\IR)$\footnote{ It is $SL(2,\mathbb{R})$ CS theory in sense that the boundary Hibert-space looks like a quantization  of $SL(2,\mathbb{R})$ flat connections on the boundary.  In a recent paper  \cite{Cordova:2013cea}, the 3d-3d relation is derived from the first principal and find that the $S^3_b$ partition function corresponds to $SL(2,\mathbb{C})$ CS theory with level $k=1$. We expect there's an isomorphism between  a Hilbert-space obtained by quantizing  $SL(2,\mathbb{R})$ flat connections  and one from $SL(2,\mathbb{C})$ with $k=1$. }
Chern-Simons (CS) theory on $M$ \cite{Dimofte:2011ju,Dimofte:2011py,Terashima:2011qi,Yagi:2013fda}, respectively. From this analysis, we obtain the following non-trivial
prediction of the 3d/3d correspondence:
\begin{align}
&\textrm{Superconformal index/$S^3_b$ partition function for $T_M$   } \nn
\\
&=  \textrm{ $SL(2,\IC)/SL(2,\IR)$ Chern-Simons partition function on $M$.} \label{3d/3d correspondence}
 \end{align}
 
 One interesting class of the 3d-3d correspondence arises from the 3d duality domain wall theory \cite{Gaiotto:2008ak,
Drukker:2010jp,Hosomichi:2010vh,Terashima:2011qi,Terashima:2011xe,Dimofte:2011jd,Teschner:2012em,Gang:2012ff} associated with 4d theory $T_\Sigma$  and a duality group element $\varphi$. 
The corresponding  internal 3-manifold  $M$ is the mapping cylinder $\Sigma \times_\varphi I$, where $I=[0,1]$
is the unit interval, equipped with the cobordism $\varphi: (x,0)
\rightarrow (\varphi(x),1)$. Here $\varphi$ is an element of the mapping class group for $\Sigma$, which can be identified with duality group for $T_\Sigma$.
 Further identifying the two ends of the interval by
the cobordism $\varphi$, we obtain a mapping torus $\Sigma
\times_{\varphi} S^1$. Identifying the two ends of the interval
corresponds to gluing two global $SU(2)^{3g-3+h}$ symmetries  in the
duality wall theory coupled to $SU(2)^{3g-3+h}$ gauge symmetry in
$T_\Sigma$.  On the other hand, the mapping torus admits an 
ideal triangulation and the corresponding 3d theory can be
constructed by the DGG algorithm. Hence the mapping torus has two
different realizations of the  associated 3d SCFT. The one involving 
the duality wall theory has a clear  origin from M5-brane physics but
identifying the 3d SCFT for general $\Sigma$  is very non-trivial.
In the other one using the DGG algorithm, the
physical origin from M5-brane is unclear  but generalization to
arbitrary $\Sigma$ is quite straightforward. It  boils down to the
problem finding a triangulation of the mapping torus.

In this paper we are mainly interested in the 3d-3d correspondence
between the superconformal index for $T_M$ and $SL(2,\mathbb{C})$
Chern-Simons theory on $M$, where $M$ is
mapping cylinder or  torus whose fiber is once-punctured torus, $\Sigma_{1,1}$. 
The mapping torus $\Sigma_{1,1}\times_\varphi S^1$ will be denoted by tori($\varphi$) for simplicity. The analysis of the 3d-3d
correspondence \eqref{3d/3d correspondence} for mapping torus was done at the semiclassical level using the
 $S^3_b$ partition function in \cite{Terashima:2011xe}; see also 
 \cite{K.Nagao:2011,Kashaev:2012,Hikam:2012,Terashima:2013,Hikam:2013}
for interesting generalizations. To check
the 3d-3d correspondence at the full quantum level,  we carefully define the Hilbert-space of $SL(2,\mathbb{C})$ CS theory on $\mathbb{R}
\times \Sigma_{1,1}$\footnote{Or, simply we express ``CS theory on
$\Sigma_{1,1}$'' ignoring manifestly existing time-coordinate} and
construct  quantum  operators $\varphi \in SL(2,\mathbb{Z})$, which
turn out to be unitary operators. 
Even though several basic
ingredients of this construction were already given in  references
\cite{Dimofte:2011gm,Terashima:2011xe,Dimofte:2011jd,Dimofte:2011py},
working out the details of the Hilbert space turns out to be a
non-trivial and worthwhile task.   
We  are particularly interested in the case when the CS level is purely imaginary. 
In the case, the quantization is   studied in a relatively recent paper \cite{Dimofte:2011py}. In the paper,    the Hilbert-space is identified  as $L^2 (\mathbb{Z}\times \mathbb{Z})$ which have the same structure with 3d index. 
We  study the mapping-class group representation on the Hilbert-space which is a new and interesting object.  We show that the superconformal index for the duality wall
theory associated with $\varphi \in SL(2,\mathbb{Z})$  is indeed a
matrix element of $\varphi$ in a suitable basis of the
Hilbert-space. According to an axiom of topological quantum field
theory, the matrix element is nothing but the $SL(2,\mathbb{C})$ CS
partition function on the mapping cylinder,
and thus it provides an evidence 
for the 3d-3d correspondence \eqref{3d/3d correspondence} for the mapping
cylinder.For mapping torus, tori($\varphi$), the CS partition
function is given as a trace of  an operator  $\varphi \in
SL(2,\mathbb{Z})$.  Depending on the choice of basis of the
Hilbert-space, the expression for the $\Tr (\varphi)$ is equivalent
to the expression of superconformal index for  mapping torus theory
obtained either using the duality wall theory or using the DGG algorithm. 
It confirms the equivalence of the two descriptions for mapping torus
theory at the level of the superconformal index and also confirms  
the 3d/3d correspondence \eqref{3d/3d correspondence} for
$M=\textrm{tori}(\varphi)$. We also give some evidences for an
isomorphism between the Hilbert-space of $SL(2,\mathbb{C})$ CS
theory on $\Sigma_{1,1}$ and  the Hilbert-space canonically
associated to the boundary $S^2 \times S^1$ of  4d (twisted)
$\CN=2^*$ theory on $B^3 \times S^1$.

The content of the paper is organized as follows.
In section \ref{section : two routes}, we introduce the basic setup for the 3d geometry of the mapping torus and its ideal triangulation. We also explain the field theory realization, one as a `trace' of the duality
domain wall and the other as an outcome of the DGG algorithm based on the triangulation.
In section \ref{quantum riemann surface}, we  review  the quantization of $SL(2,\mathbb{C})$ Chern-Simon theory on
the  Riemann surface
$\Sigma_{1,1}$. For later purposes, we introduce several coordinate systems  for the
phase space and  explain the relation between them.
The A-polynomial for mapping torus is analyzed in two different ways.
In section \ref{sec: SCI/SL(2,C)}, we show that the superconformal index of $T_M$ with 
$M$ being mapping cylinder/torus is the
 $SL(2,\mathbb{C})$ CS partition function on $M$. To calculate the CS partition functions, we construct a Hilbert-space for $SL(2,\mathbb{C})$ CS theory on $\Sigma_{1,1}$.
 We further show that the two computations of the mapping torus index 
are simply related by a basis change of the Hilbert space in taking trace of $\varphi\in SL(2,\mathbb{Z})$, thereby providing a consistency check for the duality of the
two descriptions of the mapping torus theory. In section \ref{sec:
SL(2,R)/squashed S3}, we make comments on the partition function on
the squashed sphere for the theory on the mapping cylinder/torus. We
indicate many parallels between the partition function and the
superconformal index and argue that 
most of our findings in section \ref{sec: SCI/SL(2,C)} 
can be carried over to the context
of the squashed sphere partition function. 
Several computations are relegated to the appendices.  For a technical
reason, we mainly focus on general hyperbolic mapping torus which
satisfies $|\Tr (\varphi)|>2$ \cite{Gueritaud}. Extension of our
analysis to the non-hyperbolic  case seems  quite straightforward and
some examples are given in section \ref{index-wall}.

When we were finishing this work, an interesting article \cite{Dimofte:2013lba} appeared on arXiv.org, which
focuses on mapping cylinder and its  triangulation.  We expect that several expressions
for the $SL(2,\mathbb{C})$ CS partition function on mapping cylinder in our paper can be directly derived from their construction.

\section{Two routes to mapping torus field theories}\label{section : two routes}

A mapping torus is specified by a Riemann surface $\Sigma_{g,h}$ of genus $g$ with $h$ punctures and an element $\varphi$ of the mapping class group of  $\Sigma_{g,h}$. Topologically, it is
a bundle with $\Sigma$ fibered over an interval $I=[0,1]$
with $\Sigma$ at one end of the interval identified with $\varphi(\Sigma)$
at the other end. In other words,
\begin{align}
M=\Sigma \times_\varphi S^1 = \Sigma\times I/[(x,0)\sim (\varphi(x),1)]\,.
\end{align}
In this paper, we only consider the mapping torus for the once punctured torus $\Sigma_{1,1}$ whose mapping class group is $SL(2,\mathbb{Z})$.
The mapping torus associated with $\varphi \in SL(2,\mathbb{Z})$ will be denoted as $\textrm{tori}(\varphi)$.
\begin{align}
\textrm{tori}(\varphi) := \Sigma_{1,1}\times_{\varphi}S^1\;.
\end{align}

The 3d-3d correspondence \cite{Dimofte:2011ju,Dimofte:2011py} states that one can associate a  three-manifold $M$ with a 3d theory $T_M$.\footnote{When $M$ has boundary, the 3d theory $T_M$ also depends on the choice of polarization $\Pi$  for the boundary phase space $\CM_{SL(2)} (\partial M)$, the space of $SL(2)$ flat connections on $\partial M$. In a strict sense, the 3d theory should be labelled by $T_{M,\Pi}$.} Physically, $T_M$ can be thought of as a dimensional reduction of the 6d (2,0) theory of $A_1$-type on $M$.\footnote{It can be generalized to  (2,0) theory of general A,D,E type and the corresponding 3d theory $T[M,\mathbf{g}]$ is labelled by 3-manifold $M$ and Lie algebra $\mathbf{g}$ of gauge group \cite{Dimofte:2013iv}.}  The mapping torus theory,  $T_M$ with $M= \textrm{tori}(\varphi)$, has two different realizations.

In the first approach, one compactifies the 6d (2,0) theory on $\Sigma_{1,1}$ to obtain the 4d $\mathcal{N}=2^*$ theory and reduces it on $S^1$ with a twist by $\varphi$ to
arrive at $T_M$.
The mapping class group $SL(2,\mathbb{Z})$
is the group of duality transformations in the sense of the 4d theory.
From this viewpoint,
$T_M$ can be obtained by taking a proper ``trace'' action on a 3d duality wall theory associated with $\varphi$.

In the other approach, one begins by triangulating the mapping torus using
a finite number of tetrahedra. Dimofte, Gaiotto and Gukov (DGG) \cite{Dimofte:2011ju} proposed a systematic algorithm for constructing $T_M$ when
the triangulation for $M$ is known. One can construct $T_M$ by applying the DGG algorithm to the known information on the triangulation
of $M = \textrm{tori}(\varphi)$.

\subsection{Duality wall theory} \label{duality wall theory}

Following \cite{Gaiotto:2008ak,Terashima:2011qi}, we use the notation $T[SU(2),\varphi]$ to denote the 3d theory living on the duality wall between
two copies of 4d $SU(2)$ $\mathcal{N}=2^*$ theory associated with an element $\varphi$ of the duality group $SL(2,\mathbb{Z})$.

\begin{table}[htbp]
   \centering
   \begin{tabular}{@{} l|c|c @{}} 
      \toprule

       & $q_1\;\;\; q_2 \;\;\;q_3 \;\;\;q_4\;\;\;\phi_0$    \\
      \midrule
      $U(1)_{\rm gauge}$ &$1\;\;\;\; 1 \;-1\; -1\;\;\;0$   \\
       $U(1)_{\rm bot}$&$1\;-1 \;\;\;1\;-1\;\;\;0$     \\
        $U(1)_{\rm punct}$& $\;\half \;\;\;\;\half  \;\;\;\;\half \;\;\;\;\half\;-1$ &    \\
         $U(1)_{\rm top}$ & $ 0\;\;\;\;0\;\;\;\;0\;\;\;\;0\;\;\;\;0$ &  \\
      \bottomrule
   \end{tabular}
   \caption{$T[SU(2)]$ theory. $U(1)_{\rm top}$ denotes the topological $U(1)$ charge, $J_{\rm top} = *dA_{U(1)_{\rm gauge}}$.}
   \label{tsu(2)-charge}
\end{table}

\begin{figure}[htbp]
   \centering
   \includegraphics[scale=0.56]{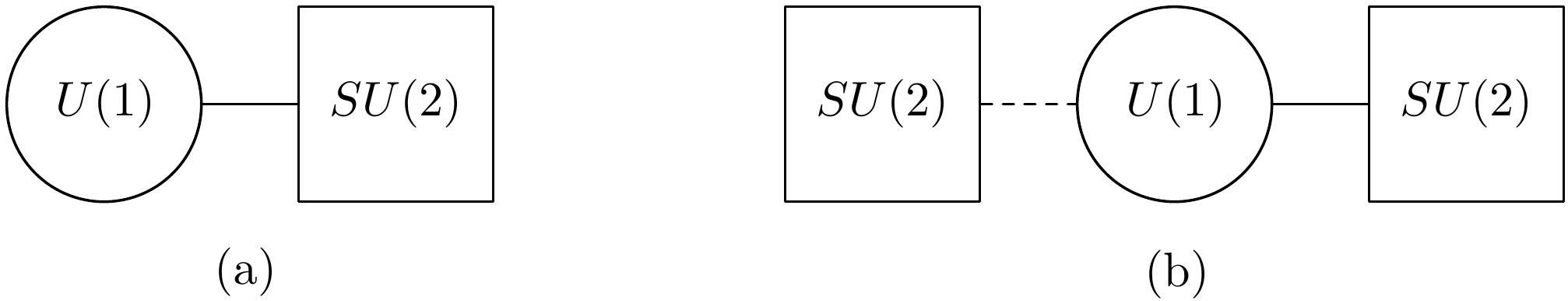} 
   \caption{Quiver diagrams for $T[SU(2)]$}
   \label{fig:quiver}
\end{figure}

We begin with the simplest case, $T[SU(2),S]$, often shortened to $T[SU(2)]$.
It is the 3d $\mathcal{N}=4$ SQED with two fundamental hyper-multiplets. Let the four chiral fields in the two hyper-multiplets be $q_1 , q_2, q_3,q_4$ and the adjoint chiral field in the vector multiplet be $\phi_0$.  The theory has global symmetry $SU(2)_{\rm bot}\times SU(2)_{\rm top}\times U(1)_{\rm punct}$ compatible with 3d $\mathcal{N}=2$ supersymmetries.  The charge assignments for chiral fields under the Cartan subalgebras of the gauge and global symmetries are summarized in Table~\ref{tsu(2)-charge}.
$U(1)_{\rm top}$ denotes the topological symmetry whose conserved charge is  a monopole charge for the $U(1)_{\rm gauge}$.
In the infrared (IR) limit, the $U(1)_{\rm top}$ is known to be enhanced to $SU(2)_{\rm top}$.
The quiver diagram for $T[SU(2)]$ is presented in Figure~\ref{fig:quiver}(a).
To emphasize that there is an additional quantum $SU(2)$ symmetry, one sometimes draws the quiver diagram as in Figure~\ref{fig:quiver}(b).

Let us consider the generalization to $T[SU(2), \varphi]$ for an arbitrary $\varphi \in SL(2,\mathbb{Z})$.
Firstly, multiplying $T^k$ to $S$ corresponds to adding Chern-Simons action of level $k$ for background gauge fields coupled to the $SU(2)$ global symmetries. Explicitly, one obtains the $T[SU(2),\varphi=T^{k}ST^{l}]$ theory by coupling the $T[SU(2)]$ theory with background gauge fields through the CS action of level $k$ for $SU(2)_{\rm top}$ and of level $l$ for $SU(2)_{\rm bot}$.
Secondly, multiplication of two mapping class elements $\varphi_1$ and $ \varphi_2$ corresponds to `gluing' $SU(2)_{\rm bot}$ in $T[SU(2),\varphi_1]$ with $SU(2)_{\rm top}$ in $T[SU(2),\varphi_2]$,
where `gluing' means gauging the diagonal subgroup. In ultraviolet (UV) region, no $SU(2)_{\rm top}$ symmetry is visible and the gluing procedure can't be implemented. To make the gluing procedure sensible in UV region, one need to consider a dual description for the $T[SU(2)]$ theory which allows the $SU(2)$ symmetry visible in UV. Some examples of these dual description is given in appendix \ref{dual description for T[SU(2)]}. Nevertheless, the gauging procedure for supersymmetric partition function can be implemented regardless of UV description choices since the partition function does not depend on the choice.  
Since $S$ and $T$
generate all elements of $SL(2,\mathbb{Z})$,
one can construct all $T[SU(2),\varphi]$ theories
by repeatedly using the field theory operations described above.

As a consistency check, we can examine the $SL(2,\mathbb{Z})$ structure of the $T[SU(2),\varphi]$ theory constructed above. $SL(2,\mathbb{Z})$ is generated
by $S$ and $T$ subject to the two relations,
 \begin{align}
 S^4 = (ST)^3 = I \;.
 \end{align}
In the next sections, we will check the equivalence between $T[SU(2), S^4 \varphi]$ and $T[SU(2),\varphi]$  by computing supersymmetric quantities for two theories. On the other hand,
the same computations indicate that $T[SU(2),(ST)^3 \varphi]$ can be identified
with $T[SU(2),\varphi]$ only after an extra twist, namely,
\\
\\
\centerline{\hskip -3cm \sl $T[SU(2),(ST)^3 \varphi] \;\;  = \;\; T[SU(2),\varphi]$ + $CS$ term with $k=\half$ }
\centerline{\hskip 4cm \sl for background gauge field coupled to $U(1)_{\rm punct}$}
\\
\\
In terms of the 4d $\mathcal{N}=2^*$ theory, the above relation says that $(ST)^3$ induces a $\theta$-term, $\Tr  (F_{\rm punct}\wedge F_{\rm punct})$ for the background gauge field coupled to the $U(1)_{\rm punct}$ symmetry which rotates an adjoint hyper.

In the context of 3d-3d correspondence, the duality wall theory is associated to a 3-manifold $\Sigma_{1,1}\times_\varphi I $ called mapping cylinder \cite{Hosomichi:2010vh,Terashima:2011qi}. Topologically, a mapping cylinder is a direct product of $\Sigma_{1,1}$ and interval $I=[0,1]$. At two ends of the interval (`top' and `bottom'), there are two  boundary Riemann surfaces denoted as $\Sigma^{\textrm{top}}_{1,1}$ and  $\Sigma^{\textrm{bot}}_{1,1}$. In 3d-3d correspondence, global symmetries of $T_M$
are related to the boundary phase space of $M$,
\begin{align}
(\textrm{Rank of global symmetry in $T_M$}) = \half \textrm{dim}_\mathbb{C}\left[ \CM_{SL(2,\mathbb{C})}(\partial M) \right]\;.
\end{align}
The two boundary  phase spaces $\CM_{SL(2)}(\Sigma^{\textrm{top}}_{1,1})$ and  $\CM_{SL(2)}(\Sigma^{\textrm{bot}}_{1,1})$ are related to $SU(2)_{\textrm{top}}$ and $SU(2)_{\textrm{bot}}$ symmetries, respectively. The phase space associated to the `cusp' boundary  made of the puncture on the Riemann surface is related to the $U(1)_{\textrm{punct}}$ symmetry.
\begin{figure}[htbp]
   \centering
   \includegraphics[scale=.25]{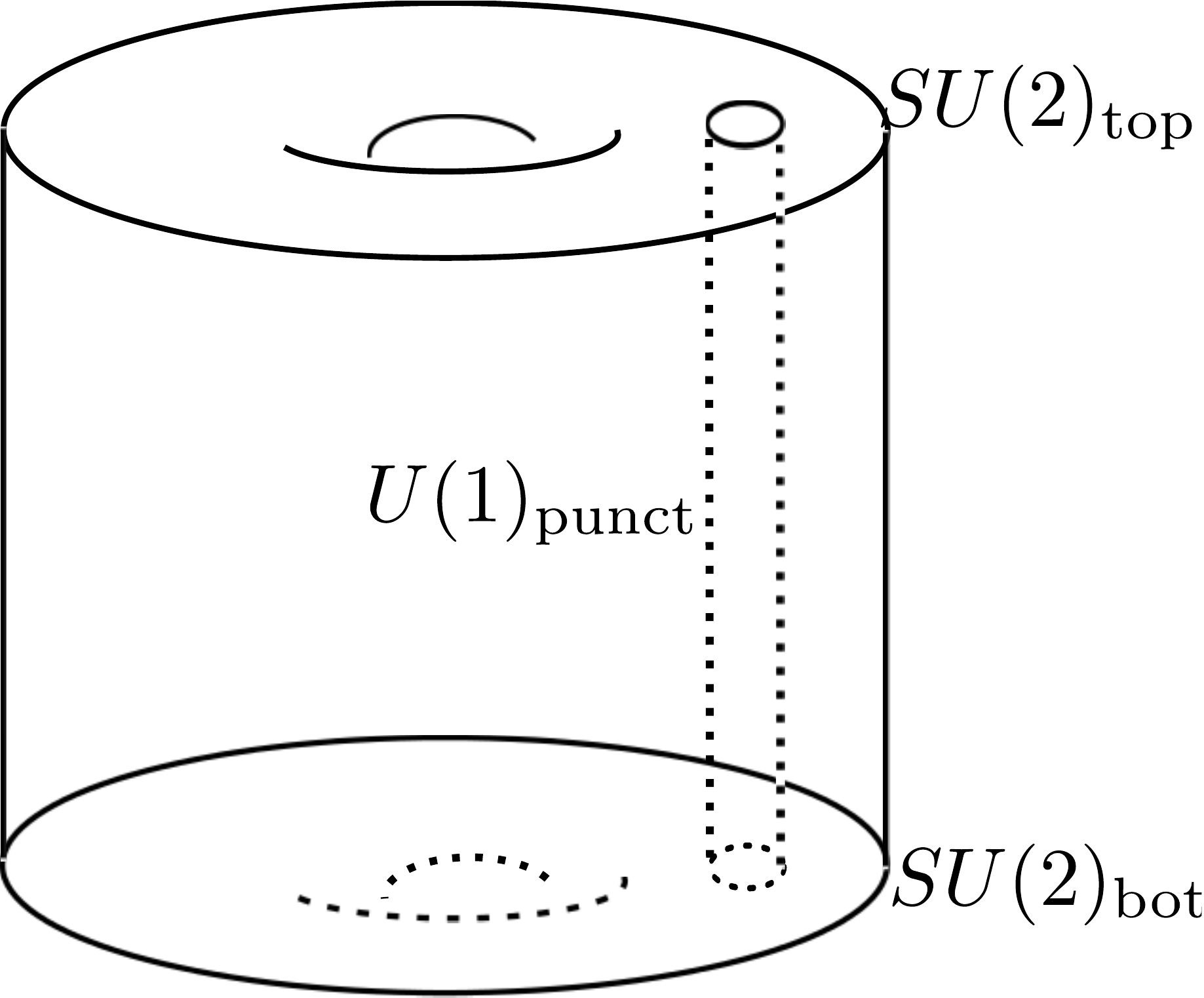} 
   \caption{Mapping cylinder $\Sigma_{1,1}\times_\varphi I$. A global symmetry in the duality wall theory is associated to each component of boundary. }
   \label{fig:LR2}
\end{figure}

The mapping torus, $\textrm{tori}(\varphi) := \Sigma_{1,1}\times_\varphi S^1$, can be obtained by gluing the two boundary Riemann surfaces, $\Sigma^{\textrm{top}}_{1,1}$ and  $\Sigma^{\textrm{bot}}_{1,1}$.   In the duality wall theory,  the gluing amounts to gauging the diagonal subgroup of the two $SU(2)$ global symmetries. The theory obtained by gluing two $SU(2)$'s in a duality wall theory $T[SU(2),\varphi]$ will be denoted as $\Tr (T[SU(2),\varphi])$. From the above discussion, we found a concrete realization of $T_M$ with $M=\textrm{tori}(\varphi)$ in terms of the duality wall theory. The theory will be denoted as $T^{T[SU(2)]}_{\textrm{tori}(\varphi)}$ ,
\begin{align}
T^{T[SU(2)]}_{\textrm{tori}(\varphi)} = \Tr (T[SU(2),\varphi])\;.
\end{align}
%


\subsection{Tetrahedron decomposition}
In \cite{Dimofte:2011ju}, Dimofte, Gaiotto, Gukov (DGG) proposed a powerful algorithm to construct $T_M$ for a broad class of 3-manifolds $M$.
We briefly review the DGG algorithm here.
The basic building block of a hyperbolic 3-manifold is the ideal tetrahedron $\Delta$.
The corresponding 3d theory, $T_{\Delta}$, is a theory of a free chiral field with a background CS action with level $- \frac{1}{2}$.
If a 3-manifold $M$
can be triangulated by a finite number of tetrahedra,
$T_M$ can be obtained by ``gluing'' copies of $T_\Delta$ accordingly.
Schematically,
\begin{align}
M=\left(\bigcup_{i=1}^N \Delta_i\right)/\sim \quad \Rightarrow  \quad T_{M} = \left(\bigotimes_{i=1}^N T_{\Delta_i}\right)/\sim\;.
\end{align}
\paragraph{Geometry of tetrahedra}
An ideal tetrahedron has six edges and four vertices.  
To the three pairs of diagonally opposite edges, we assign edge parameters $(z,z',z'')$ which are the exponential of complexified dihedral angles $(Z,Z',Z'')$ of the edges.
\begin{align}
z=\exp (Z)\; \;\textrm{with}\;\; Z= (\textrm{torsion})+i (\textrm{angle}) \,,
\quad {\rm etc.}
\end{align}
Using the equivalence between  equation of motion for hyperbolic metrics  and $SL(2)$ flat connections  on a 3-manifold,  these edge variables can be understood in terms of  either  hyperbolic structure or $SL(2)$ flat connection on a tetrahedron. Although latter interpretation is more physically relevant, the former   is more geometrically intuitive.
The hyperbolic structure of an ideal tetrahedron is determined by the edge parameters $(Z,Z',Z'')$ subject to the conditions
\begin{align}
Z+Z'+Z''=\pi i +\frac{\hbar}2 \,, \quad
e^{Z}+e^{-Z'} -1 = 0 \,.
\end{align}
The first condition defines the so-called boundary phase space
with the symplectic form $\O = \frac{1}{i\hbar}dZ \wedge dZ'$.
The second condition defines a Lagrangian submanifold of the boundary phase space.
Due to the first condition, the second condition is invariant under
the cyclic permutation $(z\goto z' \goto z''\goto z)$.

The ideal triangulation requires that
all faces and edges of the tetrahedra should be
glued such that
the resulting manifold is smooth everywhere except
for the cusp due to the vertices of ideal tetrahedra.
In particular, we have the smoothness condition
at each internal edge,
\begin{align}
C_I = \sum_{j=1}^{N} \left( c_{Ij} Z_i + c'_{Ij} Z'_j + c''_{Ij} Z''_j  \right) = 2\pi i  + \hbar \,.
\label{internal-edge}
\end{align}
The coefficients $c_{Ij}$, $c'_{Ij}$, $c''_{Ij}$ take values in $\{ 0, 1, 2 \}$.

When all the edges of $\Delta_i$ are glued,
all but one of the gluing condition \eqref{internal-edge}
give independent constraints, since the sum of all constraints, $\sum_I C_I=(2\pi i+\hbar )N$
trivially follows from $Z_i + Z'_i + Z''_i = \pi i +\frac{\hbar}2$.
The resulting manifold $M$ has a cusp boundary, composed of the truncated
ideal vertices, which is topologically a torus $\mathbb{T}^2$.
The two cycles of the torus, `longitude' and `meridian', describe the boundary phase space of $M$. The logarithmic variables for the two cycles, $V = \log (\ell)$ and $U = \log m$, are some linear combinations of $(Z_i, Z'_i, Z''_i)$.

\paragraph{Ideal triangulation of the mapping torus}

The mapping torus $\textrm{tori}(\varphi) =\Sigma_{1,1}\times_{\varphi} S^1$ is known to be hyperbolic when $|\textrm{Tr}(\varphi)|> 2$. The tetrahedron decomposition of
these  mapping torus is given explicitly in \cite{Gueritaud}.
Any $\varphi \in SL(2,\mathbb{Z})$ satisfying $|\textrm{Tr}(\varphi)|>2$ admits
a unique decomposition of the following form,
\begin{align}
\varphi = F\left( \varphi_N \varphi_{N-1}\ldots \varphi_2 \varphi_1 \right) F^{-1} \,,  \quad \varphi_i = L \textrm{ or } R\;.
\label{LR-decomp}
\end{align}
where we use the following convention for the $SL(2,\IZ)$ generators,
\footnote{Our convention for the $SL(2,\IZ)$ generators is the same as in \cite{Dimofte:2011ju,Dimofte:2011py}
but is the opposite from \cite{Dimofte:2011jd,Gueritaud,Garb:2013}.}
\begin{align}
L=
\begin{pmatrix}
1 & 1 \\ 0 & 1
\end{pmatrix} \,,
\;\;
R=
\begin{pmatrix}
1 & 0 \\ 1 & 1
\end{pmatrix} = T \,,
\;\;
S=
\begin{pmatrix}
0 & -1 \\ 1 & 0
\end{pmatrix} \,;
\;\;
L=S^{-1}T^{-1}S \,,
\;\;
S = L^{-1} R L^{-1} \,.
\label{sl2z-convention}
\end{align}
The overall conjugation by $F$ is immaterial in
the definition of the mapping torus and can be neglected.

According to \cite{Gueritaud}, to each letter $L$ or $R$ appearing in \eqref{LR-decomp} one can associate a tetrahedron with edge parameters $(Z_i, Z^\prime_i, Z^{\prime\prime}_i)$, or equivalently, $(z_i, z'_i, z''_i) = (e^{Z_i}, e^{Z'_i}, e^{Z''_i})$. The index $i$ runs from 1 to $N$ with cyclic identification, $N+1 \sim 1$. $L$ and $R$ generate `flips' on the triangulation of $\Sigma_{1,1}$. Each flip corresponds to a tetrahedron (see figure 2 in \cite{Gueritaud}).
There are $N$ tetrahedra in total and $3N$ edge parameters.   In the mapping torus, all the edges of tetrahedra are glued and there are $N-1$ independent internal edge conditions.  How the internal edges are glued together is determined by the decomposition   \eqref{LR-decomp} of $\varphi$.
Taking account of the $N$ equations $Z_i +Z'_i+Z''_i  = i \pi +\frac{\hbar}{2}$ for each $i$, there are in total $2N-1$ linear constraints on $3N$ edge parameters.
These $2N-1$ constraints can be solved by parameterizing $3N$ edge parameters by $N+1$ variables ($W_i , V$)  as shown in \eqref{tetra-gluing}.

\begin{equation}
   \begin{array}{c|cccc}
   \;   \varphi_{i} \varphi_{i-1} \;   & Z_i & Z'_i & Z''_i & U_i
      \\[5pt]
      \hline
      \\[-10pt]
      LL &  \; i \pi   +\frac{\hbar}2-W_i  \;
         & \;W_i- \displaystyle{\frac{W_{i-1}+W_{i+1}}{2}}  \;
         & \;\displaystyle{\frac{W_{i-1}+W_{i+1}}{2}}  \;
         & \; 0 \;
           \\[15pt]
      RR & \;i \pi   +\frac{\hbar}2-W_i \;
         & \; \displaystyle{\frac{W_{i-1}+W_{i+1}}{2}} \;
         & \; W_i- \displaystyle{\frac{W_{i-1}+W_{i+1}}{2}} \;
         & \; 0  \;
         \\[15pt]
       LR & \; i \pi   +\frac{\hbar}2-W_i \;
          & \; \displaystyle{\frac{W_i +W_{i-1}-W_{i+1}-V-\pi i}{2}} \;
          & \; \displaystyle{\frac{W_i -W_{i-1}+W_{i+1}+V+\pi i}{2}} \;
          & \; -\displaystyle{\frac{W_i}{2}} \;
          \\[15pt]
       RL & \; i \pi   +\frac{\hbar}2-W_i \;
          & \; \displaystyle{\frac{W_i -W_{i-1}+W_{i+1}+V+\pi i}{2}} \;
          & \; \displaystyle{\frac{W_i +W_{i-1}-W_{i+1}-V-\pi i}{2}} \;
          & \; +\displaystyle{\frac{W_i}{2}} \;
          \\[10pt]
   \end{array}
   \label{tetra-gluing}
\end{equation}

\begin{figure}[htbp]
   \centering
   \includegraphics[scale=1.2]{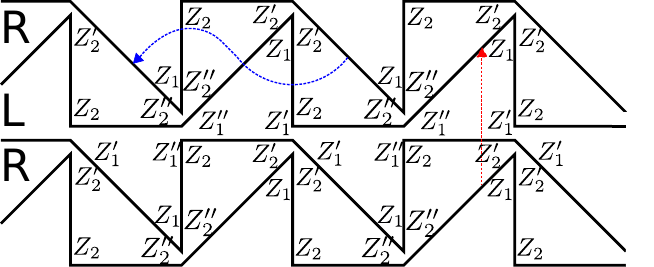} 
   \caption{Triangulation of boundary torus of tori($LR$). Four triangles for each letter, $L$ or $R$, come from  `small'  boundary triangles in  ideal tetrahedron associated to the letter. See figure 3 in \cite{Gueritaud}.}
   \label{fig:LR}
\end{figure}

\noindent
The particular form of the linear combination depends on
the ordering of letters in \eqref{LR-decomp}.
The reparametrization is `local' in the sense that
the expressions for $(Z_i,Z'_i,Z''_i)$ involve $W_i$ and $W_{i\pm 1}$ only.  Among the remaining $N+1$ variables, as was explained below eq.~\eqref{internal-edge}, two are identified as `longitude' and `meridian'.    In \eqref{tetra-gluing},  $\ell=e^V$ is the longitude variable, while the meridian variable $m=e^{U}$ is the product of all $m_i=e^{U_i}$, $U = \sum U_i $.
 As an example, tetrahedron decomposition of a mapping torus with $\varphi=LR$  is given  in figure \ref{fig:LR}.  In the case $\varphi = LR$, it is known that the mapping torus becomes the figure eight knot complement in $S^3$.

 From the figure, the two internal edges are identified as
 \begin{align}
 &C_1 = Z_1+Z_2+2 Z''_1+ 2 Z''_2  = 2\pi i +\hbar\;,\nn
 \\
 &C_2  = Z_1+Z_2 +2 Z_1'+ 2 Z_2' = 2\pi i +\hbar \;. \label{internal edges for LR}
\end{align}
Longitudinal (horizontal  blue line) and meridian (vertical red line) variables from the figure are
\begin{align}
V = Z_2''- Z_2'\;, \quad U= Z_1''-Z_2'\;. \label{long/merd for LR}
\end{align}
Using \eqref{tetra-gluing}, we will parameterize
\begin{align}
&Z_1 = i \pi +\frac{\hbar}2 - W_1\;, \; Z_1' = \frac{W_1 +V+i\pi}{2}\;, \; Z_1''= \frac{W_1- V-i\pi}2\;, \nn
\\
&Z_2 = i \pi +\frac{\hbar}2 - W_2\;, \; Z_2' = \frac{W_2 -V-i\pi}{2}\;, \; Z_2''= \frac{W_1+ V+i\pi}2\;. \nn
\end{align}
In this parametrization, the internal edge conditions \eqref{internal edges for LR} are automatically satisfied. From \eqref{tetra-gluing}, the meridian variable $U$  is $\frac{W_1 - W_2}2$  which is the same as the meridian variable in  \eqref{long/merd for LR} via the above parametrization. But the longitudinal variable  $V$ in \eqref{long/merd for LR} become $V+i \pi $ via the parametrization. The discrepancy $i \pi$ is subtle and the factor can be absorbed by simple redefinition of $V$ in \eqref{tetra-gluing}. As we will see in section \ref{sec: SCI/SL(2,C)}, however, the  variable $V$  in \eqref{tetra-gluing} has a more direct  meaning in the duality wall theory.

\paragraph{Field Theory} We will give a very brief summary of the construction of $T_{M, \Pi}$ from the tetrahedron decomposition data for $M$; see \cite{Dimofte:2011ju,Dimofte:2011py} for details. For each tetrahedron with a polarization choice $\Pi_\Delta$, we take a copy of the 3d theory    $T_{\Delta, \Pi_\Delta}$.  For the polarization $\Pi_{\Delta} =  \Pi_Z$ in which we take $(Z, Z'')$ as (position, momentum),  the
theory (often called ``the tetrahedron theory") is a free $\mathcal{N}=2$ chiral theory with a background CS term for the $U(1)$ global symmetry at CS level $-\half$.
The 3d theory $T_{M,\Pi_M}$ associated a 3-manifold $M= \cup_{i=1}^N \Delta_i/\sim$ and its boundary polarization  $\Pi_M$  can be constructed in three steps. First, we start with a direct product of $N$ tetrahedron theories,
\begin{align}
T_{\{\Delta_i\},\{\Pi_{Z_i}\}} = \otimes_{i=1}^N T_{\Delta_i,\Pi_{Z_i}}
\end{align}
Then, we perform a polarization transformation $\tilde{\Pi}  = g \circ\{\Pi_{Z_i}\}$ such that all internal edges and positions in $\Pi_M$ become position variables in  $\tilde{\Pi}$.  In the field theory, the polarization transformation corresponds to an $Sp(2N,\IZ)$ action\footnote{It generalizes Witten's $SL(2,\mathbb{Z})$ action \cite{Witten:2003ya}.} involving the $U(1)^N$ global symmetries in $T_{\{\Delta_i\},\{\Pi_{Z_i}\}}$.
\begin{align}
T_{\{\Delta_i\},\tilde{\Pi}} = g\circ T_{\{\Delta_i\},\{\Pi_{Z_i}\}} \;.
\end{align}
Finally, we impose the internal edges conditions \eqref{internal-edge} by adding the superpotential $\CW= \sum \CO_I$ which breaks the global $\otimes_I U(1)_I$ symmetries of $T_{\{\Delta_i\},\tilde{\Pi}}$ associated to internal edges $C_I$. This completes the consturction of $T_{M,\Pi}$:
\begin{align}
T_{M,\Pi_M} &= T_{\{\Delta_i\},\tilde{\Pi}} \textrm{ with superpotentail }\CW = \sum \CO_I \;.
\end{align}
Applying this general algorithm to the case $M= \textrm{tori}(\varphi)$ using the tetrahedron decomposition data described above, we give a description for the mapping torus theory. We will denote the mapping torus theory by $T^{\Delta}_{\textrm{tori}(\varphi)}$.

\section{Quantization of  Chern-Simons theory on Riemann Surface} \label{quantum riemann surface}
In this section, we will explain the classical phase space and its quantization
that are relevant to a calculation of Chern-Simons (CS) partition function on a mapping torus/cylinder. Most parts of this section are reviews of known results but are included   to make the paper self-contained. More details can be found, {\it e.g.}, in \cite{Dimofte:2011gm,Dimofte:2011jd}.

For a compact gauge group $G$, the Chern-Simons action on a 3-manifold $M$ is
\begin{align}
I_{CS} = \frac{k}{4\pi }\int_M  \Tr (A\wedge dA + \frac{2}3 A\wedge A \wedge A)\;,
\end{align}
where $k$ is a quantized CS level.
This is one of the most famous example of topological quantum field theory (TQFT).
When $M$ is a mapping torus, 
the CS theory can be canonically quantized on $\Sigma_{1,1}$ regarding the $S^1$ direction as time.
The  phase space $\mathcal{M}_G$\footnote{Since we are mainly focusing  on the case $\Sigma = \Sigma_{1,1}$ throughout the paper, we  simply denote $\CM_{G} (\Sigma)$ (phase space associated Riemann surface $\Sigma$ and gauge group $G$)  by $\CM_G $ when $\Sigma = \Sigma_{1,1}$.  We also sometimes omit the subscript $G$ when it is  obvious in the context.} canonically associated to $\Sigma_{1,1}$ is \cite{Elitzur:1989}
\begin{align}
\mathcal{M}_ G  (\Sigma_{1,1})= \{\textrm{Flat  $G$-connections on $\Sigma_{1,1}$ with fixed puncture holonomy $P$} \}/\sim\;. \label{flat connection moduli space}
\end{align}
where $\sim$ denotes the 
gauge equivalence.    The conjugacy class of gauge holonomy $P$ around a puncture  is fixed as a boundary condition.  The symplectic form $\O_G$ on $\CM_G$ derived from the CS action is
\begin{align}
\Omega_{G} =\frac{k}{4\pi} \int_\Sigma \textrm{Tr}( \delta A \wedge \delta A)\;. \label{symplectic form for unitary}
\end{align}
 One can geometrically quantize the classical phase space and obtain a Hilbert-space $\mathcal{H}_G (\Sigma_{1,1})$. Following an axiom of  general  TQFTs (see, {\it e.g.},  \cite{Atiyah:1988}), the CS partition function on the mapping cylinder
 with gauge group $G$ can be computed  as
\begin{align}
Z_{\Sigma_{1,1}\times_\varphi  I}(x_{\rm bot}, x_{\rm top})   = \langle x_{\rm bot} |\varphi |  x_{\rm top}\rangle \;, \label{CS partition function as matrix element}
\end{align}
which depends on the boundary conditions   $\big{(}x_{\rm bot}, \varphi ( x_{\rm top})\big{)}$ on two boundary $\Sigma_{1,1}$'s .    The CS partition function on the mapping torus  with gauge group $G$ can be computed  as
\begin{align}
Z_{\textrm{tori}(\varphi)}(G)   = \textrm{Tr} (\varphi)\; \textrm{over $\CH_G (\Sigma_{1,1})$}. \label{CS partition function as trace}
\end{align}
Here $\varphi$ is an operator acting on the Hilbert space $\CH_G$ obtained from quantizing  a mapping class group element $\varphi$ which generates a coordinate transformation on $\mathcal{M}_G$.

When the gauge group is  $G= SL(2,\mathbb{C})$, the CS level  becomes  complex variables $t=k+i s$ and $\tilde{t} = k - i s$. $k$ should be  an integer for  consistency of the quantum theory   and unitarity  requires $s \in \mathbb{R}$ or $s \in i \mathbb{R}$ \cite{Witten:91}.
\begin{align}
I_{CS} = \frac{t}{8\pi }\int_M  \Tr (A\wedge dA + \frac{2}3 A\wedge A \wedge A)+\frac{\tilde{t}}{8\pi }\int_M  \Tr (\bar{A}\wedge d\bar{A} + \frac{2}3 \bar{A}\wedge \bar{A} \wedge \bar{ A})\;.
\end{align}
The induced symplectic form from the CS action is
\begin{align}
\Omega_{SL(2,\mathbb{C})} =\frac{t}{8\pi }\int_\Sigma \textrm{Tr}( \delta A \wedge \delta A) + \frac{\tilde{t}}{8\pi} \int_\Sigma \textrm{Tr}(\delta \bar{A} \wedge \delta\bar{A})\;.
\end{align}
In \cite{Dimofte:2011py}, the superconformal index  $I_{T_M}$ for the theory $T_M$ was  claimed to be equivalent to the $SL(2,\mathbb{C})$ CS partition function on $M$ with $t=- \tilde{t} = i s \in i \mathbb{R}$.
\begin{align}
I_{T_M} = Z_M(SL(2,\mathbb{C}))\;, \;\textrm{with  $q:=e^{\hbar } :=e^{\frac{4\pi}s}$}\;. \label{3d-3d dictionary for SCI}
\end{align}
Here, $q$ is a fugacity variable in the superconformal index to be explained in section \ref{sec: SCI/SL(2,C)}.   Using the above map, the  symplectic form  becomes
\begin{align}
\Omega_{SL(2,\mathbb{C})} := \frac{i}{2\hbar}\int_\Sigma \textrm{Tr}(\delta A\wedge \delta A) - \frac{i}{2\hbar}\int_\Sigma \textrm{Tr}(\delta \bar{A}\wedge \delta \bar{A})\;. \label{SL(2,C) symplectic form}
\end{align}

\subsection{Classical phase  space and its coordinates} \label{sec: Classical phase space}
  In this subsection, we will review coordinate systems for the phase space $(\mathcal{M},\O)_{SL(2,\mathbb{C})}$ on once-punctured torus $\Sigma_{1,1}$ and the action of $\varphi \in SL(2,\mathbb{Z})$ on these coordinates. We also consider a phase space $(\CM, \O)^{\rm knot}_{SL(2,\mathbb{C})}$ which  is canonically associated to the cusp boundary $\mathbb{T}^2$ of mapping torus. For mapping cylinder, the boundary phase space is given by, at least locally, $\CM (\Sigma_{1,1})^2 \times \CM^{\rm knot}$.

\paragraph{Loop  coordinates} Generally speaking, the moduli space of flat connections on manifold $M$ with gauge group $G$ is  parametrized by holonomy variables up to conjugation. In other words,
\begin{align}
\CM_G (M) = \textrm{Hom}(\pi_1 (M),   G)/\textrm{conj}. \label{general flat moduli space}
\end{align}
The fundamental group for $\Sigma_{1,1}$ is
\begin{align}
\pi_1 (\Sigma_{1,1}) = \{  A , B,P  | A B A^{-1}B^{-1} = P \}\;.
\end{align}
Here $A,B$ are two cycles of the torus and $P$ denotes the loop around the puncture.
Thus $\CM_{SL(2,\mathbb{C})}(\Sigma_{1,1})$ is given by
\begin{align}
\CM_{SL(2,\mathbb{C})} = \{ A, B\in SL(2,\mathbb{C}) | ABA^{-1}B^{-1} = P \}/\textrm{conj}\;. \label{MSL(2,C) in terms of holonomies}
\end{align}
Conjugacy class of the holonomy $P$ around the puncture is fixed by the following condition
\begin{align}
\textrm{eigenvalues of $P$} = \{\ell , \ell^{-1} \} \;.
\end{align}
 Loop coordinates  $(W,H,D)$ on $\CM_{SL(2,\mathbb{C})}$ are defined as  trace of these holonomy variables
\begin{align}
W = \Tr(A)\; , \quad H= \Tr(B)\; , \quad D = \Tr(AB)\;.
\end{align}
They are not independent and subject to the follwoing constraint:
\begin{align}
 W^2 + H^2 + D^2 - WHD +\ell +\ell^{-1}-2=0 \;.
 \end{align}
  Anticipating close relations to gauge theory observables,
we call the three loop coordinates Wilson loop ($W$), `t Hooft loop ($H$) and dyonic loop ($D$).
\footnote{$(W,H,D)$ here are the same as $(-x, -y, -z)$
in \cite{Dimofte:2011jd}. See eq.(2.13) and
Figure 3 of \cite{Dimofte:2011jd}.}

\begin{figure}[htbp]
   \centering
   \includegraphics[scale=.8]{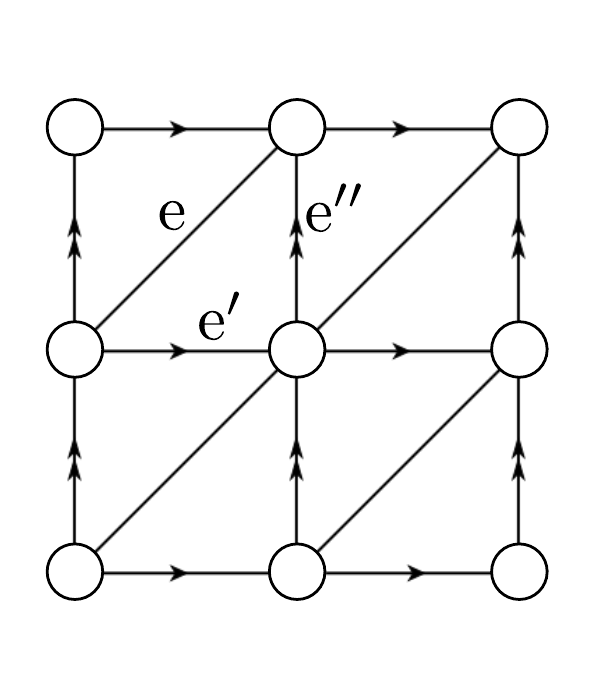} 
   \caption{Triangulation of once-punctured torus, $\Sigma_{1,1}$, represented as $(\mathbb{R}^2 - \mathbb{Z}^2 )/\mathbb{Z}^2$. 
 Each circle in the figure denotes an
   image of the puncture in the covering space $\IR^2$.}
   \label{fig:shear}
\end{figure}

\paragraph{Shear coordinates} The shear coordinates $(\sqrt{t},\sqrt{t'},\sqrt{t''})$ are associated to the three edges appearing in the ideal triangulation of $\Sigma_{1,1}$
depicted in Figure~\ref{fig:shear}.
\noindent
Naively, the shear coordinates represent the partial holonomy eigenvalues along a path  crossing each edge. For more precise description of the shear coordinates,
 see, {\it e.g.}, \cite{Fock:1993pr,Fock:1997}. Following the description in the references, the holonomy $A,B$ can be expressed in terms of the shear coordinates as follows,
\begin{align}
&A(t,t',t'') = \CE(t)\CV \CE^{-1}(t')\CV^{-1} =\left(\begin{array}{cc}\sqrt{tt'} &  \sqrt{t t'} \\ \sqrt{\frac{t'} t} & \frac{1}{\sqrt{tt'}}+\sqrt{\frac{t'}t}\end{array}\right) \nn
\\
&B(t,t',t'') = \CE(t)\CV^{-1} \CE^{-1}(t'')\CV =\left(\begin{array}{cc}\sqrt{\frac{t}{t''}}+ \sqrt{tt''} &  \sqrt{\frac{t}{t''}} \\ \frac{1}{\sqrt{t t''}} &  \frac{1}{\sqrt{t t''}}\end{array}\right) \,,\label{A,B matrices}
\end{align}
where
\begin{align}
\CE(z):= \left(\begin{array}{cc}0 & z^{1/2} \\- z^{-1/2} & 0\end{array}\right)\;, \quad  \CV:= \left(\begin{array}{cc}1 & 1 \\-1 & 0\end{array}\right)\;.
\end{align}
The relation $ABA^{-1}B^{-1} = P$ in \eqref{MSL(2,C) in terms of holonomies} holds provided that
\begin{align}
\sqrt{t}\sqrt{ t'}\sqrt{ t''} = \sqrt{-\ell} \;.
\label{conditions on shears}
\end{align}
This relation states  that products of  partial holonomies around the three edges  give the square root of holonomy around a puncture.
The logarithmic shear variables $(T,T',T'')$ are defined as
\begin{align}
(e^{T/2},e^{T'/2},e^{T''/2}) = (\sqrt{t},\sqrt{t'},\sqrt{t''})\;.
\end{align}
Note that the matrices ($A, B$) in eq.~\eqref{A,B matrices} are invariant under  individual shifts of $T,T',T''$ by $4\pi i $.\footnote{Under  shifts by $2\pi i$, $A, B$ remains invariant up to a sign. Thus for $G=PSL(2,\mathbb{C})  =  SL(2,\mathbb{C})/\langle \pm 1 \rangle$, the periodicity for each of $(T,T',T'')$ is  $2\pi i$.}
\begin{align}
T \sim T+4\pi i\;, \; T' \sim T'+4\pi i \;, \; T''  \sim T''+4\pi i \;.  \label{periodicity in shear}
\end{align}
In the logarithmic variables, the condition \eqref{conditions on shears} become
\begin{align}
T+T'+T''= i \pi +V \;. \label{conditions on logarithmic shears}
\end{align}
From this, we see $V$ is also periodic variable with periodicity $4 \pi i$.
The symplectic form \eqref{SL(2,C) symplectic form} takes a simple form in the shear coordinates. %
\begin{align}
\Omega_{SL(2,\mathbb{C})} &=- \frac{i}{2\hbar } \frac{dt}{t}\wedge \frac{dt'}{t'} + \frac{i}{2\hbar } \frac{d\bar{t}}{\bar{t}}\wedge \frac{d\bar{t'}}{\bar{t'}} =- \frac{i}{2\hbar} dT\wedge dT' + \frac{i}{2\hbar} d\bar{T}\wedge d\bar{T'} \nn
\\
&= (\textrm{cyclic permutation of } T\rightarrow T' \rightarrow T'')\;. \label{symplectic form in shear}
\end{align}
The $SL(2,\IZ)$ generators act on the shear coordinates as follows.
\begin{equation}
\begin{array}{c|ccc}
\;\;\varphi\;\; & \sqrt{t} \mapsto & \sqrt{t'} \mapsto & \;\; \sqrt{t''} \mapsto \;\;
\\[5pt]
\hline
\\[-10pt]
\;\;S\;\; & \;\; \displaystyle{\frac{1}{\sqrt{t}}} \;\; & \;\; \displaystyle{\frac{\sqrt{t''}}{1+ t^{-1}}} \;\; & \sqrt{t'}(1+t)
\\[12pt]
\;\;L\;\; & \;\; \displaystyle{\frac{1}{\sqrt{t''}}} \;\; & \;\;  \displaystyle{\frac{\sqrt{t'}}{1+ t''^{-1}}} \;\; & \sqrt{t}(1+t'')
\\[12pt]
\;\;R\;\; & \;\; \displaystyle{\frac{1}{\sqrt{t'}}} \;\;  & \;\; \displaystyle{\frac{ \sqrt{t}}{1+ t'^{-1}}} \;\; & \sqrt{t''}(1+t')
\end{array}
\label{c-shear-transf}
\end{equation}

\paragraph{Fenchel-Nielson coordinates}

We adopt the modified Fenchel-Nielson (FN) coordinates defined in \cite{Dimofte:2011jd}.
Classically, the FN coordinates $(\lambda,\tau):=(\exp \Lambda, \exp \mathcal{T})$ and the shear coordinates are related by
\begin{align}
\sqrt{t} = \frac{i(\tau^{-1/2} -\tau^{1/2})}{\l -\l^{-1}} \,, \;\;
\sqrt{t'} =  \frac{i(\l -\l^{-1})}{\l^{-1}\t^{1/2} - \t^{-1/2} \l} \,, \;\;
\sqrt{t''} =  \frac{i\sqrt{\ell}(\l^{-1}\t^{1/2} - \t^{-1/2} \l)}{\t^{-1/2} -\t^{1/2}} \,.
\label{shear2FN}
 \end{align}
The FN coordinates are defined up to Weyl-reflection $\mathbb{Z}_2$, whose generator $\sigma$ acts as
\begin{align}
\sigma \; : \; (\Lambda, \mathcal{T}) \; \rightarrow \; (-\Lambda,- \mathcal{T})\;. \label{Weyl-reflection}
\end{align}
Note that the Weyl reflection leaves the shear coordinates
invariant in the relation \eqref{shear2FN}.

\paragraph{Phase space $(\CM,\Omega)^{\textrm{knot}}_{SL(2,\mathbb{C})}$}
So far we  have considered a phase space associated to the Riemann
surface $\Sigma_{1,1}$ in the  $\textrm{tori}(\varphi)$. There is
another important phase space in the computation of the CS partition
function on  $\textrm{tori}(\varphi)$ which is associated to the
cusp boundary $\mathbb{T}^2= \partial (\textrm{tori}(\varphi))$. We
denote this phase space by $\CM^{\textrm{knot}}_{SL(2,\mathbb{C})}
:= \CM_{SL(2,\mathbb{C})}(\mathbb{T}^2)$ and parametrize it by the
(logarithmic) holonomy variables $(U,V)$\footnote{The eigenvalues
for the longitudinal holonomy is given by $(\ell , \ell^{-1}) =(e^V,
e^{-V}). $ On the other hand, for meridian holonomy, the eigenvalues
are   $(m^{\half} ,m^{-\half})=(e^{\frac{U}2}, e^{-\frac{U} 2}) $. }
along the two cycles of $\mathbb{T}^2$. The  symplectic form
$\Omega^{\textrm{knot}}$ is given by
\begin{align}
\Omega^{\textrm{knot}}_{SL(2,\mathbb{C})} = \frac{1}{i\hbar} dU \wedge dV-\frac{1}{i\hbar} d\bar{U} \wedge d\bar{V}\;. \label{Symplectic for knot}
\end{align}

\begin{figure}[htbp]
   \centering
   \includegraphics[scale=.6]{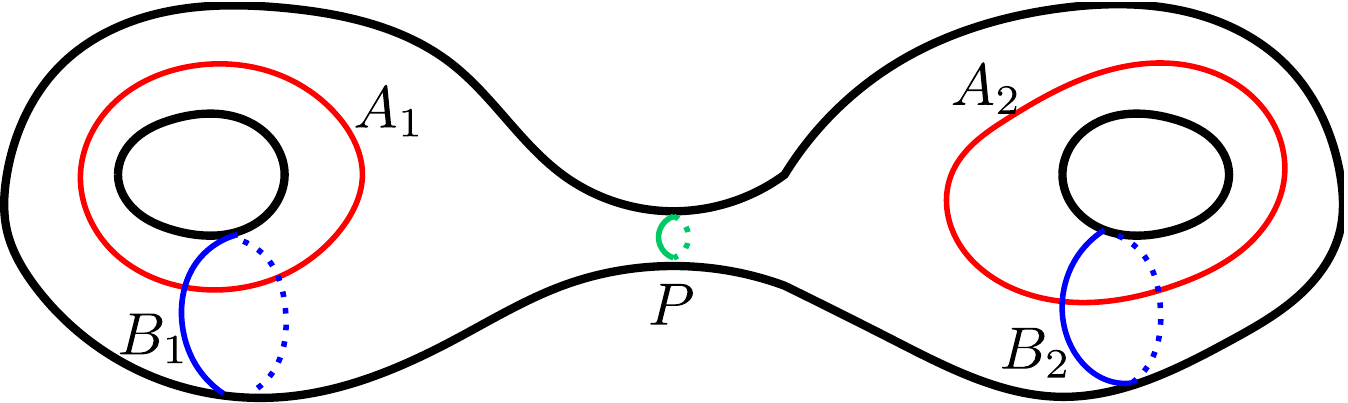} 
   \caption{Generators of $\pi_1 (\Sigma_{2,0})$}
   \label{fig:genus2}
\end{figure}

\paragraph{Boundary phase space of mapping cylinder}
The boundary is the genus two Riemann surface without puncture,  $\partial (\Sigma_{1,1} \times_\varphi I) =  \Sigma_{2,0}$, which can be obtained by gluing punctures in two once-punctured tori. The fundamental group for $\Sigma_{2,0}$ is
\begin{align}
\pi_1 (\Sigma_{2,0}) = \{ A_1, B_1, A_2, B_2 | A_1 B_1 A_1^{-1}B_1^{-1} A_2 B_2 A_2^{-1} B_2^{-1} = 1 \}\;.
\end{align}
The non-trivial cycles $(A_i, B_i)$s on $\Sigma_{2,0}$ are depicted in Figure \ref{fig:genus2}.
\\
Thus, by eq.~\eqref{general flat moduli space}
\begin{align}
\mathcal{M}_{SL(2,\mathbb{C})} (\Sigma_{2,0})= \{ A_1, B_1, A_2, B_2  \in SL(2,\mathbb{C})| A_1 B_1 A_1^{-1}B_1^{-1} A_2 B_2 A_2^{-1} B_2^{-1} = 1 \}/{\rm conj} \;.
\end{align}
The corresponding phase space $\CM(\Sigma_{2,0})$  can be sliced by a constant $P:=A_1 B_1 A_1^{-1}B_1^{-1}=(A_2 B_2 A_2^{-1}B_2^{-1})^{-1}$ surface.
In the  slice, the phase space  locally looks like a $GL(1,\mathbb{C})$ bundle of two copies of $\CM(\Sigma_{1,1})$ with the same $\ell=e^{V}$, an eigenvalue of $P$.  The $GL(1,\mathbb{C})$ fiber direction  corresponds to  opposite $GL(1,\mathbb{C})$ conjugation action  on   a representative elements $(A_1, B_1)$ and $(A_2, B_2)$ of the two $\CM(\Sigma_{1,1})$'s, where  the $GL(1,\mathbb{C})\subset SL(2,\mathbb{C})$ is the stabilizer subgroup of  $P$.
Locally,  the boundary phase space looks like $\CM(\Sigma_{1,1})^2 \times \CM^{\rm knot}$ where $\CM^{\rm knot}$ is parameterized by   conjugacy class of $P$ (or equivalently $V$) and the $GL(1,\mathbb{C})$ fiber direction.
In total, $\textrm{dim}_\mathbb{C} \left[\CM_{SL(2,\mathbb{C})} \big{(}\partial (\Sigma_{1,1}\times_\varphi I )\big{)}\right]  = 6$. As is obvious from the  construction of $\CM \big{(}\partial (\Sigma_{1,1}\times_\varphi I )\big{)}$, the  $V$ variable in $\CM \big{(}\partial (\Sigma_{1,1}\times_\varphi I )\big{)}$ can be  identified with the  puncture variable $V$ in $\CM(\Sigma_{1,1})$. Considering the procedure of gluing two boundary $\Sigma_{1,1}$ components in $\Sigma_{1,1}\times_\varphi I$ to form a mapping torus $\Sigma_{1,1}\times_\varphi S^1$, the $V$ variable  can also be identified with the longitudinal variable $V$ in $\CM^{\rm knot} = \CM \big{(}\partial (\Sigma_{1,1}\times_\varphi S^1) \big{)}$.\footnote{
But there is a subtle difference between the two $V$ variables. The  $V$ in $\CM^{\rm knot}$ is periodic with period $ 2 \pi i $, while the $V$ in $\CM(\Sigma_{1,1})$ has period $4\pi i $.  This discrepancy  may be  due to an  additional $\mathbb{Z}_2$ quotient on $V$ during gluing the two $\Sigma_{1,1}$'s. The $\mathbb{Z}_2$ action could be identified by carefully analyzing how flat connection moduli space changes  during the gluing procedure.}

\subsection{A-polynomial} \label{A-polynomial for mapping torus}

Consider a 3-manifold $M$ with boundary $\partial M$. Obviously, the moduli space of $SL(2,\mathbb{C})$ flat connections on $M$ can be
thought of as a submanifold of the moduli space on the boundary.
\begin{align}
\CM(M) \subset \CM(\partial M)\;.
\end{align}
In fact, the submanifold is Lagrangian with respect to the symplectic form \eqref{symplectic form for unitary}.
For  a knot-complement $M$, the moduli space $\CM(M)$ is  $(\mathbb{C}^*)^2/\mathbb{Z}_2$  parametrized by `longitude' and 'meridian' variable $(\ell, m)$ modulo a Weyl-reflection $(\ell , m)\sim (\ell^{-1},m^{-1})$.  In this case the Lagrangian submanifold  is given by  the vanishing locus of the so-called ``A-polynomial", $A(\ell, m)$ \cite{Gukov:2003na}.  Mapping torus is one  example  of knot-complement.  In this section we will analyze the A-polynomial for mapping torus from two different approaches and show their equivalence.

\paragraph{From tetrahedron decomposition}
In \eqref{tetra-gluing}, we presented a solution to the gluing conditions for the mapping torus by parametrizing the $3N$ edge parameters $(Z_i,Z'_i,Z''_i)$
for the $N$ tetrahedra in terms of $N+1$ parameters ($W_i,V$). The A-polynomial for the mapping torus can be obtained by imposing additional $N$ non-linear constraints, $e^{Z_i}+e^{-Z'_i}  -1 = 0$,
and eliminating all $W_i$'s in favor of $\ell=  e^V$ and $m = \prod e^{U_i}$.

For instance, consider the simplest example, $\varphi= LR$ ($\varphi_2 = L, \varphi_1 = R$).
From \eqref{tetra-gluing}, we find ($w_i := e^{W_i}$)
\begin{align}
&z_1 = -\frac{1}{w_1}\,, \quad
z'_1 = \sqrt{w_1 (-\ell)}\,, \quad
z''_1 = \sqrt{\frac{w_1}{(-\ell)}}\,,
\nonumber
\\
&z_2 = -\frac{1}{w_2}\,, \quad
z'_2 = \sqrt{\frac{w_2}{(-\ell)}}\,, \quad
z''_2 = \sqrt{w_2 (-\ell)}\,,
\qquad
m = \sqrt{\frac{w_2}{w_1}} \,.
\end{align}
For the boundary phase spaces of the two tetrahedra,
the equations for the Lagrangian submanifolds  ($z'+ (z'')^{-1} -1 =0$) are
\begin{align}
\sqrt{w_1(-\ell)} + \sqrt{\frac{(-\ell)}{w_1}} -1=0\,, \quad
\sqrt{\frac{w_2}{(-\ell)}} +\frac{1}{\sqrt{w_2(-\ell)}} -1=0\;.
\end{align}
Eliminating $w_1$ and $w_2$ in favor of $\ell$ and $m$, we obtain the A-polynomial for the mapping torus with $\varphi=LR$:
\begin{equation}
A(\ell,m) = \ell +\ell^{-1} -( m^{-2} - m^{-1} -2 - m + m^2) = 0 \; ,
\end{equation}
which coincide with  A-polynomial for figure eight knot complement; see \cite{Gukov:2003na}.
\paragraph{From  Lagrangian submanifold for mapping cylinder} The Lagrangian submanifold for a mapping torus is a `diagonal' subspace of  a Lagrangian submanifold for the corresponding mapping cylinder \cite{Dimofte:2011jd}. As explained above, the boundary phase space for the mapping cylinder contains a product of two phase spaces associated with two $\Sigma_{1,1}$'s  at the two ends of the interval $I$,
\begin{align}
\mathcal{M}^{\textrm{bot}}(\Sigma_{1,1})\times \mathcal{M}^{\textrm{top}}(\Sigma_{1,1})\subset \CM \big{(}\partial  (\Sigma_{1,1}\times_\varphi I )\big{)}   \;,
\end{align}
where we labelled the two Riemann surface at the two ends of the interval $I$ by `top' and `bot'(bottom). The phase space $\CM^{\textrm{bot,top}}$  can be parametrized by two copies of shear coordinates $(t,t',t'')_{\textrm{bot,top}}$ with common $\ell$. The Lagrangian submanifold  $\CL_{\varphi}$ for mapping cylinder is \cite{Dimofte:2011jd}
\begin{align}
\CL_\varphi = \{ t_{\textrm{top}} - \varphi_*(t)_{\textrm{bot}} =0 , t'_{\textrm{top}} - \varphi_*(t')_{\textrm{bot}} =0,t''_{\textrm{top}} - \varphi_*(t'')_{\textrm{bot}} =0\} \subset \mathcal{M}^{\textrm{bot}} \times \mathcal{M}^{\textrm{top}}\;. \label{Langrangian submanifold for mappying cylinder}
\end{align}
Here $\varphi_*( O)$ is a coordinate for $\CM_{SL(2,\mathbb{C})}$ which is related to $O$ by a mapping class group element $\varphi \in SL(2,\mathbb{Z})$. For example, $\varphi_* (t) =1/t'' $ when $\varphi= L$ as one can see in \eqref{c-shear-transf}. Among three equations between  braces, only two equations are independent and the remaining one is automatically satisfied due to the relation $tt't'' = -\ell$.  Then, the Lagrangian submanifold   for mapping torus is
\begin{align}
\CM ( \textrm{tori}(\varphi)) = \CL_{1} \cap \CL_{\varphi}\;.
\end{align}
$ \CL_{1}$ denotes a diagonal subspace of $\mathcal{M}^{\textrm{bot}} \times \mathcal{M}^{\textrm{top}}$ which can be interpreted as the Langrangian submanifold for mapping cylinder with $\varphi=1$; see eq.~\eqref{Langrangian submanifold for mappying cylinder}.

However, the above construction for the Langrangian submanifold of
$\textrm{tori}(\varphi)$ is incomplete since all the algebraic
equations  depend on $\ell$ but not on $m$. We need an additional
algebraic relation involving $m$. As we saw above, $U=\log m$ is
conjugate to $V = \log \ell$. Anticipating the consequences of
quantization, which promotes $m$ to a shift operator ($m: \ell
\rightarrow q\ell$), we propose the following prescription for $m$;
see \eqref{[m, L] and [m,R]}.
\begin{align}
&m = 1\; , \quad \textrm{for $\varphi=R$} \nn
\\
&m = \frac{\sqrt{t}_{\textrm{top}}}{\sqrt{t}_{\textrm{bot}}}\;, \quad \textrm{for $\varphi=L$}\;. \label{equations involving m}
\end{align}
Consider general $\varphi$ which can be written as product of $L$ and $R$,
\begin{align}
\varphi = \varphi_{N}\cdots \varphi_1\;, \quad \varphi_i = L \textrm{ or }R\;.
\end{align}
For each letter $\varphi_i$, we assign a mapping cylinder $\Sigma_{1,1}\times_{\varphi_i} I$ whose boundary phase space associated to two boundary $\Sigma_{1,1}$'s is parameterized by $(t_i, t'_i, t''_i)_{\textrm{bot},\textrm{top}}$.
To  glue $\Sigma^{\textrm{bot}}_{1,1}$ of $i$-th mapping cylinder with  $\Sigma^{\textrm{top}}_{1,1}$ of $i+1$-th mapping
cylinder, we parametrize
\begin{align}
&(t_i, t'_i, t''_i)_{\textrm{bot}} =(t_{i+1}, t'_{i+1}, t''_{i+1})_{\textrm{top}} := (t_{i+1} , t_{i+1}', t_{i+1} '')\;.
\end{align}
$i$ is a cyclic parameter running from 1 to $N$, $N+1 \sim 1$. In the parametrization,
the Lagrangian in \eqref{Langrangian submanifold for mappying cylinder} becomes
\begin{align}
\CL_{\varphi_i} (t^*_i ,t^*_{i+1}) = \left\{\begin{array}{ll} \{ \sqrt{t_i} - \frac{1}{\sqrt{t''_{i+1}}}=0, \; \sqrt{t''_i}  - \sqrt{t_{i+1}} (1+t''_{i+1}) =0\}\;, &\quad \varphi_i = L \\
\{ \sqrt{t_i } - \frac{1}{\sqrt{t'_{i+1}}}=0, \;\sqrt{t'_i}  - \sqrt{t_{i+1}}/ (1+1/t'_{i+1}) =0\} \;, & \quad \varphi_i = R \label{Lag for L, R-1}
 \end{array}\right.
\end{align}
Equation involving $m$ for mapping cylinder  $\Sigma_{1,1}\times_\varphi I $ is
\begin{align}
&m= \prod_{i=1}^N m_i\;, \quad  \textrm{where} \nn
\\
&m_i  = \left\{\begin{array}{ll}1 \;,&  \quad \textrm{$\varphi_i$ = R}\\  \frac{\sqrt{t_i}}{\sqrt{t_{i+1}}} \;,& \quad \textrm{$\varphi_i$ = L}\end{array}\right. \label{Lag for L, R-2}
\end{align}
Since $m$ act as a shift operator, $m$ for $\varphi_N \ldots \varphi_1$ is   product of $m_i$ for each $\varphi_i$.
Then, the A-polynomial   for tori($\varphi$) is given by solving  all equations in  \eqref{Lag for L, R-1},\eqref{Lag for L, R-2} in terms of $(\ell , m)$.

\paragraph{Equivalence of two approaches}
In the above, we explained two ways of calculating the A-polynomial for the mapping torus.
The equivalence of the two approaches  can be explicitly shown by i) finding a map between  $(w_i)$ variables in the first approach    and   $(t^*_i)$ variables in the second approach and ii)  showing that the equations \eqref{Lag for L, R-1}, \eqref{Lag for L, R-2}
in the second approach are either trivially satisfied or mapped into equations in the first approach. The map we found is
\begin{equation}
\begin{array}{c|ccc}
\;\; \varphi_{i-1} \;\; & t_{i} & t'_{i} & t''_{i}
\\[5pt]
\hline
\\[-10pt]
\; L \; & \;\;\; \displaystyle{\frac{1}{w_{i}}} \;\;
        & \; (-\ell) \displaystyle{\frac{w_{i}}{w_{i-1}}}\; & \; w_{i-1} \;
\\[12pt]
\; R\;  & \;\;\; \displaystyle{\frac{1}{w_{i}}} \;\;
        & \; w_{i-1} \; & \; (-\ell) \displaystyle{\frac{w_{i}}{w_{i-1}}}\;
\end{array}
\label{shear-w}
\end{equation}
From the fact that a single flip ($\varphi_{i-1}$ =$L$ or $R$) generate a tetrahedron whose edges  are  originated from edges in a triangulation of $\Sigma_{1,1}$, identification of    $\frac{1}{w_i}$ (which is $-z_i$) with $t_i$    is understandable.  For $\varphi_i=L$, the map \eqref{shear-w} ensures that the transformation rule
$\sqrt{t_{i}} = 1/\sqrt{t''_{i+1}}$ in \eqref{Lag for L, R-1}
is trivially satisfied. The other transformation rule,
$\sqrt{t''_{i}} = \sqrt{t_{i+1}} (1+t_{i+1}'')$,
is equivalent to the constraint $z''_i + (z_i)^{-1} - 1 = 0$.
Similarly, for $\varphi_i=R$, the transformation rule
$\sqrt{t_{i}} = 1/\sqrt{t'_{i+1}}$ in \eqref{Lag for L, R-1}
is trivially satisfied, while
$\sqrt{t'_{i}} = \sqrt{t_{i+1}}/(1+1/t'_{i+1})$
is equivalent to $z''_i + (z_i)^{-1} - 1 = 0$.
Finally, we note that the meridian variable $m$ can be written
in terms of the shear coordinates as
\begin{align}
m= \frac{ \prod_{\varphi_{i} \varphi_{i-1} =RL} \sqrt{w_i} }{ \prod_{\varphi_{i} \varphi_{i-1} = LR} \sqrt{w_i}}
&= \prod_{\varphi_i =L} \frac{\sqrt{w_{i+1}}}{\sqrt{w_{i}}}
= \prod_{\varphi_{i} =L} \frac{\sqrt{t_{i}}}{\sqrt{t_{i+1}}}\; ,
\end{align}
which is the same as the last equation in \eqref{Lag for L, R-2}.

Thus we have proved the classical equivalence of the two approaches using the A-polynomial. This classical equivalence was already observed in  \cite{Terashima:2011xe}. In section \ref{Z(Tr)=Z(Delta)}, we will prove the equivalence at the quantum level by computing the $SL(2,\mathbb{C})$ CS partition function from the two approaches and confirming an exact agreement. The quantized $(\ell, m)$ variables, denoted as  ($\fl, \sm$), act  as difference operators on the CS partition function. The quantum A-polynomial  $A(\fl ,\sm;q)$ annihilates the CS partition  function. Taking the classical limit, $q\rightarrow 1$, we obtain the A-polynomial, $A(\ell, m)$ discussed in this section.

\subsection{Quantization for $G=SL(2,\IC)$ \label{qsl2c} }

In this section, we will quantize the classical phase spaces $(\mathcal{M},\Omega)_{SL(2,\mathbb{C})}$ and $(\mathcal{M},\Omega)^{\rm knot}_{SL(2,\mathbb{C})}$. By quantization of the classical phase space $(\mathcal{M},\Omega)$, we mean finding the following maps:
\begin{align}
&\textrm{Classical phase space }\mathcal{M}  \quad \rightarrow \quad \textrm{Hilbert-space }\CH \nn
\\
&\textrm{Observables (functions of coordinate $x_i$)  $O(x_i)$ }  \quad  \rightarrow  \quad \textrm{Operators $\sO(\sx_i)$ acting on } \CH \nn
\\
&\textrm{Poisson bracket $\{O_1, O_2  \} = O_3 $} \quad \rightarrow  \quad \textrm{Commutation relation $[\sO_1, \sO_2] = \sO_3$} \;.
\end{align}
The two main ingredients of quantization are the Hilbert-space $\CH$ and operators $\{\sO \}$ acting on it. In this section, we will focus on the latter.  Operators and their commutation relations can be considered before constructing a concrete Hilbert-space.
The construction of the Hilbert-space will be given in section \ref{sec: SCI/SL(2,C)} .

\paragraph{Extended shear operators}
After quantization, the shear coordinates $T,T',T''$ (and its conjugations) become operators
\begin{align}
&(T,T',T'')\rightarrow (\sT_+, \sT'_+,\sT''_+)\;, \nn
\\
&(\bar{T},\bar{T}',\bar{T}'')\rightarrow (\sT_-, \sT'_-,\sT''_-)\;.
\end{align}
The commutation relations for these shear operators follow
from the symplectic form \eqref{symplectic form in shear},
\begin{align}
&[\sT_\pm, \sT'_\pm ] = [\sT'_\pm, \sT''_\pm ] = [\sT''_\pm, \sT_\pm ] =\pm  2\hbar \,.
\\
& [\sT^*_\pm , \sT^{**}_\mp] =0\;, \quad \textrm{for any }*,**\;.
\end{align}
The exponentiated operators $(\sqrt{\st} = e^{\half \sT}, \sqrt{\st'} = e^{\half \sT'}, \sqrt{\st''} = e^{\half \sT''})$ satisfy
\begin{align}
\sqrt{\st} \sqrt{\st'} = q^{ \half} \sqrt{\st'}\sqrt{\st} \,,
\quad
\sqrt{\st'}\sqrt{\st''} = q^\half \sqrt{\st''}\sqrt{\st'} \,,
\quad
\sqrt{\st''} \sqrt{\st'} = q^\half \sqrt{\st'}\sqrt{\st''} \, .
\end{align}
Recall that the quantum parameter $q$ is defined as $q:=e^{\hbar}$.
From here on, we will ignore the subscript $(\pm)$ and all expressions will be for $(+)$ operators unless otherwise stated. The same expressions hold for the $(-)$ operators upon replacing $q$ by $q^{-1}$.

Quantizing \eqref{conditions on shears}, shear operators are subject to the following central constraint.
\begin{align}
\sqrt{\st} \sqrt{\st'} \sqrt{\st''} = q^{ \frac{1}4}\sqrt{-\ell}\;.
\end{align}
 In the logarithmic shear operators, the constraint becomes
\begin{align}
\sT + \sT' + \sT'' = V +i\pi   \;.
\end{align}
In the literature, the variable $V$ is usually regarded as a central charge since they are focusing on the Riemann surface $\Sigma_{1,1}$ itself where $V$ is a fixed parameter.  But, when considering mapping cylinder or torus,  we need  to elevate $V$  to a quantum operator $\sV$ and introduce its conjugate operator $\sU$ satisfying
\begin{align}
[\sU, \sV] = \hbar\;, \label{uv-comm}
\end{align}
since $V$  appears as  a dynamical variable (a coordinate for the
boundary phase space).
Then, the central constraint is promoted to an operator relation
\begin{align}
\sT + \sT' + \sT'' = \sV +i\pi \,.
\label{q-cent}
\end{align}
Since $\sV$ originates from the central constraint, it is natural to
assume that
\begin{align}
[ \sV , \sT ] = [ \sV, \sT' ] = [\sV, \sT'' ] = 0\,.
\end{align}
We cannot require that $\sU$ commute with all three shear
coordinates; that would contradict with \eqref{uv-comm} and
\eqref{q-cent}. The best we can do is to demand that $\sU$ commutes
with two of the shear coordinates and to determine the last
commutator with \eqref{uv-comm} and \eqref{q-cent}. For instance,
\begin{align}
[ \sU, \sT ] = [ \sU, \sT']= 0 \quad \imp \quad [\sU,\sT''] = \hbar \,.
\end{align}
Alternatively, we may choose
\begin{align}
&[ \sU', \sT' ] = [ \sU', \sT'']= 0 \quad \imp \quad [\sU',\sT] = \hbar \,,
\nn \\
&[ \sU'', \sT'' ] = [ \sU'', \sT]= 0 \quad \imp \quad [\sU'',\sT'] = \hbar \,.
\end{align}
The three choices are related by simple canonical transformations,
\begin{align}
\sU' = \sU - \half \sT' \,, \quad
\sU'' = \sU + \half \sT \,.
\label{uuu}
\end{align}
Among these choices, $\sU$ (instead of $\sU', \sU''$)  is identified as quantum counterpart of the classical `meridian` variables $U$ in $\CM^{\textrm{knot}}$. $\sV$ is identified as  a quantum counterpart of `longitudinal' variable $\sV$ in $\CM^{\textrm{knot}}$.

Now, we  can give a more precise meaning to the expression in \eqref{CS partition function as matrix element}, \eqref{CS partition function as trace}. If we consider $V$ as a fixed  parameter, the trace of $\varphi$ will be a function on the  parameter,  $\textrm{Tr}(\varphi)(V) $.  On the other hand,  the CS partition function $Z_{\textrm{tori}}$ should be understood as a wave-function in the Hilbert space $\mathcal{H}^{\textrm{knot}}$ associated to a choice of polarization $\Pi = (\CX, \CP)$,
 \begin{align}
 Z_{\textrm{tori}(\varphi)} (x) =\;_{\Pi}\langle \CX=x |   Z_{\textrm{tori}(\varphi)}\rangle\;, \quad |   Z_{\textrm{tori}(\varphi)}\rangle \in \CH^{\textrm{knot}}\;.
 \end{align}
Here $_\Pi \langle \CX =x|$ denotes a position eigenstate in $\Pi$ polarization.  A more precise statement of \eqref{CS partition function as trace}  is that  the  function $\textrm{Tr}(\varphi) (V)$ is a wave-function in the polarization $\Pi = (\sV, \sU )$: 
\begin{align}
 \textrm{Tr}(\varphi) (V) =\; _{\Pi}\langle \CX = V | Z_{\textrm{tori}(\varphi)}\rangle\;,
\end{align}
Using the quantum operator $\sV$, the above can be written as (in the choice $\Pi = (\sV, \sU )$)
\begin{align}
\textrm{Tr}(\varphi) (V)  &= \; _{\Pi} \langle \CX =  V | \Tr(\varphi)(\sV) | \CP =0 \rangle_\Pi \nn
\\
& = \; _{\Pi} \langle \CX =  V | \Tr(\varphi)(\sV) | \sU =0 \rangle\;.
\end{align}
Thus,  we  find the following polarization-independent expression,
\begin{align}
 |Z_{\textrm{tori}(\varphi)}\rangle =\Tr(\varphi)| \sU =0 \rangle  \in \CH^{\rm knot}\;.
\end{align}
Similarly, for mapping cylinder, the precise meaning of \eqref{CS partition function as matrix element} is
\begin{align}
|Z_{\Sigma_{1,1}\times_\varphi I} \rangle = \varphi |\sU =0\rangle  \in \CH (\Sigma_{1,1}) \otimes \overline{\CH(\Sigma_{1,1})} \otimes \CH^{\rm knot}\;.
\end{align}
Recall that the boundary phase space of mapping cylinder is locally
$  \CM(\Sigma_{1,1})^2 \times \CM^{\rm knot} $ and quantization  of
the phase space gives a Hilbert-space $\CH (\Sigma_{1,1}) \otimes
\overline{\CH(\Sigma_{1,1})} \otimes \CH^{\rm knot}$.
$\overline{\CH}$ denote a dual Hilbert-space and the structure $\CH
\otimes \overline{\CH}$ is due to the two oppositely oriented
boundary Riemann surfaces.

Since $(\sqrt{t},\sqrt{t'},\sqrt{t''}, \ell)$ and  $(\sqrt{\bar{t}},\sqrt{\bar{t}'},\sqrt{\bar{t}''}, \bar{\ell})$ are related by complex conjugation,  it is natural to  define the adjoint of their  quantum counterparts as follows
\begin{align}
(\sqrt{\st}, \sqrt{\st'} ,\sqrt{\st''}, \fl := e^\sV)_\pm^\dagger = (\sqrt{\st}, \sqrt{\st'} ,\sqrt{\st''}, \fl )_\mp\;. \label{adjoin of shear and ell}
\end{align}

\paragraph{SL$(2,\IZ)$ action}

Under the action of the generators of SL$(2,\IZ)$ in \eqref{sl2z-convention},
the transformation rule for the quantum shear coordinates can be summarized
as follows.

\begin{equation}
\begin{array}{c|ccc}
\;\;\varphi\;\; & \sqrt{\st} \mapsto & \sqrt{\st'} \mapsto & \;\; \sqrt{\st''} \mapsto \;\;
\\[5pt]
\hline
\\[-10pt]
\;\;S\;\; & \;\; \displaystyle{\frac{1}{\sqrt{\st}}} \;\; & \;\; \sqrt{\st''}\displaystyle{\frac{1}{1+q^{\half} \st^{-1}}} \;\; & \sqrt{\st'}(1+q^{\half}\st)
\\[12pt]
\;\;L\;\; & \;\; \displaystyle{\frac{1}{\sqrt{\st''}}} \;\; & \;\;  \sqrt{\st'}\displaystyle{\frac{1}{1+q^{\half} \st''^{-1}}} \;\; & \sqrt{\st}(1+q^{\half}\st'')
\\[12pt]
\;\;R\;\; & \;\; \displaystyle{\frac{1}{\sqrt{\st'}}} \;\;  & \;\;  \sqrt{\st}\displaystyle{\frac{1}{1+q^{\half} \st'^{-1}}} \;\; & \sqrt{\st''}(1+q^{\half}\st')
\end{array}
\label{q-shear-transf}
\end{equation}
%
%
After quantization, the $SL(2,\mathbb{Z})$ transformation $\varphi$
becomes an operator acting on  Hilbert-space. Operator $\varphi$ can
be expressed in terms of $(\sT,\sT',\sT'')_\pm$. For $\varphi=\sL$
and $\sR$, the operator is determined by the following conditions
\begin{align}
 & \sL \cdot\sqrt{\st_\pm} =\frac{1}{\sqrt{\st''_{\pm}}} \cdot \sL \;, \quad  \sL\cdot \sqrt{\st''_\pm} -  \sqrt{\st_\pm}(1+q^{\pm 1/2}\st^{\prime\prime}_\pm) \cdot \sL=0 \;, \nonumber
 \\
 & \sR \cdot\sqrt{\st_\pm} =\frac{1}{\sqrt{\st'_{\pm}}} \cdot \sR \;, \quad  \sR\cdot \sqrt{\st''_\pm} -  \sqrt{\st''_\pm}(1+q^{\pm 1/2}\st'_\pm) \cdot \sR=0 \;.
 \label{conditions for L}
\end{align}
The solution for the operator equation can be given as follows
 \begin{align}
 &\sL =\sL_{+} \sL_{-}  = \left( \prod_{r=1}^{ \infty} \frac{ 1+ q^{ r- \half} ( \st''_{+})^{-1} }{ 1+ q^{ r- \half} (\st''_{-})^{-1} } \right)
 \exp \left[ - \frac{1}{ 4 \hbar}  \left( (\sT''_{+} + \sT_{+})^2 - (\sT''_{-} + \sT_{-} )^2 \right) \right] \;,  \nn
 \\
 & \sR = \sR_{+} \sR_{-} = \left(  \prod_{r=1}^{ \infty}  \frac{ 1+ q^{r- \half}  \st'_{-}  }{ 1+ q^{ r- \half} \st'_{+} } \right) \exp \left[  \frac{1}{ 4 \hbar} \left( (\sT'_{+} + \sT_{+})^2 - (\sT'_{-} + \sT_{-} )^2 \right) \right] \;. \label{L,R operators}
 \end{align}
From the solution, we find that ($\sm := e^{\sU} $)
\begin{align}
& \sm_\pm \cdot \sL \cdot \sm_\pm^{-1}  =\frac{1}{\sqrt{\st_\pm}}\cdot \sL \cdot \sqrt{\st_\pm} \nn
\\
&\sm_\pm \cdot \sR \cdot \sm_\pm^{-1} = \sR \;. \label{[m, L] and [m,R]}
\end{align}
We use that  $\sm_\pm  \cdot \st''_\pm \cdot \sm_\pm^{-1} = q^{\pm 1} \st''_\pm$. This give a derivation  of \eqref{equations involving m}. Note that these operators are all unitary; see \eqref{adjoin of shear and ell}. Since all $SL(2,\mathbb{Z})$ elements can be constructed by multiplying $\sL,\sR$ and their inverses, we can easily see that  all $\varphi \in SL(2,\mathbb{Z})$ are unitary operators. As we will see in section \ref{sec: SCI/SL(2,C)}, this unitarity  is closely related to $SL(2,\mathbb{Z})$ duality invariance of the supeconformal index for 4d $\mathcal{N}=2^*$ theory.

\paragraph{Shear vs Fenchel-Nielson}

Quantization of the FN coordinates can be summarized as
\begin{align}
\hat{\tau} = e^{\hat{\mathcal{T}}}\,,
\;\;
\hat{\lambda} = e^{\hat{\Lambda}}
\,, \quad
[ \hat{\mathcal{T}}  , \hat{\Lambda} ] = \hbar
\,, \quad
\hat{\tau} \hat{\lambda} = q \hat{\lambda} \hat{\tau} \,.
\label{FN-op}
\end{align}
The relation to quantum shear coordinates was given in \cite{Dimofte:2011jd}.
\begin{align}
&\sqrt{\st} = \frac{i}{\hat{\l} -\hat{\l}^{-1}}(\hat{\tau}^{-\half} -\hat{\tau}^{\half}) \,,
\nn \\
&\sqrt{\st'} =  \frac{i}{q^{-\frac{1}{4}}\hat{\l}^{-1}\hat{\t}^{\half} - q^{\frac{1}{4}}\hat{\t}^{-\half} \hat{\l}}(\hat{\l} -\hat{\l}^{-1}) \,,
\nn \\
&\sqrt{\st''} =  \frac{i q^{\frac{1}4}\fl^{\half}}{\hat{\t}^{-\half} - \hat{\t}^{\half}}
(q^{-\frac{1}{4}}\hat{\l}^{-1}\hat{\t}^{\half} - q^{\frac{1}{4}}\hat{\t}^{-\half} \hat{\l}) \,.
\label{shear2FN-q}
 \end{align}
\paragraph{Loop vs Fenchel-Nielson}
Quantizing the loop coordinates $(W, H, D)$ yields \cite{Dimofte:2011jd}
\begin{align}
&\sW = \hat{\lambda} +\hat{\lambda}^{-1} \;,  \nn
\\
&\sH= \frac{q^{-\frac{1}{4}} \fl^{\half} \hat{\lambda}  - q^{ \frac{1}{4}} \fl^{-\half} \hat{\lambda}^{-1}  }{\hat{\lambda}-\hat{\lambda}^{-1}} \hat{\tau}^{-\half}
+ \frac{q^{\frac{1}{4}} \fl^{-\half}\hat{\lambda}- q^{-\frac{1}{4}} \fl^{\half} \hat{\lambda}^{-1}}{\hat{\lambda} - \hat{\lambda}^{-1}} \hat{\tau}^{\half}\;. \nn
\\
&\sD= \frac{q^{-\frac{1}{4}} \fl^{\half} \hat{\lambda}  - q^{\frac{1}{4}} \fl^{-\half} \hat{\lambda}^{-1}  }{\hat{\lambda}-\hat{\lambda}^{-1}} q^{-\frac{1}4} \hat{\lambda}\hat{\tau}^{-\half}
+ \frac{q^{\frac{1}{4}} \fl^{-\half}\hat{\lambda}- q^{-\frac{1}{4}} \fl^{\half} \hat{\lambda}^{-1}}{\hat{\lambda} - \hat{\lambda}^{-1}}q^{-\frac{1}4} \hat{\lambda}^{-1} \hat{\tau}^{\half}\;. \label{loop2FN-q}
\end{align}
%



\section{Superconformal index/$SL(2,\mathbb{C})$ CS partition function} \label{sec: SCI/SL(2,C)}
 The superconformal index for 3d SCFTs with global symmetry $U(1)^N$ is defined as \cite{Kim:2009wb,Imamura:2011su,Krattenthaler:2011da,Kapustin:2011jm}
\begin{align}
I(q, m_i , u_i ) = \textrm{Tr} (-1)^F q^{\half R+j_3} \prod_{i=1}^N u_i^{H_i}
\end{align}
where the trace is taken over Hilbert-space $\mathcal{H}_{\{m_i\}}$ on $S^2$, where  background monopole
fluxes $\{m_i\}$ coupled to global symmetries $U(1)^N$ are turned on. $R$ and $j_3$ denote  $U(1)$ R-charge and spin on  $S^2$ respectively.
$\{u_i\}$ are  fugacity variables for the $U(1)^N$ whose generators are denoted by $\{ H_i\}$.  It is often useful to
express the index in a charge basis $(m_i, e_i)$ instead of $(m_i, u_i)$,
\begin{align}
I(m_i , u_i ) = \sum_{e_i} I(m_i, e_i) u_i^{e_i} \;.
\end{align}
In the charge basis, the $Sp(2N,\mathbb{Z})$ transformation \cite{Witten:2003ya} on 3d SCFTs with $U(1)^N$ global symmetry acts linearly. For two 3d SCFTs,  $\mathcal{T}$ and $g\cdot \mathcal{T}$, related by $g\in Sp(2N,\mathbb{Z})$, the generalized indices for the two theories are related as \cite{Dimofte:2011py}
\begin{align}
I_{g\cdot \mathcal{T}} (m,e) = I_{\mathcal{T}} (g^{-1}\cdot (m,e))\;.
\end{align}

In section \ref{section : two routes}, we gave two alternative descriptions for mapping torus theories which we
denote by $T^{\textrm{T[SU(2)]}}_{\textrm{tori}(\varphi)}$ and $T^{\Delta}_{\textrm{tori}(\varphi)}$.
The two descriptions give seemingly different expressions for the index.
We will denote the index for $T^{\textrm{T[SU(2)]}}_{\textrm{tori}(\varphi)}$ and $T^{\Delta}_{\textrm{tori}(\varphi)}$  by $I^{\textrm{T[SU(2)]}}_{\textrm{tori}(
\varphi)}$ and   $I^{\Delta}_{\textrm{tori}(\varphi)}$, respectively. By proving
\begin{align}
I^{\textrm{T[SU(2)]}}_{\textrm{tori}(
\varphi)} = I^{\Delta}_{\textrm{tori}(\varphi)} \label{equality for two mapping tori index}
\end{align}
for general $\varphi$ with $|\Tr (\varphi)|>2$,  we will confirm
the equivalence of two descriptions   at the quantum level. The 3d-3d
correspondence \cite{Dimofte:2011py} predicts that the index  is the
same as the $SL(2,\mathbb{C})$ CS partition function on the mapping
torus, $Z_{\textrm{tori}(\varphi)}(SL(2,\mathbb{C}))$.
\begin{align}
I_{\textrm{tori}(\varphi)} =  Z_{\textrm{tori}(\varphi)}(SL(2,\mathbb{C}))\;.
\end{align}
There are also two  independent ways of calculating $
Z_{\textrm{tori}(\varphi)}(SL(2,\mathbb{C}))$ depending on the way
of viewing the 3-manifold tori($\varphi$). Viewing tori($\varphi$)
as a 3-manifold obtained by gluing  tetrahedra, the CS partition can
be calculated using a state integral model developed in
\cite{Dimofte:2011gm}. Let's denote the CS partition function
obtained in this way by $Z^{\Delta}_{\textrm{tori}(\varphi )}
(SL(2,\mathbb{C}))$. It was shown in \cite{Dimofte:2011py} that the
$SL(2,\mathbb{C})$ CS partition function on $M$ obtained from the
state integral model is always  the same  as superconformal index
for $T_M$ theory obtained from gluing tetrahedron theories,
$\CT_{\Delta}$'s.
  Thus, it is already proven that
\begin{align}
Z^{\Delta}_{\textrm{tori}(\varphi)} (SL(2,\mathbb{C})) =  I^{\Delta}_{\textrm{tori}(\varphi)}\;. \label{ZDelta=IDelta}
\end{align}
Another way of calculating the $SL(2,\mathbb{C})$ CS partition function is using the canonical quantization of the CS theory
on $\Sigma_{1,1}$ viewing the $S^1$ direction in tori($\varphi$) as a time direction.  The partition function obtained in
this approach will be denoted as $Z^{\textrm{Tr}(\varphi)}_{\textrm{tori}(\varphi)} (SL(2,\mathbb{C}))$.  As  mentioned
in section \ref{quantum riemann surface},
\begin{align}
Z^{\textrm{Tr}(\varphi)}_{\textrm{tori}(\varphi)} (SL(2,\mathbb{C}))  = \textrm{Tr} (\varphi) \; \textrm{on $\CH_{SL(2,\mathbb{C})}$} \;. \label{CS partition from trace}
\end{align}
We will  show that two approaches are equivalent
\begin{align}
Z^{\textrm{Tr}(\varphi)}_{\textrm{tori}(\varphi)} (SL(2,\mathbb{C})) = Z^{\Delta}_{\textrm{tori}(\varphi)} (SL(2,\mathbb{C}))\;,
\end{align}
by expressing the trace  in \eqref{CS partition from trace} using  a basis of $\CH _{SL(2,\mathbb{C})}$ called `SR basis'. On the other hand, by expressing the trace in a basis called `FN basis', we will show that
\begin{align}
Z^{\textrm{Tr}(\varphi)}_{\textrm{tori}(\varphi)} (SL(2,\mathbb{C})) =  I^{\textrm{T[SU(2)]}}_{\textrm{tori}(
\varphi)} \;.
\end{align}
Since the trace is independent of basis choice, the proof of
\eqref{equality for two mapping tori index} now follows from
the known proof of \eqref{ZDelta=IDelta}. Further, by showing that the matrix element
of $\varphi$ in the FN basis is the same as the superconformal index
for duality wall theory $T[SU(2),\varphi]$ we also confirm the 3d-3d
dictionary \eqref{3d-3d dictionary for SCI} for mapping cylinder.

\subsection{Duality wall theory : $I^{T[SU(2)]}_{{\rm tori}(\varphi)}$} \label{index-wall}
In this section, we will calculate the superconformal indices for duality wall theories $T[SU(2),\varphi]$ and mapping torus
theories $T^{T[SU(2)]}_{\textrm{tori}(\varphi)} =\Tr (T[SU(2),\varphi]) $.
First, consider the case $\varphi= S$. The $T[SU(2)]\equiv T[SU(2),S]$ theory is
explained in detail in section \ref{duality wall theory} and summarized in table \ref{tsu(2)-charge}.
The generalized superconformal index for the theory can be obtained by the using general prescriptions in \cite{Kim:2009wb,Imamura:2011su,Kapustin:2011jm},\footnote{Throughout this paper, the Cantor integral $\oint \frac{du}{2\pi i u} I(u)$ will be interpreted as  picking up the coefficient of $u^{0}$ by regarding $I(u)$ as an element in a ring $\mathbb{Z}[u^{1/p},u^{-1/p}]$ with a positive integer $p$. }
\begin{align}
&I_{\varphi=S}(m_b,u_b,m_t, u_t;m_\eta,u_\eta)
=\sum_{m_s\in \mathbb{Z}+m_b+\half m_\eta}\oint \frac{du_s}{2\pi i u_s} I^{(0)} I^{(1)} \;. \label{index formula-1}
\end{align}
Our notations for the fugacity and flux variables appearing in the index are summarized in Table~\ref{tsu(2)-2}.
\begin{table}[htbp]
   \centering
   \begin{tabular}{@{} l|c|c|c|c @{}} 
      \toprule

       & $q_1\;\;\; q_2 \;\;\;q_3 \;\;\;q_4\;\;\;\phi_0$  & fugacity & flux  \\
      \midrule
      $U(1)_{\rm gauge}$ &$1\;\;\;\; 1 \;-1\; -1\;\;\;0$  & $u_s$ &  $m_s$ \\
       $U(1)_{\rm bot}$&$1\;-1 \;\;\;1\;-1\;\;\;0$    & $u_b$& $m_b$ & \\
        $U(1)_{\rm punct}$& $\;\half \;\;\;\;\half \;\;\;\;\half \;\;\;\;\half \;-1$ &  $u_\eta$ & $m_\eta$  &  \\
         $U(1)_{\rm top}$ & $ 0\;\;\;\;0\;\;\;\;0\;\;\;\;0\;\;\;\;0$ & $ u_t$ & $m_t$  & \\
      \bottomrule
   \end{tabular}
   \caption{Fugacity, background monopole flux variables for symmetries in $T[SU(2)]$ theory }
   \label{tsu(2)-2}
\end{table}

\noindent
$I^{(1)}$ is the Plethystic exponential (PE) of the single letter indices from chiral-multiplets
\begin{align}
I^{(1)} &=\textrm{PE}[\sum_{\epsilon_1,\epsilon_2 =\pm 1}\frac{  ( u_b^{\epsilon_1}  u_\eta^\half u_s^{\epsilon_2} q^{1/4} -u_b^{-\epsilon_1}  u_\eta^{-\half} u_s^{-\epsilon_2} q^{3/4})q^{\frac{1}{2}|\epsilon_1 m_b+\half m_\eta+ \epsilon_2 m_s|} }{1-q}+\frac{q^{\frac{1}{2}+\half |m_\eta|}}{1-q}(u_\eta^{-1}-u_\eta)]\;, \nonumber
\\
&=
\prod_{ l =0}^{ \infty}
\left(
\prod_{ \epsilon_1, \epsilon_2 = \pm 1}
\frac{ ( 1- ( u_b^{ \epsilon_1} u_\eta^\half u_s^{ \epsilon_2})^{-1} q^{ \frac{3}{4} + \frac{1}{2} | \epsilon_1 m_b + \half m_{ \eta} + \epsilon_2 m_s|+l} )}
{(1-( u_b^{ \epsilon_1} u_\eta^\half u_s^{ \epsilon_2}) q^{ \frac{1}{4} + \frac{1}{2} | \epsilon_1 m_b +\half m_{ \eta} + \epsilon_2 m_s|+l} )}
\right)
\frac{
( 1- u_\eta q^{ \frac{1}{2} + \half |m_{ \eta}| + l })
}{(1- u_\eta^{-1} q^{ \frac{1}{2} + \half  |m_{ \eta}| + l } )}
\, ,
\nonumber
\end{align}
where
\begin{align}
\textrm{PE}[f(q, u_s, u_\eta,u_b)]  = \exp \left[ \sum_{n=1}^\infty \frac{1}n f(q^n, u_s^n, u_\eta^n, u_b^n) \right] \,.
\end{align}
The above index  can be rewritten as in \eqref{TSU2 index using tetrahedron index} which is free from absolute values of magnetic fluxes.  We assign conformal dimension $\Delta$ for chirals  as follows
\begin{align}
\Delta (q_i)= \frac{1}2\,, \quad \Delta(\phi_0) =1\,,
\end{align}
which is canonical for 3d $\mathcal{N}=4$ SCFTs.\footnote{However,
general $R$ charge assignments can be easily incorporated.}
$I^{(0)}$ collects all contributions from classical action and zero-point shifts
\begin{equation}
I^{(0)} = u_t^{2m_s} u_s^{2m_t} q^{\epsilon_0} u_\eta^{F_{\eta,0}} u_b^{F_{b,0}} u_s^{F_{s,0}}(-1)^{{\rm sgn}}\;.\nonumber
\end{equation}
The $u_t^{2s} u_s^{2m_t}$ term originates from the BF-term which couples background gauge field for $U(1)_{\rm top}$ to the field strength of $U(1)_{\rm gauge}$.  The zero-point contributions, $ \epsilon_0$, $F_{\eta,0}, F_{b,0}, F_{ s,0}$ are given by \cite{Imamura:2011su}
\begin{align}
&\epsilon_0 :=\frac{1}8 \sum_{\epsilon_1, \epsilon_2 = \pm} (|\epsilon_1 m_b+\half m_\eta+\epsilon_2 m_s|)\;,\nonumber
\\
&F_{\eta,0} := \half |m_\eta|-\frac{1}4 \sum_{\epsilon_1, \epsilon_2 = \pm} (|\epsilon_1 m_b+\half m_\eta+\epsilon_2 m_s|)\;,\nonumber
\\
&F_{b,0} :=-\frac{1}2 \sum_{\epsilon_1, \epsilon_2 = \pm} \epsilon_1 (|\epsilon_1 m_b+\half m_\eta+\epsilon_2 m_s|)\;,\nonumber
\\
&F_{s,0} :=-\frac{1}2 \sum_{\epsilon_1, \epsilon_2 = \pm} \epsilon_2 (|\epsilon_1 m_b+\half m_\eta+\epsilon_2 m_s|) \;.\nonumber
\end{align}
The subtle sign factor
\begin{align}
{\rm sgn} := 2m_b+\half (m_\eta+ |m_\eta|)+\sum_{\epsilon_1, \epsilon_2 = \pm}\frac{1}2 (|\epsilon_1 m_b+\half m_\eta+\epsilon_2 m_s|)\;,
\end{align}
is chosen for the index to satisfy the so-called self-mirror property \cite{Tong:2000ky, Hosomichi:2010vh, Gang:2012ff}.%
\begin{align}
I_{\varphi=S}(m_b,u_b,m_t,u_t;m_\eta,u_\eta) =I_{\varphi=S}(m_t,u_t,m_b,u_b;-m_\eta,u_\eta^{-1})\;.
\end{align}
 This sign factor (or more generally phase factor) always appears in the computation of 3d generalized index and lens
 space partition function \cite{Imamura:2012rq}. To the best our knowledge,  a systematic method for fixing the subtlety has not been developed
 yet, though it has survived numerous tests. 

In our normalization, the background monopole charges $(m_b, m_t)$ are half-integers and $(m_\eta)$ is an integer.
\begin{align}
&2m_b, 2m_t, m_\eta \in \mathbb{Z} \;. \nonumber
\end{align}
Note that the summation range, $ m_s \in {\mathbb Z} + m_b +\half  m_{ \eta}$, is to satisfy the following Dirac quantization conditions,
\begin{align}
& \pm m_b + \half m_{ \eta} \pm  m_s \ \in {\mathbb Z}\; .
\end{align}

Multiplying  $\varphi \in SL(2, \mathbb{Z})$ by $T^{k}$ amounts to turning on a Chern-Simons term with level $k$ for the background gauge field of $U(1)_{\rm bot}$ or $U(1)_{\rm top}$. It  affects the index as follows
\begin{align}
I_{T^{k} \cdot \varphi} =( u_b)^{2km_b} I_{\varphi}\;, \quad I_{\varphi \cdot T^{k}} = (u_t)^{2k m_t} I_{\varphi}\;. \label{index formula-2}
\end{align}
Here the phase factors $(u_b)^{2km_b}$ and $(u_t)^{2k m_t}$ come from the classical action for the added CS term. The theory $T[SU(2),\varphi_2\cdot \varphi_1]$ is obtained by gauging the diagonal subgroup of $SU(2)_{\textrm{top}}$ from $T[SU(2),\varphi_2]$ and  $SU(2)_{\textrm{bot}}$ from $T[SU(2),\varphi_1]$. Accordingly, the index is glued by $\odot$ operation defined below under the $SL(2,\mathbb{Z})$ multiplication
\begin{align}
&I_{\varphi_2 \cdot \varphi_1}(m_b,u_b,m_t,u_t;m_\eta,u_\eta)  =(I_{\varphi_2} \odot I_{\varphi_1})(m_b,u_b,m_t,u_t;m_\eta,u_\eta) \nn
\\
&  := \sum_{n\in \frac{1}2 \mathbb{Z}} \oint [dv]_n I_{\varphi_2}(m_b,u_b,n,v;m_\eta,u_\eta)  I_{\varphi_1}(n,v,m_t,u_t;m_\eta,u_\eta)\;. \label{odot gluing rule in T[SU(2)] theories}
\end{align}
The integration measure $[dv]_m$ comes from the index for a $\mathcal{N}=2$ vector multiplet for the diagonal subgroup of the two $SU(2)$'s,
\begin{align}
 &\oint [dv]_n := \oint \frac{dv}{2\pi i v} \Delta(n,v)\;,  \nonumber
 \\
 &\Delta(n,v):=  \frac{1}2 (q^{\frac{n}2} v - q^{-\frac{n}2} v^{-1})(q^{\frac{n}2} v^{-1} - q^{-\frac{n}2} v) \;. \label{su(2) measure}
\end{align}
In terms of the charge basis for $U(1)_{\rm punct}$,  the  operation $\odot$ is given by
\begin{align}
&(I_{\varphi_2} \odot I_{\varphi_1})(m_b,u_b,m_t,u_t;m_\eta,e_\eta) \nn
\\
&  = \sum_{e_{\eta,1}e_{\eta,2}}  \sum_{n} \oint [dv]_n  \delta(e_\eta -\sum_{k=1}^2 e_{\eta,k})I_{\varphi_2}(m_b,u_b,n,v;m_\eta,e_{\eta,1})  I_{\varphi_1}(n,v,m_t,u_t;m_\eta,e_{\eta,2})\;.
 \label{index formula-3}
\end{align}
Using the $\odot$ operation, one can write
\begin{align}
&I_{T^{k_1} \cdot \varphi  \cdot T^{k_2}} =\overbrace{ I_{T} \odot \ldots I_{T}}^{k_1}\odot I_{\varphi} \odot \overbrace{ I_{T} \odot \ldots I_{T} }^{k_2}\;, \; \textrm{where} \nn
\\
&I_{\varphi=T}(m_b, u_b, m_t, u_t ;m_\eta,u_\eta) = u_b^{2m_b}\frac{ \delta_{m_b, m_t}\delta(u_b-u_t)}{\Delta(m_b, u_b)} \;. \label{mapping cylinder index for T}
\end{align}
Using  prescriptions in eq.~\eqref{index formula-1}, \eqref{index formula-2} and \eqref{index formula-3}, one can calculate the index $I_{\varphi}$ for general $T[SU(2),\varphi]$. The $SL(2,\mathbb{Z})$ structure is encoded  in the index. We find that
\begin{align}
I_{S^2 \cdot \varphi} = (-1)^{ m_\eta} I_{\varphi} \;, \quad I_{(ST)^3 \cdot \varphi} = u_\eta^{\half m_\eta}  I_{\varphi} \;, \label{SL(2,Z) in index}
\end{align}
by calculating the index in $q$-series expansions. In appendix \ref{T[SU(2)] classical}, these $SL(2,\mathbb{Z})$ structure will be analyzed by studying  classical difference equations for $I_{\varphi}$. The factor $u_\eta^{\half m_\eta}$ can be interpreted  as a CS term for background gauge field coupled to $U(1)_{\rm punct}$.

Finally, the index $I^{T[SU(2)]}_{\textrm{tori}(\varphi)} (m_\eta, u_\eta)$ for the mapping torus theory $\Tr(T[SU(2),\varphi])$ is given by
\begin{align}
I^{\textrm{T[SU(2)]}}_{\textrm{tori}(\varphi)} (m_\eta, u_\eta) =\sum_{n\in \frac{1}2 \mathbb{Z}} \oint [du]_n I_{\varphi}(m,u,m,u;m_\eta,u_\eta)\;. \label{mapping torus index from duality wall}
\end{align}
For $\varphi = S$ the mapping torus index becomes extremely simple (checked in $q$ expansion)
\begin{align}
I^{\textrm{T[SU(2)]}}_{\textrm{tori}(S)} (m_\eta, u_\eta) = \left\{\begin{array}{ll} (-1)^{\half m_\eta}\;, & \quad m_\eta\in 2\mathbb{Z} \\0 \;,  &  \quad  m_\eta \in 2\mathbb{Z}+1\end{array}\right.
\end{align}
This may imply that the corresponding  $T_{\textrm{tori}(S)}$ theory  is a topological theory. See  \cite{Ganor:2010md,Ganor:2012mu} for related discussion.
The mapping torus index is also simple for $\varphi = R^{-1}L= -T^{-1}ST^{-1}S$,
\begin{align}
I^{\textrm{T[SU(2)]}}_{\textrm{tori}(R^{-1}L)} (m_\eta, e_\eta) =  (-1)^{e_\eta}\delta_{m_\eta ,-3 e_\eta}\;.
\end{align}
Actually the mapping torus with $\varphi  = R^{-1}L$ is trefoil knot complement in $S^3$ \cite{Garb:2013} and the above index is identical to the corresponding index  in \cite{Dimofte:2011py} computed by gluing two tetrahedron indices,   up to  a  polarization  difference $(\CX,\CP)_{\rm here} = (-\CP + \pi i , \CX)_{\rm there}$. Refer to section \ref{index-tetra} for how polarization change affects the index.  For $\varphi= LR = ST^{-1}S^{-1}T$, the mapping torus  index is
\begin{align}
&I^{\textrm{T[SU(2)]}}_{\textrm{tori}(LR)}( m_\eta  = 0,u_\eta)= 1-2q + 2 (u_\eta +\frac{1}{u_\eta}) q^{3/2} - 3q^2 + (2 +u_\eta^2+\frac{1}{u_\eta^2})q^3
+\cdots\;,
\nn
\\
&I^{\textrm{T[SU(2)]}}_{\textrm{tori}(LR)}( m_\eta  =\pm 1,u_\eta)= -(2-u_\eta - \frac{1}{u_\eta}) q -(2- u_\eta - \frac{1}{u_\eta}) q^2
+\cdots\;, \nn
\\
&I^{\textrm{T[SU(2)]}}_{\textrm{tori}(LR)}(m_\eta  =\pm 2,u_\eta)= (u_\eta + \frac{1}{u_\eta}) q^{1/2} -q -q^2 -(u_\eta + \frac{1}{u_\eta}) q^{5/2} \cdots\;, \nn
\\
&I^{\textrm{T[SU(2)]}}_{\textrm{tori}(LR)}(m_\eta  =\pm 3,u_\eta)= -(u_\eta +\frac{1}{u_\eta}-u_\eta^2 - \frac{1}{u_\eta^2})q^2 + \cdots\;, \nn
\\
&I^{\textrm{T[SU(2)]}}_{\textrm{tori}(LR)}(m_\eta  =\pm 4,u_\eta)= (u_\eta^2 +\frac{1}{u_\eta^2})q-(u_\eta +\frac{1}{u_\eta})q^{5/2} + q^3 + \cdots\;.
 \label{LR index}
\end{align}
Note that only integer powers of $u_\eta$ appear in the mapping torus indices. This is  true for any mapping torus index with
$\varphi$ being products of $L$ and $R$.\footnote{Due to the second property in \eqref{SL(2,Z) in index},  we need to specify decomposition of $L, R$ in terms of $S,T$ for the index computation. If not, the index is defined  only  up to an overall factor $u_\eta^{\half m_\eta \mathbb{Z}}$.  Throughout this paper we use the simplest decomposition, $R=T$ (instead of   $T(ST)^{3n}$ with $n \neq 0$) and $L = S^{-1}T^{-1}S$, in the index computation. In this choice, $I^{T[SU(2)]}_{\textrm{tori}(\varphi)}$ and  $I^{\Delta}_{\textrm{tori}(\varphi)}$ are the same  without any polarization change as we will see in section \ref{Z(Tr)=Z(T[SU(2)])}. }
On the other hand, for mapping cylinder indices, half-integer powers of $u_\eta$ may appear. For example, the index for $T[SU(2),\varphi = LR]$ is
\begin{align}
 \frac{u_b+ u_b u_t^2 }{u_t u^\half_\eta} q^{\frac{1}4} + \frac{u_b (1+u_t^2)(1+u_t^4)}{u_t^3 u_\eta^{\frac{3}2}} q^{\frac{3}4}+\ldots
\end{align}
when $(m_b , m_t , m_\eta)= (\half, 0,0)$. The disappearance of $u_\eta^{\mathbb{Z}+\half}$ in  mapping torus index is closely related to the fact that periodicity of longitudinal variable $V$ become half after making mapping torus from mapping cylinder, as mentioned in the last paragraph in section \ref{sec: Classical phase space}. We will come back to this point in section \ref{index : Hilbert-space} during the construction of $\CH^{\textrm{knot}}$.

\subsection{Tetrahedron decomposition : $I^{\Delta}_{{\rm tori}(\varphi)}$} \label{index-tetra}

In this section, we will explain how to calculate the superconformal index $I^{\Delta}_{{\rm tori}(\varphi)}$ for the theory
$T^\Delta_{\textrm{tori}(\varphi)}$. In section \ref{section : two
routes}, we briefly reviewed the construction of $T_M$ from the
tetrahedron decomposition data of $M$. Using this construction
and well-developed algorithms
\cite{Kim:2009wb,Imamura:2011su,Kapustin:2011jm} for calculating  the
superconformal indices for general 3d theories, we can calculate the
superconformal indices for $T_M$. The procedure of
calculating indices  from  tetrahedron gluing is well explained in \cite{Dimofte:2011py} and the
procedure is shown to be equivalent to the procedure of calculating
$SL(2,\mathbb{C})$ CS partition function  using the state integral model
developed \cite{Dimofte:2011gm}. First, we will  review the
procedure of calculating the indices for general $T_M$ from
the tetrahedron gluing data for $M$. Then, we will apply the general
procedure to $M={\rm tori}(\varphi)$ with $|\Tr (\varphi)|>2$.

Suppose that $M$ can be decomposed into $N$ tetrahedra $\{ \Delta_i \}$ $(i=1,\ldots, N)$ with proper gluing conditions $\sim$, $M=(\bigcup_i \Delta_i)/\sim$. For each tetrahedron $\Delta_i$
we  assign a ``wave-function" (index) $\CI^{\Pi_i}_{\Delta}(m_i,e_i)$ which depends on the  choice of polarization $\Pi_i =(\mathcal{X}_i , \mathcal{P}_i)$ of the tetrahedron's boundary phase-space $\CM(\partial \Delta_i)$. Recall that the phase space $\CM(\partial \Delta)$ is a 2 dimensional  space represented by three edge parameters $Z,Z',Z''$ with the constraint.
\begin{align}
Z+Z'+Z'' = \pi i +\frac{\hbar}2 \;.
\end{align}
The symplectic form on the phase space is
\begin{align}
\frac{1}{i\hbar} dZ\wedge dZ' - \frac{1}{i\hbar} d\bar{Z}\wedge d\bar{Z}'= (\textrm{cyclic permutation in $Z, Z', Z''$})\;.
\end{align}
For the  choice of polarization $\Pi=\Pi_Z:=(Z,Z'')$, the index  is given as \cite{Dimofte:2011py} (see also \cite{Garb:2012})
\begin{align}
\CI^{\Pi_Z}_{\Delta}(m, \zeta)=  \sum_{e\in \mathbb{Z}} \CI^{\Pi_Z}_{\Delta}(m,e) \zeta^e =\prod_{r=0}^\infty \frac{1-q^{r-\frac{m}2+1} \zeta^{-1}}{1-q^{r-\frac{m}2}\zeta } \,.
 \label{tetrahedron index}
 \end{align}
The index can be understood as an element of $\mathbb{Z}[\zeta, \zeta^{-1}]((q^{\half}))$ by expanding the index in $q$.  An element of $\mathbb{Z}[\zeta, \zeta^{-1}]((q^{\half}))$ contains only finitely many negative powers of $q$ and  each coefficient  is written as Laurent series in $\zeta$ . 
In the infinite product, there is an ambiguity when $r=m$ where a factor $\frac{1}{1-\zeta}$ appear. We formally interpret the factor as
\begin{align}
\frac{1}{1-\zeta}= \sum_n \zeta^n \in \mathbb{Z}[\zeta, \zeta^{-1}]\;.
\end{align}
The domain of $(m,e)$ in the tetrahedron index $\CI^{\Pi_Z}_\Delta(m,e)$ is
\begin{align}
m,e\in \mathbb{Z}\;.
\end{align}
For later use, we will extend the range of the function $\CI_{\Delta}$ to  $\{ (m,e) \in \mathbb{Q} + i \pi \mathbb{Q}+ \frac{\hbar}2 \mathbb{Q}\} $.  For  $(m,e) \in \mathbb{Q}$,  we define the function as
\begin{align}
\CI_{\Delta} (m,e) :=\left\{\begin{array}{ll} \CI^{\Pi_Z}_{\Delta}(m,e) \;, &  \quad (m,e) \in \mathbb{Z}^2\\ 0 \;,  & \quad   \textrm{$(m,e) \in \mathbb{Q}^2 - \IZ^2 $}\end{array}\right. \label{function IDelta}
\end{align}
For general $(m+\alpha,e+\beta)$  with $(m,e) \in \mathbb{Q}$ and $(\alpha, \beta)  \in i \pi \mathbb{Q}+  \frac{\hbar}2 \mathbb{Q}$, the function is determined by \eqref{function IDelta}  and the following additional relation
\begin{align}
\CI_{\Delta} \big{(}m+\a, e+\b  \big{)} :=e^{e \a - m \b} \CI_{\Delta}(m,e) \;.\label{affine shift in tetrahedron index}
\end{align}
Under the polarization change from $\Pi=(\mathcal{X},\mathcal{P})$ to $\tilde{\Pi}=(\tilde{\mathcal{X}}, \tilde{\mathcal{P}})$, related by the following $SL(2,\mathbb{Q})$ and affine shifts\footnote{In  \cite{Dimofte:2011py},  the $SL(2,\mathbb{Z})$ polarization change in CS theory on a tetrahedron $\Delta$ is identified with Witten's $SL(2,\mathbb{Z})$ action on the tetrahedron theory $T_\Delta$. Witten's $SL(2,\mathbb{Z})$ action can be extended to $SL(2,\mathbb{Q})$ by including   charge rescaling of $U(1)$ global symmetry.}
\begin{align}
\left(\begin{array}{c} \tilde{\mathcal{X}} \\ \tilde{\mathcal{P}}\end{array}\right) = g\cdot \left(\begin{array}{c} \mathcal{X}\\ \mathcal{ P} \end{array}\right) + (\pi  i + \frac{\hbar}2)\left(\begin{array}{c} \alpha_m \\ \alpha_e \end{array}\right) \;, \label{tetra polarization change}
\end{align}
the tetrahedron index transforms as \cite{Dimofte:2011py}
\begin{align}
&\CI^{\tilde{\Pi}}_{\Delta}(m,e) =(- q^{1/2})^{m \alpha_e - e\alpha_m} \CI^{\Pi}_{\Delta}(g^{-1} \cdot (m,e)) \;. \label{SL(2,Q)+affine transformation on tetra index}
\end{align}
Under the $SL(2,\mathbb{Q})$ transformation, the domain of  charge $(m,e)$ also should be transformed. The domain in the transformed polarization $\tilde{\Pi}$ is determined by  demanding  $g^{-1}\cdot (m,e)$  is in an allowed domain in the original polarization $\Pi$.
The transformation rule can be written  as
\begin{align}
\CI^{\tilde{\Pi}}_{\Delta}(\tilde{m},\tilde{e}) = \CI^{\Pi}_\Delta (m,e)\;, \; %
\quad
\left(\begin{array}{c} \tilde{m} \\ \tilde{e}\end{array}\right) = g\cdot \left(\begin{array}{c} m\\ e \end{array}\right) + (\pi  i + \frac{\hbar}2)\left(\begin{array}{c} \alpha_m \\ \alpha_e \end{array}\right) \;. \label{tetra polarization change-2}
\end{align}
Shifts by a linear combination of $i \pi$ and $\frac{\hbar}2$ in the arguments of the function $\CI_{\Delta}$ is defined  in eq.~\eqref{SL(2,Q)+affine transformation on tetra index}.
Comparing  \eqref{tetra polarization change} with \eqref{tetra polarization change-2}, we may identify
\begin{align}
(m,e)\simeq (\CX, \CP)\;, \label{Polarization change equivalence}
\end{align}
as far as  transformation rules under polarization changes are concerned.
Three choices of polarization, $\Pi_Z = (Z,Z''), \Pi_{Z'} = (Z' ,Z)$ and  $\Pi_{Z''}=(Z'', Z')$  of a single tetrahedron are related to one another by discrete symmetries of tetrahedron. Demanding the  conditions $\CI^{\Pi_Z}_\Delta = \CI^{\Pi_{Z'}}_\Delta = \CI^{\Pi_{Z''}}_\Delta$, we obtain the following triality relations on $\CI_\Delta$:
\begin{align}
\CI_{\Delta} (m,e) =(-q^{1/2})^{-e} \CI_{\Delta}(e,-e-m) = (-q^{1/2})^m \CI_{\Delta}(-e-m,m)  \;. \label{Triality}
\end{align}
Another useful identity for the tetrahedron index is
\begin{align}
\CI_{\Delta}(m,e) = \CI_{\Delta}(-e,-m)\;. \label{identity for tetrahedron index}
\end{align}
The above identities on $\CI_{\Delta}(m,e)$ are valid only when $(m,e)\in \mathbb{Z}^2$.

The gluing conditions for $M=(\bigcup_i \Delta_i)/\sim$ can be specified by expressing linearly independent internal edges $C_I$ ($I=1,\ldots, k \leq N-1$) in terms of linear combination (and shifts) of $\mathcal{X}_i, \mathcal{P}_i$ variables.
\begin{align}
&C_I = \sum_{j=1}^N (c^x_{Ij} \CX_j+c^{p}_{Ij} \CP_j ) + a_I = 2\pi i +\hbar \;\; \textrm{with coefficients $\{ c^{x}_{Ij} ,c^p_{Ij}, a_I\}$}. \label{internal edge conditions}
\end{align}
The boundary phase space $\CM(\partial M)$  is given by a symplectic reduction
\begin{align}
\CM(\partial M) = \prod_{i=1}^N \CM(\partial \Delta_i)//\{ C_I = 2\pi i +\hbar \}.
\end{align}
The dimension of $\CM(\partial M)$ is  $2d =2(N-k)$ and we choose a polarization $\Pi_{\partial M}=(\mathcal{X}_\alpha,\mathcal{P}_\alpha)|_{\alpha=1}^{d}$ for the boundary phase space as
\begin{align}
\CX_\alpha = \sum_{j=1}^N (X^x_{\alpha j }\CX_j + X^p_{\alpha j} \CP_j )+ a_\alpha\;, \quad \CP_\alpha = \sum_{j=1}^N (P^x_{\alpha j }\CX_j + P^p_{\alpha j} \CP_j) + b_\alpha\;, \label{external polarization conditions}
\end{align}
with coefficients $\{ X^x_{\alpha j} , X^p_{\alpha j}, P^{x}_{\alpha j }, P^p_{\alpha j} , a_\alpha, b_\alpha \}$ which guarantee that $[\CX_\alpha, \CP_\beta] = -  \hbar  \delta_{\alpha \beta}$ and $[\CX_\alpha, C_I ] =[\CP_\alpha, C_I] = 0 $ for all $\alpha, \beta, I$.

Now  let's explain how to calculate the index $I^\Delta_M$ for $T_M$,  or equivalently the $SL(2,\mathbb{C})$ CS partition function $Z^{\Delta}_M (SL(2,\mathbb{C}))$, from  the tetrahedron gluing data $\{C_I , \CX_\alpha, \CP_\alpha\}$ for $M$ explained above. In the polarization  $\Pi_{\partial M}= (\CX_\a , \CP_\a)$, the index for $T_M$ theory with $M=\bigcup_{i=1}^{N} \Delta_i /\sim$ is  given as
\begin{align}
I_M(m_\alpha, e_\alpha) =\sum_{(m_*, e_*)\in \mathbb{Z}} \delta^{2d} (\ldots)\delta^{k}(\ldots)\prod_{i=1}^{N} \CI^{\Pi_i}_{\Delta}(m_i, e_i)
\end{align}
Here, the Knocker delta functions $\delta^{2d}(\ldots)$ and $\delta^{k}(\ldots)$ come from external polarization choice \eqref{external polarization conditions} and internal edge gluing conditions \eqref{internal edge conditions}, respectively. In view of
the identification \eqref{Polarization change equivalence}, these constraints can be translated  into constraints on  charge variables $(m_i, e_i)$.
\begin{align}
&\delta^{2d}(\ldots) = \prod_{\alpha=1}^d\delta (m_\alpha - \sum_{j=1}^N (X^{x}_{\alpha j} m_j +X^{p}_{\alpha j} e_j) -a_\alpha ) \delta (e_\alpha - \sum_{j=1}^N( P^{x}_{\alpha j} m_j +P^{p}_{\alpha j} e_j )-b_\alpha ) \;, \nn
\\
&\delta^{k}(\ldots) = \prod_{I=1}^k \delta \big{(}\sum_{j=1}^N (c^{x}_{Ij} m_j +c^{p}_{Ij}e_j)+a_I -2\pi i - \hbar \big{)} \;.
\end{align}
Solving the $2d+k$ Knonecker deltas on $2N$ variables $(m_i, e_i)$, we have remaining $k$ variables to be summed. Although  the procedure described here looks different from the description in  \cite{Dimofte:2011py}, one can easily check that they are equivalent.

For each tetrahedron $\Delta_i$ in $\textrm{tori}(\varphi) = \bigcup \Delta_i/\sim $,  we choose the following polarization
\begin{align}
\Pi_i  =\Pi_{Z'}\;, \; \textrm{for all $i$}\;.\label{polarization choice in tetrahedron gluing}
\end{align}
Under this choice, tetrahedron gluing rule \eqref{tetra-gluing} can be written in terms of $(\CX_i , \CP_i)$ as in  Table~\ref{tetra-gluing 2}.
\begin{table}[htbp]
   \centering
   \begin{tabular}{@{} l|c|c|c|c|c @{}} 
      \toprule
     $\varphi_i \varphi_{i-1}$  &  $(L,L)$  & $(L,R)$ & $(R,R)$ & $(R,L)$  \\
      \midrule
            $\mathcal{X}_i$ & $\frac{-W_{i+1}+2W_i -W_{i-1}}2$  &$\frac{-W_{i+1}+W_i +W_{i-1}-\sV-i\pi}2$ &  $\frac{W_{i+1}+ W_{i-1}}2$ &$\frac{W_{i+1}+ W_i-  W_{i-1}+\sV+i\pi}2$  \\
               \midrule
      $\mathcal{P}_i$ & $i \pi +\frac{\hbar}2 - W_i$  &  $i \pi +\frac{\hbar}2 - W_i$  &  $i \pi +\frac{\hbar}2 - W_i$& $i \pi +\frac{\hbar}2 - W_i$  \\
      \midrule
      $\sU_i$ & $0$   & $-\frac{W_i}2 $  &  0 &$\frac{W_i}2$ \\
      \bottomrule
   \end{tabular}
   \caption{Tetrahedron gluing   for tori($\varphi$) in the polarization \eqref{polarization choice in tetrahedron gluing}.}
   \label{tetra-gluing 2}
\end{table}

\noindent
In this polarization choice, the index for the $i$-th tetrahedron in  the mapping torus is given by $\CI_{\Delta}\big{(}{\cal X}_i  ,{\cal P}_i \big{)}$. The $i$-th tetrahedron's  position/momentum variables $(\CX_i , \CP_i)$  are thought of as magnetic/electric charge of $\CI_{\Delta_i}$ via  \eqref{Polarization change equivalence}. They  are parametrized by the variables $W_*$.  The index for mapping torus can be constructed by multiplying all
the indices from each tetrahedron and summing over all   $W_i$
variables modulo a `meridian' condition $\sU=\sum_i \sU_i (W_*)$.
The condition say that a particular linear combination of $W_*$ is
fixed to  be a  meridian variable $\sU$.
\begin{align}
I^{\Delta}_{\textrm{tori}(\varphi)} (\sV,\sU)= \sum_{ W_i \in \mathbb{Z} } \delta \big{(} \sU-\sum_i \sU_i(W_*) \big{)}\prod_{i=1}^{N} \CI_{\Delta}\big{(}\mathcal{X}_i (W_*), \mathcal{P}_i(W_*) \big{)} \label{mapping torus from tetra}
\end{align}
The factors $(i \pi + \frac{\hbar}2)$ in the argument of  $\CI_\Delta$ can be understood from \eqref{affine shift in tetrahedron index}.
The cusp boundary   variables $(\sU,\sV)$  are in $(\mathbb{Z},  \mathbb{Z})$. As an example, for $\varphi=LR$
\begin{align}
&I^{\Delta}_{\textrm{tori}(LR)} (\sV,\sU)  \nonumber
\\
&= \sum_{W_1,W_2\in \mathbb{Z}}  \delta(\sU-\frac{-W_1+ W_2}2)(q^{\half})^{-\half (W_1+W_2)} \CI_{\Delta}(\frac{W_1-\sV}2,-W_1 ) \CI_{\Delta}(\frac{W_2+\sV}2,-W_2) \;. \nonumber
\end{align}
To list a few non-vanishing results, we have
\begin{align}
&I^{\Delta}_{\textrm{tori}(LR)} (0,0)= 1-2q-3q^2+2q^3+\ldots \nonumber
\\
&I^{\Delta}_{\textrm{tori}(LR)} (0,\pm 1)= 2q^{3/2} - 4q^{7/2}+\ldots \nonumber
\\
&I^{\Delta}_{\textrm{tori}(LR)} (\pm 1,\pm 1)= q+q^2-2q^3+\ldots \nonumber
\\
&I^{\Delta}_{\textrm{tori}(LR)} (\pm 1,\pm 2)= q^3 +3q^4 \ldots \nonumber
\end{align}
Comparing these indices with \eqref{LR index}, we find the following non-trivial
agreement,
\begin{align}
I^{\Delta}_{\textrm{tori}(LR)} (\sV,\sU) :=  I^{\textrm{T[SU(2)]}}_{\textrm{tori}(LR)} (m_\eta = \sV, e_\eta = \sU)\;.
\end{align}
In section \ref{equality between three SL(2,C)/index ptns}, we will show that $I^{\Delta}_{\textrm{tori}(\varphi)} = I^{T[SU(2)]}_{\textrm{tori}(\varphi)}$ for general $\varphi$ with $|\Tr(\varphi)|>2$.

\subsection{Hilbert-spaces $\CH_{SL(2,\mathbb{C})}$ and $\CH^{\textrm{knot}}_{SL(2,\mathbb{C})}$  } \label{index : Hilbert-space}

In this section, we  quantize the classical phase spaces $(\mathcal{M},\O)_{SL(2,\mathbb{C})}$ and $(\mathcal{M},\O)^{\rm knot}_{SL(2,\mathbb{C})}$ studied  in section \ref{quantum riemann surface} and construct the Hilbert-spaces $\mathcal{H}_{SL(2,\mathbb{C})}$ and $\mathcal{H}^{\rm knot}_{SL(2,\mathbb{C})}$. 
We show explicitly how the quantum operators introduced in section \ref{quantum riemann surface} act on the Hilbert-spaces. Based on constructions in this section, we will calculate $SL(2,\mathbb{C})$ CS partition function on mapping cylinder/torus in section \ref{equality between three SL(2,C)/index ptns}.

\paragraph{Hilbert-space $\mathcal{H}_{SL(2,\mathbb{C})}$ } As explained in section \ref{quantum riemann surface}, the phase space $\mathcal{M}(\Sigma_{1,1})_{SL(2,\mathbb{C})}$ can be parameterized by three shear coordinates $(\sT, \sT',\sT'')$ with one linear constraint.  The symplectic form $\Omega_{SL(2,\mathbb{C})} $ on the phase space is given in \eqref{symplectic form in shear}. Rewriting the symplectic form in terms of real and imaginary parts of shear coordinates,
\begin{align}
\Omega_{SL(2,\mathbb{C})} = -\frac{1}{\hbar}d \textrm{Im}(T)\wedge d \textrm{Re}(T') -\frac{1}{\hbar}d \textrm{Re}(T)\wedge d \textrm{Im}(T') \,.
\end{align}
To obtain the Hilbert-space,  we first  need to specify a choice of  `real' polarization. We will choose the following polarization,
\begin{align}
\big{(} X_1, X_2, P_1, P_2\big{)} = \big{(} \textrm{Re}(T'),\half  \textrm{Re}(T),  -\textrm{Im}(T), 2\textrm{Im}(T') \big{)}\;.
\end{align}
In this choice of real polarization, as  noticed in \cite{Dimofte:2011py}, the momenta  are periodic variables \eqref{periodicity in shear} and thus their conjugate position variables should be quantized. Since the periods for $(P_1,P_2)$ are $(4\pi , 8\pi)$ respectively, the correct quantization condition for $X_1,X_2$ is
\begin{align}
X_1 \in \frac{\hbar}{2}\mathbb{Z}\;, \quad X_2 \in \frac{\hbar}4 \mathbb{Z}\;.
\end{align}
  Thus position eigenstates  $| X_1, X_2 \rangle$ are labelled by integers and we will introduce charge basis $| m,e \rangle$ as
\begin{align}
| m, e \rangle  := | X_1= m \frac{\hbar}2, X_2  = e\frac{\hbar}2 \rangle   \;, \quad  \textrm{with}\; m, 2e\in \mathbb{Z} \;.
\end{align}
The shear operator $(\sT,\sT')_\pm=(2X_2 \mp  i P_2, X_1
\pm \frac{i} 2 P_2) $ acts on the basis  as
\begin{align}
\sT_\pm =e\hbar \pm  2\partial_m\;, \quad \sT'_\pm = m \frac{\hbar}2 \mp \partial_e\;.
\end{align}
The exponentiated operators act as
\begin{align}
\langle m,e| \sqrt{\st}_\pm |I \rangle  =  q^{\half e} \langle m\pm 1, e| I \rangle \;, \quad \langle m,e| \sqrt{\st'}_\pm|I\rangle =  q^{\frac{1}4 m} \langle m, e \mp \half | I\rangle \;. \label{t,t' action on (t,t') basis}
\end{align}
Using the basis, the Hilbert-space $\CH_{SL(2,\mathbb{C})}$ can be constructed as
\begin{align}
\CH_{SL(2,\mathbb{C})}= \textrm{Hilbert-space spanned by  a basis } \{| m, e \rangle \}_{ m, 2e\in \mathbb{Z}}\;.
\end{align}
%
%
One may introduce another basis called fugacity basis $\{ | m, u \rangle\}$ which is related to charge basis by Fourier expansion.
\begin{align}
| m,e  \rangle  =  \oint \frac{du}{2\pi i u} u^{e} | m,u \rangle \;. \label{fugacity basis}
\end{align}
In $\{ | m,u \rangle \}$, $m$ is integer and $u^{1/2}$ is on a unit
circle $|u^{1/2}|=1$ in complex plane. As explained in
\cite{Dimofte:2011py}, elements in this basis are position
eigenstates under the following choice of real polarization
\begin{align}
\big{(} X_1, X_2, P_1, P_2\big{)} = \big{(}\textrm{Re}(T'),\textrm{Im}(T') ,-\textrm{Im}(T),  -  \textrm{Re}(T)\big{)}\;.
\end{align}

\paragraph{Inner product on $\mathcal{H}_{SL(2,\mathbb{C})}$ } We defined the adjoint of shear operators in \eqref{adjoin of shear and ell}.   Adjoint operation depends on the inner-product structure on $\mathcal{H}_{SL(2,\mathbb{C})}$. Requiring consistency  between  \eqref{adjoin of shear and ell} and  \eqref{t,t' action on (t,t') basis}, one can uniquely determine the inner-product  on $\mathcal{H}_{SL(2,\mathbb{C})}$ up to an overall factor $\kappa$.
\begin{align}
\langle m, e|m',e'\rangle = \kappa \delta (m-m')\delta (e-e')\;. \label{inner product on T,T' basis}
\end{align}
For simplicity, we will set $\kappa =1$ by rescaling the charge basis.

\paragraph{Basis on $\CH_{SL(2,\mathbb{C})}$ associated to  polarization $\Pi$}
So far we have only considered two choices of basis, $\{| m,e \rangle \}$ and  $\{| m,u \rangle\}$  for $\CH_{SL(2,\mathbb{C})}$.
 We will introduce more bases $\{| m,e\rangle_\Pi \}$ and $\{| m,u\rangle_\Pi\}$  for $\CH_{SL(2,\mathbb{C})}$, one for each polarization choice $\Pi$ of the phase space $\CM_{SL(2,\mathbb{C})}$. Polarization $\Pi=(\mathcal{X},\mathcal{P})$ is determined by identifying position variable $\mathcal{X}$  and its conjugate momentum variable $\mathcal{P}$ satisfying the canonical commutation relation $[\mathcal{X}_\pm ,\mathcal{P}_\pm] = \mp \hbar$. A simple choice of polarization is $\Pi_{\sT, \sT'} = (\sT',\half \sT)$. The basis $\{| m,e\rangle_{\Pi=(\mathcal{X},\mathcal{P})}\}$ is defined by following conditions
\begin{align}
&_\Pi\langle m,e| e^{\mathcal{X}_\pm } =  q^{\frac{m}2} \; _\Pi\langle m,e\mp 1| \;,   \quad  _\Pi\langle m,e| e^{\mathcal{P}_\pm} =q^{\frac{e}2}\;_\Pi\langle m\pm 1,e| \;.
\label{xp-operation}
\end{align}
These conditions determine the basis $\{| m,e \rangle_\Pi\}$ up to  an overall constant which is universal to all basis.\footnote{There is no guarantee that for given polarization $\Pi$ there exist  a basis satisfying these conditions.}  In this notation, basis $| m,e\rangle$  in the  above can be understood as $| m,e \rangle_{\Pi}$ with $\Pi = \Pi_{\sT, \sT'}:=(\sT', \half \sT)$. Similarly fugacity basis $| m,u\rangle_\Pi$  associated to a polarization $\Pi$ can be defied as Fourier transformation on $ | m,e\rangle_\Pi$.
Under a linear transformation of the polarization
\begin{align}
\left(\begin{array}{c} \tilde{\mathcal{X}}\\ \tilde{\mathcal{P}}\end{array}\right)_\pm  =g\cdot \left(\begin{array}{c}\mathcal{ X}\\ \mathcal{P}\end{array}\right)  \pm \left(\begin{array}{c} \a \\ \b \end{array}\right) \;,  \quad g\in SL(2,Q)
\end{align}
the basis transforms as
\begin{align}
&_{\tilde{\Pi}} \langle m,e| =\;_\Pi \langle g^{-1}\cdot (m,e)|   e^{  m \b - e \a} \;.\label{basis polarization transformation}
\end{align}
Note that this transformation rule is equivalent to \eqref{SL(2,Q)+affine transformation on tetra index}  after identifying the index  $\CI (m,e)$ as matrix element $\langle m,e |\CI \rangle$. Under the polarization transformation, the range of charge $(m,e)$ also should be transformed accordingly.

\paragraph{SR basis} We  define  `SR basis'  as a basis  associated to a polarization $\Pi_{\SR}$,
\begin{align}
\Pi_{\SR} := ({\cal S},{\cal R}) := \half (\sT+\sT', \sT -\sT') \;. \label{SR polarization}
\end{align}
This basis will be denoted as $|m,e\rangle_{\SR}:=| m, e\rangle_{\Pi_{\SR}}$. 
From the basis transformation \eqref{basis polarization transformation}, the quantization condition for $(m,e)$ in the SR basis $|m,e\rangle_{\SR}$ is determined:
\begin{align}
m,e \in \frac{\mathbb{Z}}2\;, \quad \textrm{with a condition } m+e\in \mathbb{Z}\;. \label{range of (m,e) for SR}
\end{align}
The inner-product on SR basis takes the same form as \eqref{inner product on T,T' basis},
\begin{align}
_{\SR}\langle m,e|m',e'\rangle_{\SR}=  \delta (m-m')\delta(e-e') \;. \label{Inner product on SR basis}
\end{align}
Thus, the completeness relation in the SR basis is
\begin{align}
\mathds{1}_{\CH_{SL(2,\mathbb{C})}} = \sum_{(2m,2e)\in \mathbb{Z}: m+e \in \mathbb{Z}} |m,e\rangle_{\SR} \langle m,e|\;. \label{completeness relation in SR basis}
\end{align}
This SR basis will play a crucial role in section \ref{equality between three SL(2,C)/index ptns} in proving $ Z^{\Delta}_{\textrm{tori}(\varphi)} = Z^{\textrm{Tr}(\varphi)}_{\textrm{tori}(\varphi)} $.

\paragraph{FN basis}
We will introduce yet another basis, called FN (Fenchel-Nielsen) basis, which will play important roles  in section \ref{equality between three SL(2,C)/index ptns} in proving $I^{T[SU(2)]}_{\textrm{tori}(\varphi)} = Z^{\textrm{Tr}(\varphi)}_{\textrm{tori}(\varphi)}(SL(2,\mathbb{C})) $.   The FN charge basis $| m,e \rangle _{\FN}$ is not defined on $\CH_{SL(2,\mathbb{C})}$ but on $\widetilde{\CH}_{SL(2,\mathbb{C})}$, which will be identified with a double cover of $\CH_{SL(2,\mathbb{C})}$.
The Hilbert-space $\widetilde{\CH}_{SL(2,\mathbb{C})}$ is defined as
\begin{align}
\widetilde{\CH}_{SL(2,\mathbb{C})}= \textrm{Hilbert-space  whose basis are } \{| m, e \rangle_{\FN} \}_{ 2m, e\in \mathbb{Z}}\;.
\end{align}
FN fugacity  basis can be defined  as Fourier expansion of FN charge basis
\begin{align}
_{\FN}\langle m,e | =  \oint \frac{du}{2\pi i u} u^{-e}\; _{\FN}\langle m,u| \;.
\end{align}
In the FN basis, the FN operators $(\hat{\Lambda}, \hat{\mathcal{T}})$, introduced in  \eqref{FN-op}, \eqref{shear2FN-q} act like $(\mathcal{X},\mathcal{P})$,
\begin{align}
&_{\FN}\langle m,e| \hat{\lambda}_\pm = q^{\frac{m}2}\; _{\FN}\langle m,e\mp 1|\;, \quad  _{\FN}\langle m,e| \hat{\tau}_\pm = q^{\frac{e}2}\; _{\FN}\langle m \pm 1,e|\;, \nn
\\
&_{\FN}\langle m,u| \hat{\lambda}_\pm = q^{\frac{m}2} u \; _{\FN}\langle m,u|\;, \quad  _{\FN}\langle m,u| \hat{\tau}_\pm = e^{\frac{\hbar}2 u \partial_u}\;_{\FN}\langle m \pm 1, u |\;.
\end{align}
In terms of the fugacity basis, the inner-product on $\widetilde{\mathcal{\CH}}_{SL(2,\mathbb{C})}$ is defined as
\begin{align}
_{\FN}\langle m_1, u_1 | m_2, u_2 \rangle_{\FN} &=
\Delta(m_1,u_1)^{-1} \delta (m_1 - m_2)\delta(u_1 - u_2) \,. \label{inner product on FN basis}
\end{align}
where $\Delta(m, u)$ is the measure factor appearing in the $\odot$ operation \eqref{su(2) measure}.
The delta function $\delta(u_1,u_2)$ is defined  by following condition
\begin{align}
\oint \frac{du_1}{2\pi i u_1} \delta(u_1,u_2) f(u_2) = f(u_1)\;, \quad \textrm{for arbitrary $f(u)$.}
\label{delta-u-1}
\end{align}
The inner product \eqref{inner product on FN basis} implies the completeness relation in $\widetilde{\CH}_{SL(2,\mathbb{C})}$ in the FN basis,
\begin{align}
\mathds{1}_{\widetilde{\CH}(SL(2,\mathbb{C}))} = \sum_{m \in \mathbb{Z}/2} \oint \frac{du}{2\pi i u}\Delta(m,u) |m,u\rangle_{\FN}\langle m,u| \;. \label{completeness relation in FN basis}
\end{align}
With respect to the inner product, the adjoint of FN operators are
\begin{align}
(\hat{\lambda}_\pm )^\dagger = \hat{ \lambda}_\mp\;, \quad (\hat{\tau}_\pm)^\dagger = \frac{1}{\hat{\lambda}_\mp - \hat{\lambda}_\mp^{-1}} \hat{\tau}_\mp (\hat{\lambda}_\mp - \hat{\lambda}_\mp^{-1})\;. \label{adjoint of FN}
\end{align}
To establish an isomorphism between $\mathcal{H}_{SL(2,\mathbb{C})}$ and a subspace of $\widetilde{\CH}_{SL(2,\mathbb{C})}$, we  use the operator relation \eqref{shear2FN-q} between FN and shear operators. 
Combining \eqref{shear2FN-q} and \eqref{adjoint of FN}, one can show that
\begin{align}
(\sqrt{\st},\sqrt{\st'},\sqrt{\st''})_\pm^\dagger = (\sqrt{\st},\sqrt{\st'},\sqrt{\st''})_\mp\;, \label{adjoint of shear operator}
\end{align}
with respect to the inner product \eqref{inner product on FN basis}, 
precisely as we anticipated in \eqref{adjoin of shear and ell}.

Using these relations, one can determine the action of the SR operators $(\ss,\sr)_\pm :=(\exp (\CS_\pm), \exp(\CR_\pm))$ in the FN basis. We will consider states  $ | m,u\rangle_{\widetilde{\SR}}$  in $\widetilde{\CH}_{SL(2,\mathbb{C})}$ on which operator $(\CS,\CR)$ acts like $(\mathcal{X},\mathcal{P})$,
\begin{align}
_{\widetilde{\SR}}\langle m,u| {\ss}_\pm = q^{\frac{m}2} u^{\pm 1} \;_{\widetilde{\SR}}\langle m,u|\;, \quad  _{\widetilde{\SR}}\langle m,  u| {\sr}_\pm =e^{\frac{\hbar}2 u \partial_u } \;_{\widetilde{\SR}}\langle m+1,u | \;. \label{def of tilde SR basis}
\end{align}
As we will see in appendix \ref{basis from FN to shear}, from the above condition one can explicitly express the basis $|m,u\rangle_{\widetilde{\SR}}$  in terms of FN basis $|m,u\rangle_{\FN}$ up to overall constant. The explicit expression copied from appendix \ref{basis from FN to shear} is
\begin{align}
_{\widetilde{\SR}}\langle m,u|=\sum_{\tilde{m}} \oint \frac{d\tilde{u}}{2\pi i \tilde{u}} \Delta(\tilde{m},\tilde{u})(-q^{\half} u)^m \CI_{\Delta} (-m-\tilde{m}, u^{-1}\tilde{u}^{-1})\CI_{\Delta}(\tilde{m}-m, \tilde{u}/u) _{\FN}\langle \tilde{m},\tilde{u}| \;.
\label{SR basis in FN basis}
\end{align}
As argued  in appendix \ref{basis from FN to shear},  the
range of charge $(m,e)$ for  the  charge basis
$|m,e\rangle_{\widetilde{\SR}}$, which is related to $|m,u\rangle_{\widetilde{\SR}}$ by Fourier expansion, is the same as that of SR basis
$|m,e\rangle_{\SR}$ \eqref{range of (m,e) for SR} and the inner
product on  $|m,e\rangle_{\widetilde{\SR}}$ is also the same as that
of SR basis  \eqref{Inner product on SR basis}. Furthermore, by
definition of $|m,e\rangle_{\widetilde{\SR}}$ in \eqref{def of tilde
SR basis}, the action of shear operators are  the same on the two
basis $|m,e\rangle_{\SR}$ and $|m,e\rangle_{\widetilde{\SR}}$. Thus
one can naturally identify
\begin{align}
|m,e\rangle_{\widetilde{\SR}} \in \widetilde{\CH}_{SL(2,\mathbb{C})} \textrm{ with } |m,e\rangle_{\SR} \in \CH_{SL(2,\mathbb{C})}\;.
\end{align}
From the above identification, we can consider the SR basis $|m,e\rangle_{\SR}$ as an element
in $\widetilde{\CH}_{SL(2,\mathbb{C})}$ and  the Hilbert-space $\CH_{SL(2,\mathbb{C})}$  as a
subspace of  $\widetilde{\CH}_{SL(2,\mathbb{C})}$. The subspace is spanned by $\{ |m,e\rangle_{\widetilde{\SR}} \}$.
The Weyl-reflection operator $\sigma$ in \eqref{Weyl-reflection} acts on $\widetilde{\CH}_{SL(2,\mathbb{C})}$ as
\begin{align}
\sigma \; : \; &| m,u \rangle_{\FN} \rightarrow | -m, u^{-1}\rangle _{\FN}\;,\; \textrm{or equivalently} \nn
\\
&| m,e\rangle _{\FN} \rightarrow | -m, -e \rangle _{\FN}\;.
\end{align}
From the explicit expression  \eqref{SR basis in FN basis}, one can easily see that the SR basis $|m,u\rangle_{\SR}$ is Weyl-reflection invariant. Thus we see that
\begin{align}
\mathcal{H}_{SL(2,\mathbb{C})} \subseteq \{ \textrm{$\sigma$-invariant subspace in } \widetilde{\CH}_{SL(2,\mathbb{C})} \}\;.
\end{align}
Furthermore, it is argued in appendix \ref{basis from FN to shear} that the equality holds. 
In other words, the Weyl-reflection invariant combination of the FN basis states $\{ |m,e\rangle^S_{\FN} \}$ form a complete basis for $\CH_{SL(2,\mathbb{C})}$.
\begin{align}
|m,e\rangle^{S}_{\FN}:= \frac{1}2\left(|m,e\rangle_{\FN} + |-m,-e\rangle_{\FN}\right)\;. \label{S-FN basis}
\end{align}
\paragraph{Hilbert-space $\CH^{\textrm{knot}}_{SL(2,\mathbb{C})}$} As we have seen in section \ref{sec: Classical phase space}, the phase space $\CM^{\rm knot}$ is parametrized by `longitude' and  `meridian' variables, $\ell = e^{V}$ and $m= e^{U}$.  The  symplectic form is \eqref{Symplectic for knot}
\begin{align}
\Omega^{\textrm{knot}}_{SL(2,\mathbb{C})} = \frac{1}{i\hbar} dU \wedge dV - \frac{1}{i\hbar} d\bar{U} \wedge d \bar{V}
\end{align}
We choose the real polarization as
\begin{align}
(X_1, X_2, P_1, P_2) = (\textrm{Re}(V), \textrm{Re}(U), -2 \textrm{Im}(U), 2 \textrm{Im}(V))
\end{align}
Again, since the momenta are periodic variables, their conjugate position variables are quantized.  Considering mapping torus, the periodicity of $U$ and $V$ are $4\pi $ and $2\pi $, respectively as we saw in the last paragraph in \ref{sec: Classical phase space}. Thus the correct  quantization for $X_1, X_2$ seems to be
\begin{align}
X_1 \in \frac{\hbar}4 \mathbb{Z}\;, \quad  X_2 \in \frac{\hbar}2 \mathbb{Z}\;.
\end{align}
However,  there is an additional  quantum  $\mathbb{Z}_2$ symmetry which  shifts meridian variable $U$ by $2\pi i$ for  CS theories on knot complement (see secion 4.2.5 and section 4.2.7 in
\cite{Witten:2010cx}\footnote{The normalization of meridian variable in \cite{Witten:2010cx} is  different from ours, $m_{ours} = \sm^2_{theirs}$. In the reference, the $\mathbb{Z}_2$ symmetry is shown for knot complements in $S^3$. We expect that the $\mathbb{Z}_2$ symmetry also exists for our mapping torus case. One evidence  is that the A-polynomial analyzed in section  \ref{A-polynomial for mapping torus} is always polynomial in $m$ instead of $m^{1/2}$.}). Taking account of this quantum $\mathbb{Z}_2$ effect, the quantization condition is modified as
\begin{align}
X_1 \in \frac{\hbar}2 \mathbb{Z}\;, \quad  X_2 \in \frac{\hbar}2 \mathbb{Z} \,.
\end{align}
This is compatible with the quantization condition for $(m_\eta, e_\eta)\in \mathbb{Z}$ in the mapping torus index computation in section \ref{index-wall}.
When we consider mapping cylinder, as we already mentioned in section \ref{sec: Classical phase space}, the period for $V$ is doubled, and the correct quantization is
\begin{align}
X_1 \in \frac{\hbar}2 \mathbb{Z}\;, \quad  X_2 \in \frac{\hbar}4 \mathbb{Z}\;.
\end{align}
This quantization is also compatible with the quantization conditions for $(m_\eta, 2 e_\eta) \in \mathbb{Z}$ in mapping cylinder index computation.
Since we are also interested in the mapping cylinder index, we will use this quantization conditions in constructing $\CH^{\textrm{knot}}$.
After making mapping torus by gluing two boundary $\Sigma_{1,1}$'s,   the $SL(2,\mathbb{C})$ CS partition function vanishes automatically when $X_1 = \frac{\hbar}2 (\mathbb{Z}+\half)$ as we will see in section \ref{Z(Tr)=Z(Delta)}.
We introduce the charge basis $|m_\eta, e_\eta\rangle$ for $\CH^{\textrm{knot}}$ as
\begin{align}
|m_\eta, e_\eta \rangle := |X_1 = \frac{\hbar}2 m_\eta, X_2 = \frac{\hbar}2 e_\eta \rangle\;, \quad (m_\eta, 2e_\eta)\in \mathbb{Z}\;.
\end{align}
and we define
\begin{align}
\CH^{\textrm{knot}}_{SL(2,\mathbb{C})} = \textrm{Hilbert-space spanned by $\{ |m_\eta, e_\eta\rangle \}$}\;.
\end{align}
Using Fourier transformation, we can introduce  fugacity basis $\{ |m_\eta, u_\eta\rangle \}$,
\begin{align}
\{ |m_\eta, u_\eta\rangle \; :\; m_\eta \in \mathbb{Z}\;,\; u^{\half}_\eta=e^{i \theta}\; (0 \leq \theta <2\pi)\} \;.
\end{align}
On the charge basis, the operators $(\sV_\pm, \sU_\pm)$, quantum counterparts  of $(V, \bar{V}, U, \bar{U})$, act as
\begin{align}
\sV_\pm = \frac{\hbar}2 m_\eta \mp \partial_{e_\eta}\;, \quad \sU_\pm = \frac{\hbar}2 e_\eta \pm \partial_{m_\eta}\;.
\end{align}
In terms of the exponentiated operators $(\fl_\pm, \sm_\pm):=(e^{\sV_\pm}, e^{\sU_\pm})$,  the action is given by
\begin{align}
&\langle m_\eta, u_\eta| \fl_\pm = \langle m_\eta, u_\eta| q^{\half m_\eta}u_\eta^{\pm 1}\;,\quad \langle m_\eta, u_\eta| \sm_\pm =e^{\frac{\hbar}2 u_\eta\partial_{u_\eta}}\langle m_\eta\pm 1,e_\eta|\;, 
 \nn
\\
&\langle m_\eta, e_\eta| \fl_\pm = \langle m_\eta, e_\eta \mp 1| q^{\half m_\eta}\;,\quad \langle m_\eta, e_\eta| \sm_\pm =\langle m_\eta\pm 1,e_\eta|q^{\half e_\eta} \;.
\end{align}
The inner-product  on the Hilbert space is defined as
\begin{align}
\langle m_\eta, e_\eta |m'_\eta, e'_\eta \rangle = \delta (m_\eta-m'_\eta)\delta (e_\eta - e'_\eta) \;,
\label{inner-product H-u(1)}
\end{align}
which  ensures that
\begin{align}
\fl^\dagger_\pm = \fl_\mp\;,\quad \sm^\dagger_\pm = \sm_\pm\;.
\end{align}
The completeness relation in $\CH^{\textrm{knot}}_{SL(2,\mathbb{C})}$ is
\begin{align}
\mathds{1}_{\CH^{\textrm{knot}}_{SL(2,\mathbb{C})}} = \sum_{m_\eta} \oint \frac{d u_\eta}{2\pi i u_\eta} |m_\eta, u_\eta\rangle \langle m_\eta, u_\eta| = \sum_{m_\eta, e_\eta} |m_\eta,e_\eta\rangle \langle m_\eta, e_\eta|\;.
\end{align}
Consider  operators $\sO_i (\sV_\pm)$  constructed using only
$\sV_\pm$ but not $\sU_\pm $. As already mentioned in section 
\ref{quantum riemann surface},  these operators $\sO_i$ can be
understood as a state $|\sO_i \rangle$ in $\CH^{\textrm{knot}}$
through the following map,
\begin{align}
&\sO (\sV_\pm) \Leftrightarrow |\sO_i \rangle := \sO_i  |\sU=0\rangle  \;,\nn
\\
&\textrm{where } |\sU=0\rangle = |P_1 = 0 , X_2 =0\rangle  = \sum_{m_\eta }|m_\eta, e_\eta= 0 \rangle\;.
\end{align}
Using the basis $|m_\eta, u_\eta\rangle$, 
$\sO_i(\sV_\pm)$ can be further mapped to a ``wave-function",
\begin{align}
\sO_i(\sV_\pm )  \Leftrightarrow   \sO_i (m_\eta, u_\eta) &:=\langle m_\eta, u_\eta| \sO_i \rangle = \langle m_\eta, u_\eta| \sO_i | \sU=0\rangle\;, \nn
\\
 &= \langle m_\eta, u_\eta |\sO_i (\sV_\pm) |m_\eta, e_\eta = 0\rangle\;.
\end{align}
The function $\sO_i (m_\eta ,u_\eta)$ obtained in this way is nothing but
\begin{align}
\sO_i (m_\eta, u_\eta) = \sO_i(\sV_\pm )_{\sV_\pm \rightarrow \frac{\hbar}2 m_\eta \pm \log_q u_\eta  }\;.
\end{align}
The multiplication of two operator $\sO_1 \cdot  \sO_2$ is simply mapped to the   multiplication of two functions. The following relations also hold
\begin{align}
&\langle m_\eta, e_\eta | \sO_i   |m_\eta, e_\eta =0\rangle  = \textrm{Fourier transformation  on $u_\eta$ of } \sO_i (m_\eta, u_\eta)\;, \nn
\\
&\langle m_\eta, e_\eta |\sO_1 \ldots \sO_N |m_\eta,  0\rangle  = \textrm{Fourier transformation on $u_\eta$ of } \sO_1 (m_\eta, u_\eta ) \ldots \sO_N  (m_\eta, u_\eta) \; \nn
\\
&=\sum_{e_{\eta_1},\ldots , e_{\eta_N}} \delta (e_\eta - \sum_{i=1}^N e_{\eta,i  } ) \sO_1 (m_\eta, e_{\eta,1}) \ldots \sO_N (m_\eta, e_{\eta, N}) \nn
\\
&= \sum_{e_{\eta_1},\ldots , e_{\eta_N}} \delta (e_\eta - \sum_{i=1}^N e_{\eta,i  } ) \langle m_\eta, e_{\eta,1}| \sO_1 |m_\eta, 0 \rangle \ldots  \langle m_\eta, e_{\eta, N}| \sO_N |m_\eta, 0 \rangle\;. \label{property of O(l)}
\end{align}
These properties will be used in the below.
\subsubsection{The Hilbert-space from 4d gauge theory} \label{index: Hilbert-space from 4d}
The  AGT relation \cite{AGT}, which relates the $S^4$ partition function for  a 4d theory $T_\Sigma$\footnote{As defined in section \ref{sec : introduction}, $T_\Sigma$ denotes  a 4d theory of class S obtained from $A_1$ type of $(2,0)$ theory on a Riemann surface $\Sigma$.}  to a correlation function  in 2d Liouville theory on $\Sigma$,  can be recast  as an  isomorphism between the Hilbert-space $\CH(S^3)$ associated  with the 4d theory  on an omega-deformed four-ball $B^4$ (whose boundary is a squashed $S^3$)  and the Hilbert-space $\CH^{Liouv}(\Sigma)$ on the 2d Liouville theory \cite{Nekrasov:2010ka,Vartanov:2013ima}. Using  dualities  among 2d Liouville/Teichmuller/CS theory \cite{Chekhov:1999tn,Kashaev:1998,Teschner:2005,Teschner:2003em,Verlinde:1990} the Hilbert-space can be identified with $\CH_{SL(2,\mathbb{R})} (\Sigma)$, the Hilbert-space for $SL(2,\mathbb{R})$ CS theory on $\Sigma$. In this subsection, we  try to make  parallel stories in the `superconformal index version' of AGT relation \cite{Gadde:2009kb,Gadde:2011ik}, which relates a superconformal index for 4d  theory to a correlation function in 2d TQFT.

In the above we constructed the  Hilbert-space $\CH_{SL(2,\mathbb{C})}$ on which  operators studied in section \ref{quantum riemann surface} act.  There is an another space where the operators naturally act on. That is a space of half-indices for 4d $\CN=2^*$ theory which will be denoted as $\CH(S^2 \times S^1 )$.
As first noticed in \cite{Dimofte:2011py},  Schur superconformal index for 4d $SU(2)$ $\CN=2^*$ can be written in the following form
\begin{align}
&I_{\mathcal{N}=2^*} (q,  u_\eta) =\sum_{m\in \frac{ \mathbb{Z}}2} \oint \frac{du}{2\pi i u }\Delta(m,u) \Pi^{\dagger} (m,u; u_\eta)\Pi (m,u;u_\eta)\;, \nn
\\
&\Pi(m,u; u_\eta) := \delta (m)   \prod_{r=0}^\infty \frac{(1-q^{1+r} u^2)(1-q^{1+r})(1-q^{1+r} u^{-2})}{(1-q^{1/2+r}u_\eta u^2)(1-q^{1/2+r}u_\eta )(1-q^{1/2+r}u_\eta u^{-2})}\;.
\end{align}
The Schur index is defined by \cite{Gadde:2011ik,Gang:2012yr}
\begin{align}
I_{\CN =2^*} (q, u_\eta) :=\Tr (-1)^{F} q^{\half R + j_3} u_\eta^{H_\eta}\;,
\end{align}
where the trace is taken over   a Hilbert-space of the $\CN=2^*$ theory on $S^3$($\times$ time).  $j_3$ is a Cartan of the diagonal $SU(2)$ isometry of $SU(2) \times SU(2)$ of $S^3$ and $R$ is a Cartan of  the $SU(2)$ R-symmetry.  $H_\eta$ is a charge of global $U(1)_{\rm punct}$  which rotates the phase of an adjoint hypermultiplet. The half index $\Pi(m,u)$ can be understood as a (twisted) partition function on three ball $B^3$ (half of $S^3$) with supersymmetric boundary condition   labelled by $(m,u)$ imposed on  $SU(2)$ vector multiplet  at the boundary $S^2 \times S^1$ ($=\partial B^3 \times S^1$).
The half index  can also be interpreted as a wave-function in a Hilbert-space canonically associated to the boundary.
We will identify the half index $\Pi (m,u;u_\eta)$ as a coherent state $|0\rangle \in \CH_{SL(2,\mathbb{C})}$\footnote {$|0\rangle$ can be viewed as a state in $\CH_{SL(2,\mathbb{C})} \otimes  \CH^{\textrm{knot}}_{SL(2,\mathbb{C})}$ with $m_\eta =0$ by regarding $\Pi(u_\eta)$ as $\langle m_\eta=0, u_\eta |0\rangle$.} as follows
\begin{align}
|0\rangle := \sum_{m\in \frac{\mathbb{Z}}2}  \oint \frac{du}{2\pi i u} \Delta(m,u) \Pi(m,u; u_\eta ) |m,u\rangle_{\FN}\;.
\end{align}
Then the 4d  index is given by the norm of this state,
\begin{align}
I_{\CN =2^*} (q,u_\eta)  =  \langle 0|0\rangle\;.
\end{align}
One can `excite' the vacuum state $|0\rangle$ by acting operators studied in section \ref{quantum riemann surface}. For example, by acting loop operators $\sO_{L} = \sW,\sH, \sD $ on $|0\rangle$  one obtains a half-index $|\sO_L\rangle :=\sO_L |0\rangle  $ with insertion of the loop operators. Taking the norm of the state, we could get the 4d superconformal index for $\CN=2^*$ theory  with insertion of loop operators at both north and south poles of $S^3$.
\begin{align}
I^{L}_{\CN=2^*}(q,u) = \langle \sO_L | \sO_L\rangle\;, \quad \textrm{4d index with loop operators $L$}\;.
\end{align}
We will define the space of half-indices, $\CH(S^2 \times S^1 )$, as the set of all  half-indices  $|\sO\rangle$ obtained by acting all quantum operators $\sO (\sqrt{\st}, \sqrt{\st'},\sqrt{\st''})$  on $|0\rangle$.
\begin{align}
\CH(S^2 \times S^1 ) := \{ |\sO\rangle:= \sO \cdot |0\rangle \;:\; \textrm{for all $\sO$} \}\;.
\end{align}
It is obvious  that  $\CH(S^2 \times S^1)$ is a subspace of $\CH_{SL(2,\mathbb{C})}$. The $SL(2,\mathbb{Z})$ action is closed in the subspace $\CH(S^2 \times S^1)$. In \cite{Gang:2012ff}, the following integral relation was found
\begin{align}
&\sum_{m_t}\oint \frac{du_t}{2\pi i u_t}\Delta (m_t, u_t)I_{\varphi} (m_b, m_t, u_b, u_t;m_\eta =0 , u_\eta) \Pi(m_t, u_t;u_\eta) \nn
\\
&=\Pi(m_b, u_b;u_\eta)\;\quad \textrm{for any $\varphi \in SL(2,\mathbb{Z})$}\;.
\end{align}
In section \ref{Z(Tr)=Z(T[SU(2)])} (see eq.~\eqref{mapping cylinder in hilbert}), we will identify the duality wall theory index $I_{\varphi}$ as a matrix element of an $SL(2,\mathbb{Z})$ operator $\varphi$ acting on $\CH_{SL(2,\mathbb{C})}$. In this interpretation, the above integral relation can be rewritten in the following simple form,
\begin{align}
\varphi \cdot |0\rangle = |0 \rangle \;\quad \textrm{for anly $\varphi \in SL(2,\mathbb{Z})$}\;.
\end{align}
For general element $|\sO \rangle \in \mathcal{H}(S^2\times S^1)$,
\begin{align}
\varphi |\sO \rangle = \varphi \cdot \sO |0 \rangle = \varphi_*(\sO) \cdot\varphi |0\rangle = |\varphi_*(\sO)\rangle \in \CH (S^2 \times S^1)\;.
\end{align}
In \cite{Gang:2012yr}, it was argued that the 4d superconformal index for
$\CN=2^*$ theory is invariant under $SL(2,\mathbb{Z})$ duality. As
an example, it is checked in \cite{Gang:2012yr} that a
superconformal index with Wilson line operators is the same as  an
index with `t Hooft line operators, which is $S$-dual of the Wilson
line. This $SL(2,\mathbb{Z})$ invariance of the index implies that
every $SL(2,\mathbb{Z})$ operator $\varphi$ are unitary operators in
$\CH(S^2 \times S^1)$.
\begin{align}
&\langle \varphi_*(\sO_1) | \varphi_*(\sO_2)\rangle =  \langle \sO_1 | \varphi^\dagger \varphi |\sO_2\rangle = \langle \sO_1 |\sO_2\rangle\;, \quad\textrm{for all $\sO_1, \sO_2$} \nn
\\
&\therefore \varphi^\dagger \varphi = 1 \; \textrm{in } \CH(S^2 \times S^1)\;.
\end{align}
It is compatible with the observation in section \ref{quantum riemann surface} that every $\varphi\in SL(2,\mathbb{Z})$ are unitary in $\CH_{SL(2,\mathbb{C})}$.

Turning off the puncture variable (setting $u_\eta \rightarrow q^\half$ as  in \cite{Gadde:2011ik}) ,
the `vacuum state' $|0\rangle$ be drastically simplified
\begin{align}
|0\rangle_{\mathbb{T}^2} = |0\rangle_{u_\eta \rightarrow q^\half} = \oint \frac{du}{2\pi i u}\Delta (m=0,u)|m=0,u\rangle\;.
\end{align}
Turning off the puncture, the once-puncture torus $\Sigma_{1,1}$ becomes a torus $\mathbb{T}^2$. Then, the  phase space
$(\CM,\Omega)_{SL(2,\mathbb{C})}$ becomes much simpler (for example,
see \cite{Aganagic:2002wv}) and the corresponding Hilbert-space
$\mathcal{H}_{SL(2,\mathbb{C})}$ also becomes simpler. Note that the
above `vacuum state' $|0\rangle_{\mathbb{T}^2}$ is the same as a
vacuum state $|0_v\rangle$ in \cite{Aganagic:2002wv} obtained by
quantizing the CS theory on  $\mathbb{T}^2$.

So far we have only considered operators of the form $\sO(\sqrt{\st},\sqrt{\st'},\sqrt{\st''})$ which  depends only on $\fl$ but not on $\sm$. One can excite $|0\rangle$ by an operator $\sO$ which depends on $\sm$. These operators corresponds to surface operators \cite{Gaiotto:2012xa} coupled to $U(1)_{\rm punct} $ in 4d $\CN =2^*$ theory. This interpretation is consistent with the results  in \cite{Alday:2013kda}, which relate surface operators in 4d $T_\Sigma$ theory   with  Wilson loop along $S^1$ direction in $\Sigma\times S^1$ in the context of 2d/4d correspondence. Recall that $\sm$ is obtained by quantizing the meridian variable $m = e^{U}$ which measures the holonomy along the $S^1$ direction in $\Sigma_{1,1}\times_\varphi S^1$.

\subsection{$ Z^{\Delta}_{\textrm{tori}(\varphi)} =Z^{\textrm{Tr}(\varphi)}_{\textrm{tori}(\varphi)} = I^{\textrm{T[SU(2)]}}_{\textrm{tori}(
\varphi)} $} \label{equality between three SL(2,C)/index ptns}

\subsubsection{$Z^{\textrm{Tr}(\varphi)}_{\textrm{tori}(\varphi)}=Z^{\Delta}_{\textrm{tori}(
\varphi)}  $} \label{Z(Tr)=Z(Delta)}

In this section, we  calculate the CS partition function on  ${\rm tori}(\varphi)$ with $|\Tr(\varphi)|>2$ using the canonical quantization on $\Sigma_{1,1}$. As explained in section \ref{quantum riemann surface}, the partition function can be represented as  a trace of a $SL(2,\mathbb{Z})$ operator $\varphi$ (see \eqref{L,R operators} for $\varphi= \sL, \sR$) on the Hilbert-space $\mathcal{H}_{SL(2,\mathbb{C})}$   and the partition function will be denoted by $Z^{\textrm{Tr}(\varphi)}_{\textrm{tori}(\varphi)}$.  We compare the CS partition function with the partition function $Z^{\Delta}_{\textrm{tori}(\varphi)}$(=$I^{\Delta}_{\textrm{tori}(\varphi)}$) calculated in section \ref{index-tetra} using tetrahedron decomposition and find an exact match. Classical equivalence of the two approaches was already proven in section \ref{A-polynomial for mapping torus} by analyzing A-polynomial, see also \cite{Terashima:2011xe}.\footnote{ In \cite{Terashima:2011xe}, they consider the case $G= SL(2,\mathbb{R})$ instead of $SL(2,\mathbb{C})$. But the A-polynomial computation in section \ref{A-polynomial for mapping torus} does not depends on  weather $G=SL(2,\mathbb{C})$ or $SL(2,\mathbb{R})$.   }

 For a concrete computation of  trace of $\varphi$ on $\CH_{SL(2,\mathbb{C})}$, we need to choose a basis of the Hilbert-space. In this section, we use the SR basis introduced in section \ref{index : Hilbert-space}. In the SR basis, the matrix element for $\varphi = \sL$,$\sR$ \eqref{L,R operators} is
\begin{align}
&I^{\SR}_{\varphi=\sL} (m_2, e_2, m_1, e_1; m_\eta,u_\eta) :=  _{\SR}\langle m_2,e_2 |\sL(m_\eta, u_\eta)| m_1 , e_1\rangle_{\SR}\;, \nn
\\
&I^{\SR}_{\varphi=\sR} (m_2, e_2, m_1, e_1; m_\eta,u_\eta) := _{\SR}\langle m_2, e_2 |\sR(m_\eta,u_\eta) | m_1,e_1\rangle_{\SR} \;.
\end{align}
According to \eqref{CS partition function as matrix element},  the right hand sides are the  $SL(2,\mathbb{C})$ CS partition functions for mapping cylinders $\Sigma\times_{\varphi=\sL,\sR} I$ in the polarization where   positions are $ (\ss_{\rm bot}, \ss_{\rm top}, \fl)$ and momenta are $ (\sr_{\rm bot}, \sr_{\rm top}, \sm)$. Recall that the boundary phase space for the mapping cylinder is locally $\CM_{\rm bot}(\Sigma_{1,1}) \times \CM_{\rm top} (\Sigma_{1,1}) \times \CM^{\rm knot}$ and $(\ss, \sr)$ are shear coordinates for $\CM(\Sigma_{1,1})$ and $(\ell, m)$  are (longitude, meridian) variable for $\CM^{\rm knot}$.
Since  every operator $\varphi \in SL(2,\mathbb{Z})$  depends only on $\sV_\pm$ but not on $\sU_\pm$, $\varphi$ can be understood as function on $(m_\eta, u_\eta)$ as explained in the  paragraph just above section \ref{index: Hilbert-space from 4d}. Using  properties in \eqref{property of O(l)}, the above indices in charge basis $(m_\eta,e_\eta)$ are
 \begin{align}
&I^{\SR}_{\varphi=\sL} (m_2, e_2, m_1, e_1; m_\eta,u_\eta)   \nn
\\
&= _{\SR}\langle (m_2, e_2),(m_\eta, e_{\eta})|\sL | (m_1,e_1),( m_\eta,e_\eta=0)\rangle_{\SR} \nn
\\
&= (-1)^{e_2-e_1}\delta (-e_1+2m_2-m_1-m_\eta)\delta(e_2 -e_1+2e_\eta +m_2 -m_1)  \nn
\\
&\quad  \times q^{\frac{1}4(-e_2+m_2+e_1-m_1)} \CI_{\Delta} (\half (-e_1+m_1+e_2-m_2),m_1+e_1 )  \;,
\end{align}
and
\begin{align}
&I^{\SR}_{\varphi=\sR} (m_2, e_2, m_1, e_1; m_\eta,e_\eta) \nn
\\
&= _{\SR}\langle (m_2, e_2),(m_\eta, e_{\eta})| \sR | (m_1,e_1),( m_\eta,e_\eta=0)\rangle_{\SR} \nn
\\
&=(-1)^{e_2-e_1}(-1)^{m_2 -e_2} \delta (m_2 + m_1 - e_2 + e_1)\delta(e_\eta)  q^{ \half ( m_2 + e_1)} \CI_{\Delta} ( -e_1-m_2,m_1 +e_1 ) \;.  \label{shear index for L,R}
\end{align}
Here, the state $|(m,e),(m_\eta, e_\eta)\rangle_{\SR}$ denotes a basis state  $|m,e\rangle_{\SR}\otimes |m_\eta,e_\eta \rangle$ in $\CH_{SL(2,\mathbb{C})} \otimes \CH^{\textrm{knot}}_{SL(2,\mathbb{C})} $.  A derivation for the above formula  is  given in appendix \ref{derivation of SR indices}.  The SR basis  charges $(m_i,e_i)$ are  half-integers with an additional condition $m_i+e_i \in \mathbb{Z}$.  The puncture variables $(m_\eta, e_\eta)$ are  in $(\mathbb{Z},\mathbb{Z}/2 )$.
 For later use, we will express these indices in the following form
 \begin{align}
 &I^{\SR}_{\varphi} (m_2, e_2, m_1, e_1; m_\eta,e_\eta)=(-1)^{e_2-e_1}\delta_{\varphi}(\ldots)       \CI_{\Delta} (M_\varphi, E_\varphi) \;, \label{shear index for L,R 2}
 \end{align}
For $\varphi=\sL, \sR$
 \begin{align}
&(M, E)_{\varphi} = \big{(}\frac{-e_1+m_1+e_2-m_2}2,m_1+e_1 + \frac{\hbar}2 \big{)} \;,&  \nn
\\
&\delta_{\varphi}(\ldots) = \delta (-e_1+2m_2-m_1-m_\eta)\delta (e_2 -e_1+2e_\eta +m_2 -m_1)\;,  &\textrm{for $\varphi=\sL$  } \nn
\\
&  \textrm{and}& \nn
\\
&(M, E)_{\varphi} = \big{(}-e_1-m_2+i\pi, m_1+e_1 + \frac{\hbar}2   \big{)} \;,  &\nn
\\
&\delta_{\varphi} (\ldots) = \delta(e_\eta)\delta(m_2+m_1-e_2+e_1)\;,&\textrm{for $\varphi=\sR$}  \;.
\label{shear index for L,R 3}
 \end{align}
 Factors like $(-1)^{\ldots}(q^{1/2})^{\ldots}$ in \eqref{shear index for L,R}  is reflected  in a shift of $(M, E)$ by $i \pi \mathbb{Q}+\frac{\hbar}2 \mathbb{Q}$. Recall   our definition of $\CI_\Delta(M,E)$ in \eqref{affine shift in tetrahedron index}. The $SL(2,\mathbb{C})$ CS partition function on $\textrm{tori}(\varphi)$ is given by
\begin{align}
&Z^{\textrm{Tr}(\varphi)}_{\textrm{tori}(\varphi)} (SL(2,\mathbb{C})) (m_\eta, u_\eta) = \textrm{Tr}_{\CH_{SL(2,\mathbb{C})}}  \big{(} \varphi (m_\eta, u_\eta) \big{)}\;\;  \textrm{in fugacity basis}\;,  \nn
\\
&Z^{\textrm{Tr}(\varphi)}_{\textrm{tori}(\varphi)} (SL(2,\mathbb{C})) (m_\eta, e_\eta) = \langle m_\eta, e_\eta |\textrm{Tr}_{\CH_{SL(2,\mathbb{C})}} \varphi |m_\eta,0 \rangle\;\; \textrm{in charge basis}.
\end{align}
Any element $\varphi \in SL(2,\mathbb{Z})$ with $|\Tr\varphi|>2$   can be written as (up to conjugation)
\begin{align}
\varphi = \varphi_N \varphi_{N-1}\ldots \varphi_2 \varphi_1 \;, \quad \varphi_i = \sL \textrm{ or } \sR\;.
\end{align}
Using the completeness relation \eqref{completeness relation in SR basis}, the partition function can be written as (the subscript $_{\SR}$ is omitted to avoid clutter)
\begin{align}
&Z^{\textrm{Tr}(\varphi)}_{\textrm{tori}(\varphi)} (SL(2,\mathbb{C})) (m_\eta, e_\eta) = \langle m_\eta, e_\eta |\textrm{Tr}_{\CH_{SL(2,\mathbb{C})}} \varphi |m_\eta,0 \rangle \nn
\\
&= \sum_{\{ e_{\eta,*},m_{\eta,*},m_*,e_* \}} \langle (m_1, e_1),(m_{\eta},e_{\eta}) |\varphi_N |(m_N,e_N),(m_{\eta,N}, e_{\eta,N}) \rangle \ldots  \nn
\\
&  \qquad    \qquad \qquad  \qquad \ldots \langle (m_2,e_2),(m_{\eta}, e_{\eta,2}) | \varphi_1 |(m_1, e_1),(m_\eta , 0)\rangle \;,\nn
\\
& = \sum_{\{e_{\eta, *},m_*,e_* \}}\delta(e_\eta -\sum_{k=1}^N e_{\eta,k})\; \langle (m_1, e_1),(m_{\eta},e_{\eta,N}) |\varphi_N |(m_N,e_N),(m_{\eta}, 0) \rangle \ldots  \nn
\\
& \qquad    \qquad \qquad  \qquad  \ldots \langle (m_2,e_2),(m_{\eta}, e_{\eta,1}) | \varphi_1 |(m_1,e_1),(m_\eta ,0)\rangle \;, \nn
\\
&=\sum_{\{ e_{\eta,*}, m_*, e_* \}}\delta (e_\eta -\sum_{k=1}^N e_{\eta,k})\ldots I^{\SR}_{\varphi_{i+1}}(m_{i+2},e_{i+2}, m_{i+1},e_{i+1},m_\eta ,e_{\eta,i+1})  \nonumber
\\
&\quad \quad  \times I^{\SR}_{\varphi_{i}}(m_{i+1}, e_{i+1},m_{i}, e_{i}, m_\eta ,e_{\eta,i})  I^{\SR}_{\varphi_{i-1}}(m_{i}, e_{i},m_{i-1}, e_{i-1},m_\eta,e_{\eta,i-1}) \ldots \;.
\label{general mapping tori}
\end{align}
In the second line,  we used the fact that $\varphi_i$  depends only on $\sV_\pm$ but not on $\sU_\pm$ and the property in eq.~\eqref{property of O(l)}.   $I^{\Pi_{\SR}}_{\varphi_i}$ in the third line can be written as\footnote{The factor $(-1)^{e_{i+1}-e_i}$ in $I^{\Pi_{\SR}}_{\varphi_i}$ is ignored in this expression since $\prod_{i=1}^N (-1)^{e_{i+1}-e_i}$ =1 and thus the sign factors do not appear in the final expression for $Z^{\Tr (\varphi)}_{\textrm{tori}(\varphi)}$.}
\begin{align}
& I^{\SR}_{\varphi_i} =\delta_{\varphi_i}(\ldots)  \CI_{\Delta}(M_{\varphi_i},E_{\varphi_i}) \,,
\end{align} 
where $(M_{\varphi_i},E_{\varphi_i}, \delta_{\varphi_i}(\ldots) )$
for $\varphi_i=\sL,\sR$  are given in  \eqref{shear index for L,R 3}
with $(m_2,e_2, m_1,e_1, m_\eta, e_\eta)$ replaced by
$(m_{i+1}, e_{i+1}, m_{i},e_i, m_\eta, e_{\eta,i})$. 
The index $i$ runs cyclically from $1$ to  $N$. There are $2N+1$ Knoneker deltas in
the above expression \eqref{general mapping tori}.  Among them, $2N$
equations  come from $\delta_{\varphi_i}(\ldots)|_{i=1,\ldots, N}$.
These $2N$ equations can be solved by parametrizing
$(M_{\varphi_i},E_{\varphi_i},e_{\eta,i})|_{i=1}^{N}$ variables
($3N$ in total) in terms of $N$ variables $\{ w_i \}$
 \begin{align}
  &M_{\varphi_i} = M_i (w_{i+1},w_i, w_{i-1})\;, \nonumber
  \\
  &E_{\varphi_i} = E_i (w_{i+1},w_i, w_{i-1}) \;, \nonumber
  \\
  &e_{\eta,i}  = e_{\eta,i}(w_i)\;,
  \end{align}
where the  $M_i,E_i,e_{\eta,i}$ is given in Table 5.
\begin{table}[htbp]
   \centering
   \begin{tabular}{@{} l|c|c|c|c|c @{}} 
      \toprule

     $(\varphi_i ,\varphi_{i-1})$  &  $(L,L)$  & $(L,R)$ & $(R,R)$ & $(R,L)$  \\
      \midrule
      $M_i$ &$\frac{-w_{i+1}+2w_i -w_{i-1}}2$   & $\frac{-w_{i+1}+w_i +w_{i-1}-m_\eta-i\pi}2$ &  $\frac{w_{i-1}+ w_{i+1}}2$ &$\frac{w_{i+1}+ w_i-  w_{i-1}+m_\eta+i\pi}2$  \\
      \midrule
       $E_i$ &$i \pi +\frac{\hbar}2 -w_i$  & $i \pi +\frac{\hbar}2 -w_i$ &  $i \pi +\frac{\hbar}2 -w_i$ & $i \pi +\frac{\hbar}2 -w_i$ \\
          \midrule
      $e_{\eta,i}$ &$\frac{1}2(w_{i+1}-w_i)$   & $\frac{1}2(w_{i+1}-w_i)$  &  0 &0 \\
      \bottomrule
   \end{tabular}
   \caption{Solution for constraints from $2N$ Knonecker delta's}
   \label{tetra-gluing 3}
\end{table}

\noindent
From straightforward calculation, one can  check that these parametrizations satisfy all equations from the $2N$ Kronecker deltas.
Substituting this solution into eq.~\eqref{general mapping tori}, the CS partition function can be written as
\begin{align}
&Z^{\textrm{Tr}(\varphi)}_{\textrm{tori}(\varphi)} (SL(2,\mathbb{C})) (m_\eta, e_\eta)= \sum_{w_*}\delta \big{(}e_\eta ,\frac{1}2\sum_{\varphi_k =L} (w_{k+1}-w_k) \big{)} \prod_{i=1}^{N} \CI_{\Delta} (M_i(w_*) ,E_i(w_*)) \;.
 \label{final general index}
\end{align}
Comparing this index with the index in \eqref{mapping torus from tetra} and comparing Table \ref{tetra-gluing 2} and Table \ref{tetra-gluing 3}, we see the following identification
\begin{align}
&w_i \leftrightarrow W_i, \quad  (M_i,E_i)\leftrightarrow (X_i, P_i),\quad m_\eta \leftrightarrow \sV\;.
\end{align}
Under the identification, we see that
\begin{align}
\prod_{i=1}^{N} \CI_{\Delta} (M_i(w_*) ,E_i(w_*)) = \prod_{i=1}^{N} \CI_{\Delta}\big{(}X_i (W_*), P_i(W_*)\;.
\end{align}
The electric charge  $e_{\eta}$ for $U(1)_\eta$ is related to $\sU$ in the following way
\begin{align}
&\sU=\sum_i \sU_i = -\frac{1}2 \sum_{(\varphi_i,\varphi_{i-1}) = (L,R)}W_i +\frac{1}2 \sum_{(\varphi_i,\varphi_{i-1}) = (R,L)}W_i   \;, \nonumber
\\
&e_\eta=\sum_i e_{\eta,i} =\frac{1}2 \sum_{\varphi_i = L}  (w_{i+1}-w_i) =\sU\;,\quad  \textrm{under the identification $w_i = W_i$} \;.
\end{align}
From the above identifications, we see that
\begin{align}
Z^{\textrm{Tr}(\varphi)}_{\textrm{tori}(\varphi)} (SL(2,\mathbb{C})) (m_\eta, e_\eta)= I^{\Delta}_{\textrm{tori}(\varphi)} (\sV,\sU)|_{\sV=m_\eta, \sU=e_\eta}\;.
\end{align}
for general $\varphi$ with $|\Tr (\varphi)|>2 $. One remarkable property of $Z^{\textrm{Tr}(\varphi)}_{\textrm{tori}(\varphi)} (SL(2,\mathbb{C}))$ is that it always vanishes when $e_\eta \in \mathbb{Z}+\half$.

\subsubsection{$Z^{\textrm{Tr}(\varphi)}_{\textrm{tori}(\varphi)}= I^{\textrm{T[SU(2)]}}_{\textrm{tori}(
\varphi)}  $} \label{Z(Tr)=Z(T[SU(2)])}

\paragraph{Duality wall index  as  matrix element in FN basis}

We will argue that the mapping cylinder index $I_\varphi$  studied in section \ref{index-wall} can be written as the matrix element of $\varphi \in SL(2,\mathbb{Z})$ in the FN basis. More explicitly,
 \begin{align}
 &I_{\varphi} (m_b, u_b, m_t,u_t,m_\eta,u_\eta):=_{\FN}\langle m_b,u_b | \varphi (m_\eta, u_\eta)| m_t, u_t \rangle_{\FN} \;, \quad \textrm{or equivalently} \nonumber
 \\
 &I_{\varphi} (m_b, u_b , m_t,u_t,m_\eta,e_\eta):=_{\FN} \langle (m_b,u_b),(m_\eta,e_\eta)| \varphi | (m_t,u_t),(m_\eta, 0)\rangle_{\FN} \;. \label{mapping cylinder in hilbert}
 \end{align}
for any operator $\varphi \in SL(2,\mathbb{Z})$ acting on a Hilbert space $ \CH_{SL(2,\mathbb{C})}$.\footnote{More precisely, $I_\varphi (m_b, u_b, m_t, u_t) = \;^S _{\FN}\langle m_b, u_b |\varphi |m_t, u_t\rangle^S_{\FN}$ where $| m,u \rangle^S_{\FN}$ is a Weyl-reflection invariant combination of FN basis \eqref{S-FN basis}.   However, it does not matter since operator $\varphi$ is Weyl-reflection invariant, $_{\FN}\langle m_b, u_b|\varphi|m_t, u_t\rangle_{\FN} =\;^S _{\FN}\langle m_b,u_b|\varphi|m_t, u_t\rangle^S_{\FN} $, and we will not distinguish them.}  The right hand side is the  $SL(2,\mathbb{C})$ CS partition function for mapping cylinder in the polarization where  positions are $ (\hat{\lambda}_{\rm bot}, \hat{\lambda}_{\rm top}, \fl)$ and momenta are $ (\hat{\tau}_{\rm bot}, \hat{\tau}_{\rm top}, \sm)$.   Thus, the above statement is nothing but the 3d-3d dictionary in  \eqref{3d-3d dictionary for SCI} for $M= \Sigma_{1,1}\times_\varphi I$.
 Assuming eq.~\eqref{mapping cylinder in hilbert} holds,  the index for mapping torus theory can be represented as
 \begin{align}
 I^{T[SU(2)]}_{\textrm{tori}(\varphi)} (m_\eta,u_\eta) &= \sum_m \oint \frac{du}{2\pi i u}\Delta(m,u) I_{\varphi}(m,u,m,u,m_\eta, u_\eta)\;,  \quad \textrm{from }\eqref{mapping torus index from duality wall}\nn
 \\
 &=   \sum_m \oint \frac{du}{2\pi i u} \Delta (m,u) \;_{FN}\langle m_b, u_b| \varphi(m_\eta, u_\eta)|m_t, u_t\rangle_{FN}\;, \quad \textrm{from \eqref{mapping cylinder in hilbert}} \nn
 \\
&= \Tr_{\mathcal{H}_{SL(2,\mathbb{C})}} \varphi (m_\eta, u_\eta)\;, \quad \textrm{using } \eqref{completeness relation in FN basis}\;.\label{mapping torus in hilbert}
 \end{align}
Note that the quantity in the last line is nothing but $Z^{\textrm{Tr}(\varphi)}_{\textrm{tori}(\varphi)}(SL(2,\mathbb{C}))$. Thus the proposal  \eqref{mapping cylinder in hilbert}  automatically ensures that $Z^{\textrm{Tr}(\varphi)}_{\textrm{tori}(\varphi)}(SL(2,\mathbb{C}))= I^{\textrm{T[SU(2)]}}_{\textrm{tori}(
\varphi)} $, which is the main result of this section.
  How can we justify the proposal in \eqref{mapping cylinder in hilbert}?  
  There are two steps in the argument for the proposal.
  First we will argue  that the proposal  holds for $\varphi=\sL, \sR$ by $i)$ showing  the two sides in \eqref{mapping cylinder in hilbert} satisfy the same difference equations and $ii)$ by directly comparing the two sides in $q$-expansion.
Then, we will prove that
  \begin{align}
  \textrm{ If  the proposal \eqref{mapping cylinder in hilbert} holds for $\varphi_1, \varphi_2$, then it also holds for $\varphi = \varphi_2 \cdot \varphi_1 $}. \label{proposition on gluing}
  \end{align}
  From the two arguments, we can claim that  \eqref{mapping cylinder in hilbert} holds for general $\varphi\in SL(2,\mathbb{Z})$ which can be written as a product of $\sL$'s and $\sR$'s.

\paragraph{Proof of \eqref{proposition on gluing}} Since the second argument is much simpler to prove,  let's prove it first.
Suppose  \eqref{mapping cylinder in hilbert} holds for $\varphi_1$ and  $\varphi_2$, then
\begin{align}
 & _{\FN}\langle m_b,u_b | \varphi_2 \varphi_1  | m_t,u_t \rangle_{\FN}  \nonumber
 \\
 &=\sum_{m'} \oint \frac{du'}{2\pi i u'} \Delta (m',u')\langle m_b,u_b| \varphi_2 | m',u' \rangle \langle m',u' | \varphi_1 | m_t, u_t \rangle  \;, \quad \textrm{using } \eqref{completeness relation in FN basis} \nonumber
 \\
 &= \sum_{m'} \oint \frac{du'}{2\pi i u'} \Delta (m',u') I_{\varphi_2} (m_b, u_b, m', u'; m_\eta, u_\eta)I_{\varphi_1} (m', u',m_t, u_t;m_\eta, u_\eta) \;, \nonumber
 \\
 &= I_{\varphi_2 \varphi_1 }(m_b, u_b, m_t, u_t;m_\eta, e_\eta) \;, \quad \textrm{using } \eqref{odot gluing rule in T[SU(2)] theories}\;.
\end{align}
Thus the proposal also holds $\varphi = \varphi_2 \cdot \varphi_1$.

\paragraph{Check of \eqref{mapping cylinder in hilbert} for $\varphi= \sL, \sR$   by difference equations }
The index for $T[SU(2)]$ described in section \ref{index-wall} satisfies the following difference equations,
\\
\begin{align}
\big{(}\sW_b - (\sH^{\rm T})_t  \big{)}_\pm  \cdot I_{\varphi=S} =0\;, \quad (\sH_b - (\sW^{\rm T})_t)_\pm \cdot I_{\varphi=S}=0\;, \nn
\\
 \bigg{(}p_\eta-\big{(}\frac{1}{p^\half-p^{-\half}}(x -x^{-1}) \big{)}_b \bigg{)}_\pm \cdot I_{\varphi=S} =0\;. \label{diff for S}
\end{align}
The Wilson loop operator $\sW$ and `t Hooft operator $\sH$ are given by  (cf. \eqref{loop2FN-q})\footnote{In \eqref{loop2FN-q}, loop operators act on $\CH_{SL(2,\mathbb{C})}$. On the other hand, loop operators here are difference operators acting on a function $I(m_b, u_b, m_t, u_t; m_\eta, u_\eta)$. }
\begin{align}
&\sW_\pm = x_\pm +x_\pm^{-1} \;,  \nn
\\
&\sH_\pm= \frac{q^{\mp 1/4} x_\pm x^{\half}_{\eta;\pm}- q^{\pm 1/4} x^{-1}_{\pm} x^{-\half}_{\eta,\pm} }{x_\pm- x^{-1}_\pm} p_\pm^{-\half} + \frac{q^{\pm 1/4} x_\pm x_{\eta,\pm}^{-\half}- q^{\mp 1/4} x_\pm^{-1} x^\half_{\eta,\pm}}{x_\pm- x_\pm^{-1}} p_\pm^{\half}\;.
\end{align}
Basic operators $(x_\pm, p_\pm)$ act on the charge basis index as
\begin{align}
& x_\pm := \exp ( \frac{\hbar}2 m \mp \partial_e)\;,  \; p_\pm := \exp ( \frac{\hbar}2 e \pm \partial_m)\;.
\end{align}
On the fugacity basis index, they act as
\begin{align}
& x_\pm := \exp ( \frac{\hbar}2 m \pm \log u)\;,  \; p_\pm := \exp ( \frac{\hbar}2 \partial_{\log u} \pm \partial_m) \;.
\end{align}
Depending on the subscript $(b,t,\eta)$, they act on (`bot',`top',`punct') parameters, respectively. The notation $\sO^{\rm T}$ denotes a `transpose' of $\sO$ to be defined for each operator. For $\sW$, $\sH$, the transposed operators are
\begin{align}
\sW^{\rm T} = \sW\;, \quad \sH^{\rm T} =\sH/.\{q\rightarrow q^{-1}, p\rightarrow p^{-1}\} \;.
\end{align}
The difference equations can be simplified using `shear' operators. Shear operators
$(\sqrt{\st},\sqrt{\st'},\sqrt{\st''})_{\pm}:= (\exp(\frac{1}2 \sT),\exp(\frac{1}2 \sT'),\exp(\frac{1}2 \sT''))_\pm$ are defined as (cf. \eqref{shear2FN-q})
\begin{align}
&(\sqrt{\st})_{\pm} = \frac{i}{x_{\pm} -x_{\pm}^{-1}}(p_{\pm}^{-1/2} -p_{\pm}^{1/2}) \;,\nonumber
\\
&(\sqrt{\st'})_{\pm} =  \frac{i}{q^{\mp 1/4}x_{\pm}^{-1}p_{\pm}^{1/2} - q^{\pm1/4}p_{\pm}^{-1/2} x_{\pm}}(x_\pm -x_{\pm}^{-1}) \;, \nonumber
\\
&(\sqrt{\st''})_{\pm} = q^{\mp 1/4}x^\half_{\eta,\pm} \frac{i}{p^{-1/2}_\pm -p^{1/2}_\pm}(q^{\mp 1/4}x_{\pm}^{-1}p_{\pm}^{1/2} - q^{\pm1/4}p_{\pm}^{-1/2} x_{\pm}) \;. \label{shear operators}
 \end{align}
In terms of the shear operators, the difference equations can be written as
\begin{align}
\big{(}(\frac{1}{\sqrt{\st}})_b - (\sqrt{\st}^{\rm T})_t \big{)}_\pm \cdot I_{\varphi=S} =0\;, \quad \big{(}\sqrt{\st'} (1+q^{\half} \st)_b - (\sqrt{\st''}^{\rm T})_t \big{)}_\pm \cdot I_{\varphi=S} =0\;,  \nn
\\
 \big{(} p_\eta- i (\frac{1}{\sqrt{\st}})_b  \big{)}_\pm \cdot I_{\varphi=S} =0\;.  \label{diff for S-1}
\end{align}
For shear operators, the transposed operators are
\begin{align}
(\sqrt{\st},\sqrt{\st'}, \sqrt{\st ''})^{\rm T} =(\sqrt{\st},\sqrt{\st'}, \sqrt{\st ''}) /.  \{q\rightarrow q^{-1},  p\rightarrow p^{-1}\}\;. \label{Shear transpose}
\end{align}
For $\varphi=\sL (=S^{-1}T^{-1}S) , \sR(=T)$, the corresponding duality wall theory  indices $I_\varphi$ satisfy following difference equations.
\begin{align}
 &\big{(}(\frac{1}{\sqrt{\st''}})_b -  (\sqrt{\st}^{\rm T})_t\big{)}_\pm \cdot  I_{\varphi=\sL} =0\;, \quad  \big{(} \sqrt{\st}_{\pm}(1+q^{\pm 1/2}\st''_\pm)_b-(\sqrt{\st''}^{\rm T}_\pm)_t\big{)} \cdot I_{\varphi=\sL} =0 \;, \nonumber
 \\
 &\big{(} p_\eta -  \frac{(\sqrt{\st}^{\rm T})_t}{(\sqrt{\st})_b} \big{)}_\pm \cdot I_{\varphi=\sL} =0\;, \label{diff for L}
 \\
  &\big{(}(\frac{1}{\sqrt{\st'}})_b -  (\sqrt{\st}^{\rm T})_t\big{)}_\pm \cdot  I_{\varphi=\sR} =0\;, \quad  \big{(}\sqrt{\st''}_{\pm}(1+q^{\pm 1/2}\st'_\pm)_b- (\sqrt{\st''}^{\rm T}_\pm)_t\big{)} \cdot I_{\varphi=\sR} =0 \;, \nonumber
 \\
 &\big{(} p_\eta -  1\big{)}_\pm \cdot I_{\varphi=\sR} =0\;. \label{diff for R}
 \end{align}
One can check these difference equations by series expansion in $q$ at any desired order.
For $\varphi=\sR$, we have a closed expression \eqref{mapping cylinder index for T} for $I_\varphi$, from which we can  check that
\begin{align}
&(x_t - x_b)_\pm \cdot I_{\varphi=\sR} = 0\;, \quad (p_{\eta,\pm} -1)\cdot I_{\varphi=\sR}=0\;, \nn
\\
&\big{(} (\frac{1}{x_\pm-x_\pm^{-1}}p_\pm^{-\half} (x_\pm-x_\pm^{-1}))_t - (q^{\mp \frac{1}4}x^{-1}_\pm p^{1/2}_\pm)_b \big{)}\cdot I_{\varphi=\sR} =0\;, \label{diff for R-2}
\end{align}
by a brute-force computation. Expressing these difference equations
in terms of shear operators, we obtain the difference equations for
$I_{\varphi=\sR}$ in \eqref{diff for R}.

Among the three difference equations in each of \eqref{diff for S-1}, \eqref{diff for L} and \eqref{diff for R}, two are of the form  $\varphi_* (\sO)_b-\sO^{\rm T}_t \simeq 0$. From a purely 3d field theory point of view, there is no prior reason for that. As we will see below,  this structure of the difference equations  can be naturally understood from \eqref{mapping cylinder in hilbert}.
Another interesting property of these difference equations is that they are always in  $\pm$ pair. It is related to the factorization of 3d superconformal indices \cite{Beem:2012mb,Hwang:2012jh} and this property is not restricted on duality wall theories. How can we guess these difference equations?  Difference equations of the form $\varphi_* (\sO)_b-\sO^{\rm T}_t \simeq 0$ are largely motivated by the difference equations for $S^3_b$ partition function for $T[SU(2),\varphi]$ theory studied in \cite{Dimofte:2011jd}.  From the works \cite{Dimofte:2011ju,Dimofte:2011py}, we know that the $S^3_b$ partition function and $S^2\times S^1$ superconformal index satisfy the same form of difference equations. A direct way of obtaining the difference equations is expressing the mapping cylinder indices in terms of tetrahedron indices and using the gluing rules for   difference equations explained in \cite{Dimofte:2011gm} (see also \cite{Dimofte:2011py}). As we will see in appendix \ref{T[SU(2)] classical},  $I_{\varphi=S}$ can be expressed  by gluing 5 tetrahedron indices with two internal edges. However, the corresponding operator equations for difference equation gluing  is too complicated to solve. In appendix \ref{T[SU(2)] classical}, we consider classical Lagrangian (set of difference equations in the limit $q\rightarrow 1$) for $I_{\varphi=S}$. In the classical limit, operator equations become  equations for ordinary commuting variables that are relatively easy to solve.
In this way, we obtain the difference equations for $I_{\varphi=S}$ in the classical limit and check these exactly matches the difference equations in eq.~\eqref{diff for S} with $q=1$. We want to emphasize that a pair of difference equations involving $p_\eta$ is obtained from quantization of a  classical equation involving $p_\eta$ in  the classical Lagrangian obtained in appendix \ref{T[SU(2)] classical}. The ordering ambiguity is fixed by checking corresponding difference equation  in $q$ expansion.

Now let us consider difference equations satisfied by the matrix element in right-hand side of \eqref{mapping cylinder in hilbert}.  From the operator equations \eqref{conditions for L} for $\varphi=\sL, \sR$   and the following observations,
\begin{align}
&_{\FN}\langle m, e| \sO(\hat{\lambda}_\pm, \hat{\tau}_\pm)| I\rangle = \sO(x_\pm, p_\pm)\cdot \;_{\FN}\langle m, e|I\rangle\;, \nonumber
\\
&\langle I | \sO(\hat{\lambda}_\pm, \hat{\tau}_\pm)| m,- e\rangle_{\FN} = \sO^{\rm T}(x_\pm, p_{\pm})\cdot \langle I|m,-e\rangle_{\FN}\;,  \nn
\\
&\langle m_\eta, e_\eta|\sm_\pm \cdot  \sO \cdot  \sm_\pm^{-1} |m_\eta, e_\eta =0\rangle = p_{\eta,\pm} \cdot \langle m_\eta, e_\eta| \sO |m_\eta, e_\eta =0\rangle\;,
\end{align}
one can check that the matrix element in \eqref{mapping cylinder in hilbert} satisfies the same difference equations in \eqref{diff for L} and \eqref{diff for R} for $\varphi=\sL,\sR$.
The transposed operator $\sO^{\rm T}$ is defined by $\sO^{\rm T} = (\sO^\dagger)^*$, where the complex conjugation $*$ is given as
\begin{align}
  (c\; \hat{\lambda}_\pm^{m_1} \hat{\tau}_\pm^{m_2} \fl_\pm^{m_3}  q^{m_4})^* = c^* \hat{ \lambda}_\mp ^{m_1} \hat{\tau}_\mp^{-m_2} \fl_\mp^{m_3} q^{m_4}\;, \quad \textrm{$c$ : $c$-number}
\end{align}
Transpose of shear operators are  (using eq.~\eqref{adjoint of shear operator})
\begin{align}
(\sqrt{\st},\sqrt{\st'},\sqrt{\st''})^{\rm T} =(\sqrt{\st},\sqrt{\st'},\sqrt{\st''})/.\{ q\rightarrow q^{-1}, \hat{\tau} \rightarrow \hat{\tau}^{-1}\}\;.
\end{align}
It is compatible with \eqref{Shear transpose}.
\paragraph{Check of \eqref{mapping cylinder in hilbert} for $\varphi= \sL, \sR$   by direct computation in $q$ expansion } A more direct evidence for the proposal in \eqref{mapping cylinder in hilbert} is an explicit comparison  of both sides in $q$-expansion.
Plugging the completeness relation in the SR basis into \eqref{mapping cylinder in hilbert}, 
\begin{align}
&(\textrm{right-hand side in \eqref{mapping cylinder in hilbert}})  \nn \\
&=
\sum_{(m_1,e_1)} \sum_{(m_2, e_2)} \;_{\FN}\langle m_b, u_b | m_2, e_2 \rangle_{\SR} \langle m_2, e_2| \varphi  | m_1 , e_1 \rangle _{\SR}
\langle   m_1, e_1 | m_t, u_t  \rangle_{\FN} \, ,
\label{equiv}
\end{align}
and using the following relation between SR and FN basis in \eqref{eq:D-charge basis}
\footnote
{To compute $_{\FN}\langle m_b, u_b | m_2, e_2 \rangle_{\SR}$, we need to take the complex conjugation on the expression. In taking the conjugation, we regard $(-1)$ as $e^{ i \pi}$. Thus,
\begin{equation}
_{\FN}\langle m_b, u_b | m_2, e_2 \rangle_{\SR}
=\sum_{e \in \mathbb{Z}} (-1)^{-m_2}q^{ \half m_2} u_b^{2e+e_2 -m_2} \CI_{ \Delta} (-m_2 -m_b, e) \CI_{ \Delta} (-m_2+m_b, -e_2+m_2 -e) \, . 
\nn
\end{equation}
}
\begin{equation}
_{\SR}\langle m_1, e_1 | m_t, u_t  \rangle_{\FN}
= \sum_{ e \in \mathbb{Z} } (-1)^{m_1} q^{ \half m_1} u_t^{ - 2e - e_1+m_1}\CI_{ \Delta} (-m_1 -m_t , e) \CI_{ \Delta}(  -m_1+m_t , - e_1+m_1-e) \, ,
\nn
\end{equation}
one obtains the following expression 
\begin{align}
&_{\FN} \langle (m_b,u_b),(m_\eta,e_\eta)| \varphi | (m_t,u_t),(m_\eta, 0)\rangle_{\FN}   \nn
\\
&=
\sum_{(m_1, e_1)} \sum_{(m_2, e_2)}
\sum_{e,e' \in {\mathbb Z}}
(-1)^{m_1-m_2} q^{ \half (m_1+m_2)} u_b^{ 2e+e_2-m_2} u_t^{ -2e'-e_1+m_1}  \nn \\
 & \times \CI_{ \Delta} (-m_2-m_b,e) \CI_{ \Delta}(-m_2+m_b, -e_2+m_2-e) \CI_{ \Delta}(-m_1-m_t, e') \nn
 \\
 &\times \CI_{ \Delta}(-m_1+m_t, -e_1+m_1-e')
 _{\SR}\langle (m_2,e_2),(m_{ \eta}, e_{ \eta}) | \varphi | (m_1, e_1), (m_{ \eta}, 0) \rangle_{\SR} \, .
 \label{result-fn}
\end{align}
The summation ranges are over $m_i,e_i \in \frac{\mathbb{Z}}2$ such
that $e_i+m_i \in \mathbb{Z}$ due to the completeness relation in SR
basis. Plugging the matrix elements in eq.~\eqref{shear index for
L,R} into eq.~\eqref{result-fn}, one obtains 
\begin{align}
&_{\FN} \langle (m_b,u_b),(m_\eta,e_\eta)|  \sL  | (m_t,u_t),(m_\eta, 0)\rangle_{\FN}    \nn
\\
&=
\sum_{m_1, e_1, m_2, e_2, e,e' }
(-1)^{m_1-m_2+e_2-e_1} q^{ \frac{1}{4} (m_1+e_1+3m_2-e_2)}  u_b^{ 2e+e_2-m_2} u_t^{ -2e'-e_1+m_1} \nn \\
&  \times \delta (-e_1+2m_2-m_1-m_\eta)\delta(e_2-e_1+2e_{\eta} + m_2 - m_1 ) \nn
\\
 & \times \CI_{ \Delta} (-m_2-m_b,e) \CI_{ \Delta}(-m_2+m_b, -e_2+m_2-e) \CI_{ \Delta}(-m_1-m_t, e')  \nn \\
 & \times \CI_{ \Delta}(-m_1+m_t, -e_1+m_1-e')  \CI_{ \Delta}  ( \half(-e_1+m_1+e_2-m_2), m_1+e_1)
 \label{L-fn} \;.
\end{align}
We show some  examples of explicit evaluation of the above  formula in $q$-expansion
 \begin{align}
&_{\FN} \langle m_b,u_b| \sL (m_\eta, u_\eta)| m_t,u_t\rangle_{\FN}  \qquad \textrm{for }(m_b, m_t, m_{ \eta}) = (0,0,0)  \nn
\\
&= \sum_{e_{ \eta} \in \mathbb{Z}/2} {_{\FN} \langle (m_b,u_b),(m_\eta,e_\eta)| \sL | (m_t,u_t),(m_\eta, 0)\rangle_{\FN}}\; u_{ \eta}^{ e_{ \eta}}  \qquad \textrm{for }(m_b, m_t, m_{ \eta}) = (0,0,0) \nn \\
& = 1 + \big{(} \frac{1}{u_{ \eta}} \chi_1 (u_t)+ u_{ \eta} \chi_1 (u_b)  \big{)} q^{ \half}   + \big{(}-1- \chi_1 (u_b) - \chi_1(u_t)  + u_{ \eta}^{-2} \chi_2 (u_t) + u_{ \eta}^2 \chi_{2} (u_b) \big{)} q + O (q^{ \frac{3}{2}})
\nn \\
&_{\FN} \langle m_b,u_b| \sL (m_\eta, u_\eta)| m_t,u_t\rangle_{\FN} \qquad \textrm{for }(m_b, m_t, m_{ \eta}) = (0,0,1)  \nn \\
& = \left(  u_{ \eta}^{-1} \chi_{ \half} (u_b) \chi_{ \half} (u_t) - \chi_{ \half} (u_b) \chi_{ \half} (u_t)  \right) q + O(q^{ \frac{3}{2}})
\nn \\
&  _{\FN} \langle m_b,u_b| \sL (m_\eta, u_\eta)| m_t,u_t\rangle_{\FN}  \qquad \textrm{for }(m_b, m_t, m_{ \eta}) = (1,0,1) \nn \\
&= \frac{u_b}{ u_{ \eta}} \chi_{ \half} (u_t) q^{ \half} + \left( u_b^{-1} \chi_{ \half} (u_t) - u_b (-u_{ \eta}^{-2} + u_{ \eta}^{-1}) \chi_{ \frac{3}{2}}(u_t) \right) q^{ \frac{3}{2} }+ O(q^2)
\nn
 \end{align}
 where $ \chi_j (u)  $ is the character for $2j+1$ dimensional representation of $SU(2)$, $ \chi_j (u) := \sum_{l=-j}^{j} u^{ 2j}$. The result agrees with the index $I_{\sL}(m_b, u_b, m_t, u_t; m_\eta, e_\eta)$ obtained using the duality domain wall theory in section \ref{index-wall}.

 For $\varphi=\sR$, we will start from the index in the FN basis. In eq.~\eqref{mapping cylinder index for T}, we found that
 \begin{align}
 I_{\varphi = \sR} = _{\FN}\langle m_{b},u_b |\sR |m_t ,u_t\rangle_{\FN} = u_b^{2m_b}\delta (m_b- m_t) \frac{\delta(u_b - u_t)}{\Delta(m_t, u_t )}\;.
 \end{align}
 Performing the basis change from FN to SR, we find
 \begin{align}
 _{\SR}\langle m_2, e_2 | \sR |m_1, e_1\rangle_{\SR} = \sum_{m_b} \oint \frac{du_b}{2\pi i u_b}  u_b^{2m_b}\Delta(m_b, u_b) _{\SR}\langle m_2 ,e_2| m_b, u_b\rangle_{\FN} \langle m_b, u_b | m_1, e_1\rangle_{\SR} \;.\nn
 \end{align}
 Using \eqref{eq:D-charge basis}, the above formula can be explicitly evaluated in $q$-expansion, which matches with $I^{\SR}_{\varphi=\sR}$ in \eqref{shear index for L,R}.


\section{Squashed sphere partition function/$SL(2,\mathbb{R})$ CS partition function} \label{sec: SL(2,R)/squashed S3}

The squashed three sphere partition function of $T[SU(2), \varphi]$
has been discussed extensively in recent literature \cite{Dimofte:2011ju,Dimofte:2011jd,Terashima:2011qi,Terashima:2011xe,Vartanov:2013ima}, where its relation to the $SL(2,\IR)$ CS partition function and quantum Teichm\"uller theory
was pointed out. In this section, we review some salient features
of the these work to help clarify the similarities
and differences between the superconformal index of the previous section
and the three sphere partition function.

\paragraph{Quantum dilogarithm identities}
Before we proceed, let us take a brief digression to
review some properties of the non-compact quantum dilogarithm (QDL) function
\cite{Faddeev:1995nb,Kashaev} which plays a fundamental role throughout this section
and in comparison with section \ref{sec: SCI/SL(2,C)}.

\begin{enumerate}

\item Definition $(q_\pm =e^{2\pi i b^{\pm 2}})$:
\begin{align}
e_b(x)
= \prod_{r=1}^{\infty} \frac{1+ (q_+)^{r-\half} e^{2\pi bx}}{1+(q_-)^{\half-r}e^{2\pi x/b}}
= \exp\left( \frac{1}{4} \int_{\IR+i\e} \frac{dw}{w} \frac{e^{-2ixw}}{\sinh(wb)\sinh(w/b)} \right) \,.
\label{qd-def}
\end{align}

\item
The zeros $(z_{mn})$ and poles $(p_{mn})$ of $e_b(x)$ are located at
\begin{align}
z_{mn} = -c_b -i( m b + n b^{-1} ) \,,
\quad
p_{mn} = + c_b + i( m b + n b^{-1} ) \,,
\quad
(m,n\in \IZ_{\ge 0})\,.
\label{zero-pole}
\end{align}
with $c_b = i(b+ b^{-1})/2$.

\item
Inversion formula:
\begin{align}
e_b(x) e_b(-x) = e^{\frac{\pi i}{12} (b^2+b^{-2})} e^{\pi i x^2} \,.
\label{q-inv}
\end{align}

\item
Quasi-periodicity and difference equation:
\begin{align}
&e_b(x+ib) = (1+q_+^{\half}e^{2\pi bx})^{-1} \,e_b(x) \,,
\quad
e_b(x-ib) = (1+q_+^{-\half}e^{2\pi bx}) \,e_b(x) \,,
\label{shift-b}
\\
&e_b(x+i/b) = (1+q_-^{\half}e^{2\pi x/b})^{-1} \,e_b(x) \,,
\quad
e_b(x-i/b) = (1+q_-^{-\half}e^{2\pi x/b}) \,e_b(x) \,,
\label{shift-1/b}
\end{align}

The second identity in \eqref{shift-b} can be rewritten as
\begin{align}
\left(e^{-ib\partial_x} - q_+^{-\half} e^{2\pi b x} -1 \right) e_b(x) =
(\hat{z}''-\hat{z}^{-1} -1 ) e_b(x) = 0 \,,
\end{align}
with $\hat{z}'' \equiv e^{-ib\partial_x}$,
$\hat{z} \equiv q^{\half}e^{-2\pi bx}$ satisfying
$\hat{z}''\hat{z} = q \hat{z} \hat{z}''$.

\item
Generalized Fourier transform:
\begin{align}
\Psi_n(\a_1,\ldots,\a_n;\b_1,\ldots,\b_{n-1};w)
\equiv \int_{\IR} dx \, e^{2\pi i x(w-c_b)}
\prod_{j=1}^n \frac{e_b(x+\a_j)}{e_b(x+\b_j-c_b)} \,
\label{Psi_n}
\end{align}
with $\b_n=i0$ satisfy a number of identities, the simplest of which include
\begin{align}
&\Psi_1(\a;w) = e^{\pi i (b^2+b^{-2}+3)/12} \frac{e_b(\a)e_b(w)}{e_b(\a+w-c_b)} \,,
\\
&\Psi_2(\a_1,\a_2;\b;w) = \frac{e_b(\a_1)}{e_b(\b-\a_2)}\Psi_2(\b-\a_2;w;\a_1+w;\a_2) \,.
\label{Psi_2}
\end{align}


\end{enumerate}

\subsection{Duality wall theory \label{Z-wall} }

The partition function on the squashed three sphere, $S_b^3$, is obtained in \cite{Hama:2011ea} for general 3d ${\cal N}=2$ gauge theories.
Here $b$ is the dimensionless squashing parameter normalized such that $b=1$ corresponds to the round sphere.

Let us first consider the partition function for the mass-deformed $T[SU(2), \varphi = S]$
\begin{align}
\CZ_S( \mu, \zeta, m )
&= s_b ( - m) \int d \sigma
\frac{ s_{b} ( \mu + \sigma + \frac{m}{2} + \frac{c_b}{2} )  s_{b} ( \mu - \sigma + \frac{m}{2} + \frac{c_b}{2}  )    }
{s_{b} ( \mu + \sigma - \frac{m}{2} - \frac{c_b}{2}  ) s_{b} ( \mu - \sigma - \frac{m}{2} - \frac{c_b}{2} )  }
e^{ 4 \pi i \sigma \zeta }   \, . \label{ftn:tsu2}
\end{align}
Here, $\mu$ denotes the mass for fundamental hyper-multiplets and $ \zeta$ the FI parameter. The phase factor $e^{ 4 \pi i \sigma \zeta}$ originates from the FI term.
The double sine function $s_b(x)$ is defined as
\begin{equation}
s_b \left( c_b ( 1-r) - \sigma \right) = \prod_{m, n \geq 0} \left( \frac{mb+nb^{-1}+ i \sigma -ic_b (2-r) } {mb+nb^{-1}- i \sigma-ic_b r } \right) \,,
\label{double-sine}
\end{equation}
where $c_b = i(b+ b^{-1})/2$.
This function is related to the QDL function $e_b(x)$ by
\begin{align}
e_b(x) = e^{\frac{\pi i}{24} (b^2+b^{-2})} e^{\frac{\pi i}{2} x^2}\,.
\end{align}
To simplify \eqref{ftn:tsu2}, we used an identity $s_b (x) s_b (-x)=1$
which is equivalent to \eqref{q-inv}.

The double sine function in \eqref{double-sine} is a contribution to the one-loop determinant from a free chiral multiplet
with R-charge $r$ (for the scalar field)
which is coupled to a background $U(1)$
.
Thus $s_b (-m)$ in eq.~\eqref{ftn:tsu2} originates from the adjoint chiral multiplet of $T[SU(2)]$, and the other four $s_b$ functions are from the four fundamental chiral multiplets;
see Table~\ref{tsu(2)-charge}.

Let us first generalize \eqref{ftn:tsu2} to the partition function of $T[SU(2), T^k ST^{l}]$.
Recall that the multiplication of $T^k$ and $T^l$ elements add background CS terms with level $k$ and $l$ for the two $SU(2)$'s to the theory.
The classical contributions from the CS terms shall be multiplied to the partition function as follows,
\begin{equation}
\CZ_{T^l ST^k} ( \mu, \zeta , m)
= e^{ -2 \pi i l \mu^2 }  e^{   - 2  \pi  i k  \zeta^2 }  \CZ_S ( \mu, \zeta , m) \, .
\nn
\end{equation}
For the multiplication of $SL(2,\IZ)$ elements, $ \varphi = \varphi_2 \cdot \varphi_1  $, the partition function can be obtained by `gluing'
\begin{align}
\CZ_{\varphi_2 \cdot \varphi_1} ( \mu, \zeta , m)
&= \int [ d \nu]  \CZ_{ \varphi_2 } ( \mu, \nu ,m )  \CZ_{ \varphi_1 }( \nu, \zeta , m )
\,,
\nn
\end{align}
where $[d \nu ] = d \nu \sinh (2 \pi b \nu) \sinh ( 2 \pi b^{-1} \nu) $ is
the measure with the contribution from a vector multiplet of the gauged $SU(2)$ global symmetry.

As an application of the QDL identities,
we prove the `self-mirror' property of $\CZ_S$:
\begin{align}
\CZ_S(\mu,\z,m) = \CZ_S(\z,\mu,-m) \,.
\end{align}
For $b=1$, this property was proved earlier in
\cite{Hosomichi:2010vh}. We begin with replacing $s_b$ in
\eqref{ftn:tsu2} by $e_b$. Up to an overall normalization that may
depend on $b$ but no other parameters, we find
\begin{align}
\CZ_S(\mu, \zeta, m) =
e^{\pi i (2\mu +m -c_b)(2\z-m-c_b)-\frac{\pi i}{2}m^2}
e_b(-m) \Psi_2(m,2\mu+m;2\mu;2\z-m) \,,
\label{ftn:tsu2b}
\end{align}
where the function $\Psi_n$ was defined in \eqref{Psi_n}.
The self-mirror property follows easily from the identity \eqref{Psi_2} and \eqref{ftn:tsu2b}.


\subsection{Tetrahedron decomposition}

The computation of the $S^3_b$ partition function of $T_M$ using the
tetrahedron decomposition of $M$ was explained in \cite{Dimofte:2011ju},
which parallels the computation of the index reviewed in section \ref{index-tetra}.
In the polarization $\Pi_Z$, the partition function is given by
\begin{align}
\CZ_\Delta(x) = e_b(c_b-x) =
 \prod_{r=1}^{\infty} \frac{1-(q_+)^{r} e^{-2\pi b x}}{1-(q_-)^{1-r} e^{-2\pi x /b}},
\end{align}
where the real and imaginary part of the complex parameter $x$ correspond to
the twisted mass and the R-charge of the elementary chiral multiplet $\phi_Z$.

The $SL(2,\IZ)$ polarization change acts on $\CZ_\Delta$ as follows,
\begin{align}
&T: \CZ_\Delta(x) \;\; \goto \;\; \CZ'(x) = e^{-\pi i x^2} \CZ_\Delta(x) \,,
\nn \\
&S: \CZ_\Delta(x) \;\; \goto \;\; \CZ'(x') = \int dx e^{-2\pi i x x'} \CZ_\Delta(x) \,.
\label{TS-ZD}
\end{align}
The computation of the partition function for $T_M$ can be done in three steps
(see section 6.2 of \cite{Dimofte:2011ju}).
First, one takes $\CZ_{\Delta_i}(x_i)$, for each tetrahedron $\Delta_i$
in the triangulation and multiplies them all. Second, act with $Sp(2N,\IZ)$
to transform to a polarization in which all internal edges are ``positions".
Third, set the parameters $x_I$ corresponding to the internal edges equal to $2c_b$.

Thus, the procedure is conceptually identical to that of the superconformal index.
In practice, the partition function is slightly more difficult to deal
with because the $S$ operation in \eqref{TS-ZD} involves Fourier transformation,
whereas the $S$ operation for the index can be treated as a linear transformation
on the lattice of basis states of the Hilbert space.

\subsection{Quantization for $G= SL(2,\mathbb{R})$}

\paragraph{An (approximate) isomorphism of operator algebra}
In the previous sections, we saw how the CS theory with the non-compact gauge group $G=SL(2,\mathbb{C})$ is related to the superconformal index of 3d
field theories.
For the squashed sphere partition functions being discussed in this section,
the relevant gauge group is $G=SL(2,\mathbb{R})$.

Since $SL(2,\mathbb{R})$ is a real slice of $SL(2,\mathbb{C})$, the phase space $\CM_{SL(2,\mathbb{R})}$ is also a real slice of  $\CM_{SL(2,\mathbb{C})}$ in a suitable sense. Recall that we obtained the $(\pm)$ pair of operators after quantization because the coordinates for $\CM_{SL(2,\mathbb{C})}$ are complex variables, and that the $(+)$ operators commute with the $(-)$ operators.
In contrast, the coordinates are real variables for  $\CM_{SL(2,\mathbb{R})}$.
So, at first sight, the splitting into $(+)$ and $(-)$ operators seem unlikely.

Remarkably, as first noted in \cite{Faddeev:1995nb,Chekhov:1999tn} and brought into the present context in  \cite{Dimofte:2011jd}, the algebra of exponentiated operators for $G=SL(2,\mathbb{R})$ does factorizes into two mutually commuting subalgebras. To be explicit, we begin by introducing rescaled (logarithmic) shear operators as follows,
\begin{align}
(\sT, \sT',\sT'') = 2 \pi b (\hat{\sT} , \hat{\sT'} , \hat{\sT''})\;, 
\qquad
 [\hat{\sT}, \hat{ \sT'}] = [\hat{\sT'}, \hat{ \sT''}]= [\hat{\sT''}, \hat{ \sT}]=\frac{i}{\pi}\;.
\end{align}
The $(\pm)$ pair of exponentiated shear operators are defined by
\begin{align}
\st_\pm := \exp (2\pi b^{\pm 1} \hat{\sT})\;, \quad  \textrm{similarly for $\st'_\pm , \st''_\pm$}\;.
\end{align}
Note that
\begin{align}
[\st^{*}_+, \st^{**}_-]=0\;, \quad \textrm{for any $^*$, $^{**}$}\;.
\end{align}
The $(+)$ shear operators are nothing but the original shear operators.
The $(-)$ shear operators satisfy formally the same commutation relations 
but with the original quantum parameter $q_+:= q=e^{2\pi i b^2}$ replaced by
$q_- := \exp (2\pi i b^{-2})$.

If we consider composite operators made of integer powers of $(\st_\pm, \st'_\pm, \st''_\pm)$ only, there exists an isomorphism between the operator algebra for $G=SL(2,\mathbb{R})$  and that for $G=SL(2,\mathbb{C})$
with the understanding that $(q_\pm)_{SL(2,\mathbb{R})}$
are mapped to $(q^{\pm 1})_{SL(2,\mathbb{C})}$.
However, this isomorphism
breaks down slightly if the square-root operators
$(\sqrt{\st}, \sqrt{\st'}, \sqrt{\st''})$ are included. For instance, $\sqrt{\st}_\pm \sqrt{\st'}_\mp = \sqrt{\st}_\mp \sqrt{\st'}_\pm$ in the  $SL(2,\mathbb{C})$ case whereas $\sqrt{\st}_\pm \sqrt{\st'}_\mp =- \sqrt{\st}_\mp \sqrt{\st'}_\pm$ in the $SL(2,\mathbb{R})$ case.

The (approximate) operator isomorphism will be the key to understanding a large degree of similarity between the computations in section \ref{sec: SCI/SL(2,C)} and those
in this section.

\paragraph{Shear/SR basis for the Hilbert-space}

The Hilbert space $\CH_{SL(2,\IR)}$ is the familiar $L^2(\mathbb{R})$
for the quantum mechanics on a real line \cite{Kashaev}.
To see this, we recombine two independent shear coordinates, $(\sT, \sT')$,
to form a position-momentum pair. \footnote{The logarithmic operators $({\cal S}, {\cal R})$ are rescaled by a factor $(2\pi b)$ from those in previous sections. }
\begin{align}
[\sT,\sT'] = 2\hbar = 4\pi i b^2 \;\;
\imp \;\;
(2\pi b) {\cal S} \equiv \frac{\sT+\sT'}{2} \,, \;\;
(2\pi b) {\cal R} \equiv \frac{\sT-\sT'}{2} \,, \;\;
[{\cal R},{\cal S}] = \frac{i}{2\pi} \,.
\label{Z-shear-SR}
\end{align}
%
Unlike in section \ref{sec: SCI/SL(2,C)},
we take $\sT$, $\sT'$ to be Hermitian. Then, $\CS$ and $\CR$,  also Hermitian, can be identified with the position/momentum operators
for quantum Teichm\"ller theory originally introduced in \cite{Kashaev}.
We define the `position' and `momentum' basis in the usual manner,
\footnote{We flipped the sign of the `momentum' $\CR$ to match the convention of \cite{Dimofte:2011ju,Dimofte:2011py}.}
\begin{align}
&\langle x | \CS = x \langle x | \,,
\quad
\langle x | \CR =
\frac{i}{2\pi} \frac{\partial}{\partial x} \langle x |\,,
\quad
\langle x | x' \rangle = \d(x-x') \,,
\quad
\int_{-\infty}^{\infty} | x \rangle \langle x | =1 \,.
\nn \\
&\langle p | \CR = p \langle p | \,,
\quad
\langle p | \CS =
\frac{1}{2\pi i} \frac{\partial}{\partial p} \langle p |\,,
\quad
\langle p | p' \rangle = \d(p-p') \,,
\quad
\int_{-\infty}^{\infty} | p \rangle \langle p | =1 \,.
\end{align}
The transformation between the two bases can be performed as usual,
\begin{align}
\langle x | p \rangle = e^{-2\pi i p x} \,, \quad
\int e^{-2\pi i p (x-x')} dp = \d(x-x') \,.
\end{align}
%
The exponentiated SR operators,
\begin{align}
\mathsf{s}_\pm \equiv e^{2\pi b^{\pm} \CS} \,,
\quad
\mathsf{r}_\pm \equiv e^{2\pi b^{\pm} \CR} \,.
\end{align}
satisfy the following commutation relations,
\begin{align}
&\mathsf{r}_\pm \mathsf{s}_\pm = q_\pm \mathsf{s}_\pm \mathsf{r}_\pm \,, \quad
[\mathsf{r}_\pm , \mathsf{r}_\mp] = 0 \,, \quad
 [ \mathsf{r}_+ , \mathsf{r}_-] = 0 = [ \mathsf{s}_+ , \mathsf{s}_- ] \,.
\label{def-XP-Z}
\end{align}
If we only needed operators of the type $\mathsf{r}^a \mathsf{s}^b$ with $(a,b)\in \IZ^2$, the isomorphism between the operator algebra of this section
and that of section \ref{sec: SCI/SL(2,C)} would have been exact,
with minor modifications in the hermiticity condition and
the definition of $q_\pm$. But, the operators $\sqrt{\st}$, $\sqrt{\st'}$
forces us to include `half-integer points' $(a,b) \in \IZ/2$ with $a+b\in \IZ$ in the lattice of operators, which induces subtleties such as
$\sqrt{\st}_\pm \sqrt{\st'}_\mp =- \sqrt{\st}_\mp \sqrt{\st'}_\pm$.


In the previous section, we encountered lattices of states such as $|m,e \rangle$ as well as lattices of operators such as $\mathsf{r}^a \mathsf{s}^b$.
Here, while it is not clear how to organize the states of the Hilbert space
on a lattice, we can still untilize the lattice structure of operators.
For instance, it is useful to consider the generators of linear $SL(2,\mathbb{Z})$ polarization changes (not to be confused with the $SL(2,\mathbb{Z})$ of QTT which acts non-linearly on the shear coordinates)
\begin{align}
e^{\pm \pi i \mathcal{S}^2}
\begin{pmatrix}
\mathcal{S} \\ \mathcal{R}
\end{pmatrix}
e^{\mp \pi i \mathcal{S}^2}
=  R^{\pm 1}
\begin{pmatrix}
\mathcal{S} \\ \mathcal{R}
\end{pmatrix} \,,
\qquad
e^{\pm \pi i \mathcal{R}^2}
\begin{pmatrix}
\mathcal{S} \\ \mathcal{R}
\end{pmatrix}
e^{\mp \pi i \mathcal{R}^2}
= L^{\mp 1}
\begin{pmatrix}
\mathcal{S} \\ \mathcal{R}
\end{pmatrix} \,,
\label{pol-LR}
\end{align}
where $R$ and $L$ are as in \eqref{sl2z-convention}.

\paragraph{QDL and isomorphism of operator algebra revisited}
%
We can promote the QDL function $e_b(x)$ to an operator by substituting $\sx$ for $x$,
\begin{align}
e_b(\sx) = \prod_{r=1}^{\infty} \frac{1+(q_+)^{r-\thalf} e^{2\pi b \sx}}{1+(q_-)^{\thalf-r} e^{2\pi \sx /b}} = \prod_{r=1}^{\infty} \frac{1+(q_+)^{r-\thalf} e^{\CX_+}}{1+(q_-)^{\thalf-r} e^{\CX_-}} \equiv \CE_{q_+,q_-}(\CX_+,\CX_-) \,.
\label{QDL-pm-split}
\end{align}
We will often use the short-hand notation $\CE(\CX)$ for $\CE_{q_+,q_-}(\CX_+,\CX_-)$. Similar operators
can be defined in other polarizations of $L^2(\mathbb{R})$ by replacing $\CX_\pm$ by linear combinations of $\CX_\pm$ and $\CP_\pm = 2\pi b^{\pm 1} \mathsf{p}$
with integer coefficients. Here, we assume $[\mathsf{x},\mathsf{p}]=(2\pi i)^{-1}$.

This quantum version of the QDL function underlies essentially all non-trivial identities among various constructions. In addition, through the isomorphism of the operator algebra, the QDL function can also be used to illuminate
the parallel between this section and section \ref{sec: SCI/SL(2,C)}.
To see this, note that $\CZ_\Delta(x)$ of the tetrahedron decomposition
can be regarded as a collection of eigenvalues,
\begin{align}
\langle x | \CE(\CX_+, \CX_- )
= \langle x | \CZ_\Delta(c_b-x) \,.
\end{align}
%
Similarly, we can define the QDL operator $\CE(\CX)$ for the index computation by
\begin{align}
\CE(\CX)  = \prod_{r=1}^{\infty} \frac{1+(q_+)^{r-\thalf} e^{\CX_+}}{1+(q_-)^{\thalf-r} e^{\CX_-}} \,,
\end{align}
where $e^{\CX_\pm}$ operators are defined as in \eqref{xp-operation} and $q_\pm$ is to be understood as $q_\pm = q^{\pm 1}$.
By construction, $\CE(\CX)$ is diagonal in the fugacity basis,
\begin{align}
\langle m, u| \CE(\CX) = \CI_\Delta(-m,(-q^{\thalf})u^{-1}) \langle m, u| \,,
\end{align}
where $\CI_\Delta$ on the right-hand-side is precisely the
tetrahedron index \eqref{tetrahedron index}.
Now, let us see how the QDL identities imply identities for $\CI_{\Delta}$.
{}From the operator version of the inversion identity \eqref{q-inv}
and the $SL(2,\IZ)$ polarization change \eqref{pol-LR},
we find
\begin{align}
\sum_{e_1} (-q^{\thalf})^{e+2e_1} \CI(-m,e_1) \CI(m,e+e_1)
= \delta_{e,-m} \,.
\end{align}
This identity appeared in \cite{Dimofte:2011py} in the computation
of the index for the trefoil knot complement.
Similarly, by taking the matrix elements of the quantum pentagon relation,
\begin{align}
\CE(\CX) \CE(\CP) = \CE(\CP) \CE(\CX+\CP) \CE(\CX) \,,
\label{q-pentagon-2-copy}
\end{align}
and reshuflling the indices using the parity \eqref{identity for tetrahedron index} and triality \eqref{Triality}, we find
\begin{align}
&\CI_\Delta(m_1-e_2,e_1) \CI_\Delta(m_2-e_1,e_2)
\nn \\
&\qquad = \sum_{e_3} q^{e_3} \CI_\Delta(m_1,e_1+e_3) \CI_\Delta(m_2,e_2+e_3)
\CI_\Delta(m_1+m_2,e_3) \,.
\end{align}
This identity was used in \cite{Dimofte:2011py} to show the equivalence bewteen the two mirror descriptions of the bipyramid theory.

\paragraph{FN basis}

The isomophism of the operator algebras
allows us to use the same relation between the shear and FN operators
\eqref{shear2FN-q}.
In \cite{Kashaev}, what we might call the FN basis, in which  $\hat{\lambda}+\hat{\lambda}^{-1}$ becomes diagonal, was defined
through its relation to the SR basis.
Using the loop-FN-shear triality explained in section \ref{qsl2c},
we take the Wilson loop operators 
\begin{align}
\sW^{\pm} = (\mathsf{r}_\pm)^{-1} + \mathsf{s}_\pm + (\mathsf{s}_\pm)^{-1} \,.
\end{align}
In addition, we introduce the Dehn twist operator \cite{Kashaev},
\begin{align}
\sD = e^{2\pi i(\mathcal{S}^2-c_b^2)} e_b(\mathcal{S}-\mathcal{R}) \,.
\label{dehnx}
\end{align}
The loop operators and the Dehn twist operator commute with each other,
\begin{align}
[ \sW^+ , \sW^-] = 0\,, \quad
[ \sD, \sW^\pm ] = 0 \,.
\end{align}
Next, following \cite{Kashaev}, but slightly modifying the normalization
to conform to the convetions of \cite{Dimofte:2011jd}, we introduce the states
$| \mu)$ $(\mu \in \IR_+)$ by specifying the matrix elements,
\begin{align}
\langle x | \mu) &= \frac{e_b(\mu+x+c_b-i0)}{e_b(\mu-x-c_b+ i0)} e^{-2\pi i(x+c_b)\mu + \pi i \mu^2}
\nn \\
&= (q_+ q_-)^{-\frac{1}{24}} \CZ_\Delta(-x-\mu) \CZ_\Delta(-x+\mu) e^{-\pi i(x+c_b)^2}
\,.
\label{SR-FN-DG}
\end{align}
As proved in \cite{Kashaev}, the FN basis vectors are simultaneous eigenstates of the loop and Dehn twist operators,
\begin{align}
\sO^\pm | \mu ) = 2\cosh(2\pi b^\pm \mu) | \mu )  \equiv (\hat{\lambda}_\pm+(\hat{\lambda}_\pm)^{-1}) | \mu )
\,, \quad
\sD | \mu ) = e^{2\pi i (\mu^2 -c_b^2)} | \mu ) \,,
\label{LD-eigenx}
\end{align}
The second expression in \eqref{SR-FN-DG} makes it clear that
$\langle x |{\rm -}\mu) = \langle x |\mu)$
for all $x, \mu \in \mathbb{R}$. To avoid double-counting, we restrict the range of $\mu$ to $\mathbb{R}_+$, which reflects the $\IZ_2$ Weyl symmetry of the $T[SU(2)]$ theory.
The FN basis vectors satisfy the orthogonality and completeness relations
compatible with the results of section \ref{Z-wall}.
\begin{align}
(\mu|\nu) = \frac{\d(\mu - \nu)}{4\sinh(2\pi\mu b)\sinh(2\pi\mu/b)} \,,
\quad
\int_0^\infty 4\sinh(2\pi\mu b)\sinh(2\pi\mu/b) |\mu)(\mu| d\mu= 1\,.
\label{FN-Z-ortho-compl}
\end{align}
We close this subsection by noting that
the following hermiticity of the FN operators,
\begin{align}
(\lambda_\pm)^\dagger = \lambda_\pm \,,
\quad
(\tau_\pm)^\dagger =
\frac{1}{\lambda_\pm - \lambda_\pm^{-1}} \tau_\pm  (\lambda_\pm - \lambda_\pm^{-1}) \,,
\label{Z-FN-hermi}
\end{align}
is compatible with the measure in \eqref{FN-Z-ortho-compl}
and the hermiticity of the shear operators.

\paragraph{Proving the equivalence: $ Z^{\Delta}_{\textrm{tori}(\varphi)} =Z^{\textrm{Tr}(\varphi)}_{\textrm{tori}(\varphi)} =\CZ^{\textrm{T[SU(2)]}}_{\textrm{tori}(
\varphi)} $}

Given the close parallel between the computation of the index
and that of the partition function,
it is natural to expect that
the equality of three quantities explained in \ref{equality between three SL(2,C)/index ptns} can be carried over to this section.
The second equality
is essentially a basis change between the shear basis and the FN basis.
Since an explicit form of the basis change is known \eqref{SR-FN-DG},
the proof in \ref{equality between three SL(2,C)/index ptns}
can be repeated with little modification by using
the isomorphism of operator algebra.
The first equality is somewhat less trivial.
The Hilbert space for the partition function does not exhibit
a lattice structure on which the $SL(2,\IZ)$ polarization change \eqref{pol-LR}
linearly. Nevertheless, we expect that the proof of section \ref{Z(Tr)=Z(Delta)}
can be adapted to the current context by a suitable combination of $SL(2,\IZ)$ actions
in \eqref{TS-ZD}.

\vskip 1cm

\begin{acknowledgments}
We are grateful to Yoon Pyo Hong, Kazuo Hosomichi, Hee-Cheol Kim, Sung-Soo Kim, Bum-Hoon Lee, Kimyeong Lee, Yunseok Seo, Sang-Jin Sin, Yang Zhou and especially Sergei Gukov for helpful discussions.
The work of SL is supported in part by
the National Research Foundation of Korea (NRF) Grants
NRF-2012R1A1B3001085 and NRF-2012R1A2A2A02046739.
J.P. was supported by the National Research Foundation of Korea (NRF) grant
funded by the Korea government (MEST) with the Grants No. 2012-009117, 
2012-046278 and 2005-0049409 through the Center for Quantum Spacetime
(CQUeST) of Sogang University. JP is also supported by the POSTECH Basic
Science Research Institute Grant and appreciates APCTP for its stimulating environment
for research.

\end{acknowledgments}

\newpage

\appendix
\centerline{\large\bf Appendix}

\section{Alternative descriptions of $T[SU(2)]$} \label{dual description for T[SU(2)]}

In this section, we will discuss other possible descriptions of $T[SU(2)]$.

The first dual theory is ${\cal N}=2$ $SU(2)$ Chern-Simons theory of level $k=1$
with four fundamental and three neutral chiral multiplets, found in \cite{Teschner:2012em}.
The authors of the paper found that the squashed three sphere partition function of the mass-deformed $T[SU(2)]$ theory
can be interpreted as a partition function of the dual theory. The 3d superconformal index for the dual theory also has been shown
to coincide with the index for $T[SU(2)]$ \cite{Gang:2012ff}.
The dual theory has the following advantage.
The $SO(4) \simeq SU(2) \times SU(2)$ flavor symmetry of the dual theory, which corresponds to $SU(2)_{\rm top} \times  SU(2)_{\rm bot}$ global symmetry of $T[SU(2)]$,
is manifest in the Lagrangian.
Thus the operations on the superconformal index described in section \ref{index-wall} (i.e., adding Chern-Simons action and/or gluing)
can be incorporated at the Lagrangian level. Accordingly, the resulting theory dual to $T[SU(2), \varphi]$ has a Lagrangian description.

\begin{figure}[htbp]
   \centering
   \includegraphics[scale=0.27]{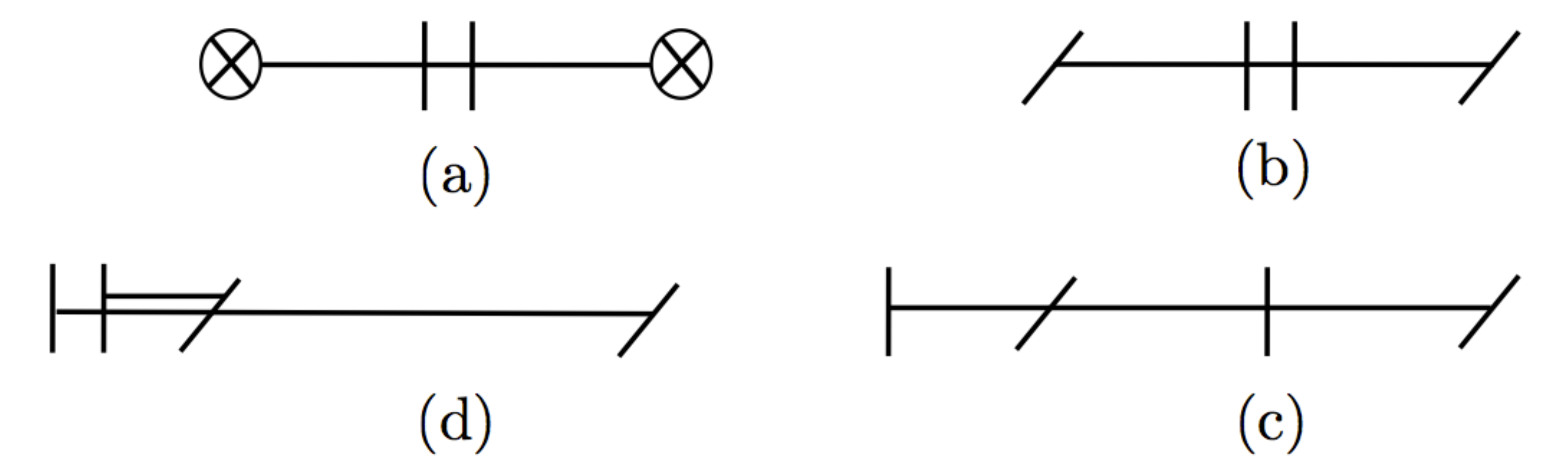} 
   \caption{Brane setups for dual theories of $T[SU(2)]$. Horizontal/vertical/tilted lines represent D3/D5/(1,1)-branes. Crossed circles represent NS5-branes.
   }
   \label{fig:brane}
\end{figure}

Other dual theories of $T[SU(2)]$ can be found from the brane set-up for $T[SU(2)]$ theory
 given in Figure~\ref{fig:brane}-(a). Taking a limit where the length between NS5-branes is very small, the theory on a D3-brane becomes a 3d $U(1)$ theory. Two D5-branes give rise to
 two fundamental hyper-multiplets. Taking T-dual transformation of $SL(2, \mathbb{Z})$ of type IIB theory on (a) results in (b), i.e., two NS5-branes are mapped to
(NS5,D5)=(1,1)-branes while D5-branes are invariant. The IR limit of (b) corresponds to $U(1)_1 \times U(1)_0 \times U(1)_{-1}$ Chern-Simons theory where subscripts denote the CS levels.
  The theory has two bi-fundamental hyper-multiplets with charges $(1,-1,0)$ and $(0,1,-1)$ under the gauge group.
   Crossing a D5-brane over the left NS5-brane in (b), a D3-brane is created between the 5-branes due to Hanany-Witten effect, as depicted in (c).
   It corresponds to $U(1)_{-1} \times U(1)_1 \times U(1)_{-1}$ theory with bi-fundamental hyper-multiplets.  Crossing the other D5-brane over NS5-brane in turn
   results in the brane set-up in (d), which corresponds to $U(1)_0 \times U(2)_{-1} \times U(1)_0 $ theory.
   Since all four theories have Lagrangian descriptions, we can use the prescription in \cite{Imamura:2011su} to write down index formulas.
   We checked that the superconformal indices as a function of fugacity for $ \epsilon +j$  coincide with one aother to several orders in fugacity.

\section{Difference equations for $T[SU(2),\varphi]$ at classical limit}\label{T[SU(2)] classical}

The index  \eqref{index formula-1} of $T[SU(2)]$
 can be constructed from tetrahedron indices $\CI_{ \Delta}$ as follows
\begin{align}
& I_{\varphi=S} (m_b, m_t, m_{ \eta}; u_b, u_t, u_\eta ) \nonumber \\
&  = (-1)^{ 2m_b} \CI_{ \Delta} ( -  m_{ \eta} , u_\eta^{-1} q^{ \frac{1}{2}}) u_ \eta^{ m_{ \eta}}
\sum_{m_s} \oint \frac{ d u_s}{ 2 \pi i u_s} u_s^{ 2m_t + 2m_s} u_t^{2m_s} u_b^{2 m_b} q^{ - \frac{1}{4} m_{ \eta}}  \nn
\\
&  \quad \times \prod_{ \epsilon_1, \epsilon_2 = \pm1}
\CI_{ \Delta} ( \epsilon_1 m_b +\half m_{ \eta} + \epsilon_2 m_s , u_b^{ \epsilon_1} u_\eta^\half u_s^{ \epsilon_2} q^{ \frac{1}{4}} ) \; . \label{TSU2 index using tetrahedron index}
\end{align}
In the charge basis, the index become
\begin{align}
&I_{\varphi =S}  (m_b, m_t, m_{ \eta};e_b, e_t, e_\eta )=
 \sum_{e_1, e_2 \in \mathbb{Z}}  \prod_{i=1}^5  \CI_\Delta( \tilde{m}_i , \tilde{e}_i)\;, \;\textrm{where} \nn
\\
 &(\tilde{m}_1, \tilde{e}_1)= (-m_{ \eta}+\frac{ \hbar}{2} , e_1 - \frac{ \hbar}{4}) \;, \;(\tilde{m}_2, \tilde{e}_2)= (m_b+\half m_{ \eta}+\frac{e_t}{2} + \frac{ \hbar}{4} , e_2 - i \pi ) \;, \nonumber \\
 &(\tilde{m}_3 ,\tilde{e}_3) =  ( -m_b+\half m_{ \eta}- \frac{e_t}{2}+ \frac{ \hbar}{4} , e_2-\frac{e_b}{2}+\frac{e_t}{2}+m_b+m_t )\;,
 \nonumber \\
 & (\tilde{m}_4,\tilde{ e}_4)= (-m_b+\half m_{ \eta}+\frac{e_t}{2} + \frac{ \hbar}{4} , e_1 - e_2 +e_{ \eta} - \frac{e_t}{2} -  m_{ \eta} - m_t+ i \pi ) \;,
 \nonumber \\
 &(\tilde{m}_5, \tilde{e}_5)= (m_b+\half m_{ \eta}- \frac{e_t}{2}+ \frac{ \hbar}{4} , e_1-e_2+ \frac{e_b}{2} +e_{ \eta} - m_b-m_{ \eta})
 \; .
\label{tetra-tsu2}
 \end{align}
We have five difference equations for the index $\prod_{i=1}^5 \CI_{ \Delta}(\tilde{m}_i, \tilde{e}_i)$, 
\begin{equation}
\tilde{x}_i^{-1} + \tilde{p}_i -1 =0 \, ,  \qquad \mbox{for } i = 1, 2,\ldots , 5\;.
\nonumber
\end{equation}
where $ \tilde{x}_i, \tilde{p}_i$ can be written in terms of $(p_1, p_2, x_b, x_t, x_\eta, p_b, p_t, p_\eta)$ from eq.~\eqref{tetra-tsu2}; see also \eqref{Polarization change equivalence}). We will focus on the classical limit, $ \hbar \to 0$ from here on. In the classical limit,
\begin{align}
& \tilde{x}_1 = \frac{1}{x_{ \eta}}, \quad \tilde{x}_2 = p_t^{ \frac{1}{2}}  x_b x^\half_{ \eta} , \quad \tilde{x}_3 = p_t^{ - \half}  x^{-1}_b x^{\half}_{ \eta},
\quad \tilde{x}_4 = p_t^{ \frac{1}{2}}  x^{-1}_b x^\half_{ \eta}  , \quad \tilde{x}_5 = p_t^{ - \frac{1}{2}}  x_b x^\half_{ \eta} \, ,
\nonumber \\
& \tilde{p}_1 = p_1, \quad \tilde{p}_2 = - p_2, \quad \tilde{p}_3 = \frac{p_t^{ \frac{1}{2}} p_2 x_b x_t}{ p_b^{ \frac{1}{2}}},
\quad \tilde{p}_4 = - \frac{ p_{ \eta} p_1}{p_t^{ \frac{1}{2}} p_2 x_{ \eta} x_t},
\quad \tilde{p}_5 =  \frac{ p_b^{ \frac{1}{2}} p_{ \eta} p_1}{ p_2 x_b x_{ \eta} } \, .
\nonumber
\end{align}
The summations $ \sum_{e_1, e_2}$ corresponds to  integrating out $p_1, p_2$ from difference equations. We use the first two equations,
$ \tilde{x}_{1} + \tilde{p}_{1}^{-1}-1=0$ and $ \tilde{x}_{2} + \tilde{p}_{2}^{-1}-1=0$ to integrate out $p_1, p_2$
\begin{equation}
p_1 = 1 - x_{ \eta} ,  \quad p_2 = \frac{1- p_t^{ \frac{1}{2}} x_b x^{\half}_{ \eta}}{ p_t^{ \frac{1}{2} } x_b x_{ \eta}^\half}
\, .
\end{equation}
Thus we are left with three equations of six variables, $  \tilde{x}_i^{-1} + \tilde{p}_i - 1=0$ for  $i = 3,4,5$ where $p_1, p_2$ in $ \tilde{x}_i, \tilde{p}_i$ are replaced by the above conditions
\begin{equation}
\{  \tilde{x}_i^{-1} + \tilde{p}_i - 1=0, \mbox{ for } i = 3,4,5. \} /. \{ p_1 \to 1 - x_{ \eta},  \ p_2 \to \frac{1- p_t^{ \frac{1}{2}} x_b x^\half_{ \eta}}{ p_t^{ \frac{1}{2} } x_b x^\half_{ \eta}} \} \, .
\end{equation}
which are classical difference equations of $I_{ \varphi = S}$.
After some algebraic manipulations, the equations can be written as
\begin{align}
& x_b + x_b^{-1} = \frac{ x^\half_{ \eta} x_t^{-1} - x_{ \eta}^{-\half} x_t}{ x_t^{-1} - x_t} p_t^{- \frac{1}{2}} + \frac{ x^{-\half}_{ \eta} x_t^{-1} - x^\half_{ \eta} x_t}{ x_t^{-1} - x_t} p_t^{ \frac{1}{2}}
\nonumber \\
& \frac{x^\half_\eta x_b - x_{ \eta}^{-\half} x_b^{-1}}{x_b-x_b^{-1}} p_b^{ - \frac{1}{2}}
+\frac{x^{-\half}_\eta x_b - x^\half_{ \eta} x_b^{-1}}{x_b-x_b^{-1}} p_b^{  \frac{1}{2}} = x_t + x_t^{-1}
\nonumber \\
& p_{ \eta}= \frac{p_t^{-\half}- p_t^\half}{x_t^{-1} - x_t}= \frac{x_b -x_b^{-1}}{p_b^{-\half} -p_b^\half}  \;.
\end{align}
which are indeed the classical limit, $\hbar \to 0$ i.e. $q \to 1$, of the difference equations of $I_{ \varphi = S}$ given in eq.~\eqref{diff for S}. These equations are the same as algebraic equations studied in \cite{Gaiotto:2013bwa} which define a moduli space of vacua for $T[SU(2)]$.   In terms of shear operators \eqref{shear operators}, the classical difference equations  are
\begin{align}
\mathcal{L}_{\varphi=S} =\{ (\sqrt{\st}^{\rm T})_t  - (\frac{1}{\sqrt{\st}})_b=0\;, \;(\sqrt{\st''}^{\rm T})_t  - \sqrt{\st'}(1+ \st)_b=0\;, \; p_\eta = - \frac{i}{\sqrt{\st}_b}\}\;.
\end{align}
For $\varphi=T$, classical difference equations are (see eq.~\eqref{diff for R-2})
\begin{align}
\mathcal{L}_{\varphi=T} =\{ x_t = x_b\;, \; p_t^{-\half } = p_b^{\half} x_b^{-1}\;, \; p_\eta =1 \}\;.
\end{align}
In terms of shear coordinates,
\begin{align}
\mathcal{L}_{\varphi=T} =\{ (\sqrt{\st}^{\rm T})_t- (\frac{1}{\sqrt{\st'}})_b=0\;, \; (\sqrt{\st''}^{\rm T})_t- \sqrt{\st''}(1+ \st')_b=0\;, \; p_\eta =1 \}\;.
\end{align}
A general $\varphi\in SL(2,\mathbb{Z})$ can be written as a product of $S$ and $T$. Thus, to obtain the classical difference equation for general $\varphi$, we only need to know the gluing rules for the classical difference equations.
For $\varphi=\varphi_2 \varphi_1$, the classical Lagrangian can be obtained by
\\
\\
1. Identify $(\sqrt{\st}^{\rm T},\sqrt{\st'}^{\rm T},\sqrt{\st''}^{\rm T})_t$ of $\varphi_2$ with  $(\sqrt{\st},\sqrt{\st'},\sqrt{\st''})_b$ of $\varphi_1$.
\\
2. If $p_\eta = p_{\eta,i} $ for $\varphi_{i=1,2}$, then $p_\eta = p_{\eta,2}p_{\eta,1}$ for $\varphi=\varphi_2\varphi_1$.
\\
3. Integrate out $(\sqrt{\st}^{\rm T},\sqrt{\st'}^{\rm T},\sqrt{\st''}^{\rm T})_t$  of $\varphi_2 $ (or equivalently $(\sqrt{\st},\sqrt{\st'},\sqrt{\st''})_b$ of $\varphi_1$)\;.
\\
\\
Using the gluing rules, the classical Lagrangian for $\varphi=S^2$ and $\varphi=(ST)^3$  become
\begin{align}
&\mathcal{L}_{\varphi=S^2} = \{ (\sqrt{\st}^{\rm T})_t - (\sqrt{\st})_b =0 \;,\; (\sqrt{\st''}^{\rm T})_t - (\sqrt{\st''})_b =0\;, \;p_\eta = -1\; \} \nn
\\
&
\mathcal{L}_{\varphi=(ST)^3} = \{ (\sqrt{\st}^{\rm T})_t - (\sqrt{\st})_b =0 \;,\; (\sqrt{\st''}^{\rm T})_t - (\sqrt{\st''})_b =0\;, \;p_\eta = \frac{-i}{\sqrt{\st_b \st'_b\st''_b }}=x_\eta^{-\half}\; \}\;.
\end{align}
In the last equation, we used the fact $\sqrt{\st\st'\st''} = -i x^\half_\eta$ in the classical limit. These classical difference equations reflect the $SL(2,\mathbb{Z})$ structure \eqref{SL(2,Z) in index} for index $I_{\varphi}$.

\section{Derivation of eq.~\eqref{shear index for L,R}} \label{derivation of SR indices}
For $\varphi = \sL$, let us first define a convenient basis $\{  |(m,e)\rangle_{\Pi_{\sL}}\}$ where
the polarization $\Pi_{\sL}$ is defined by
\begin{align}
 \Pi_{\sL} = \big{(}{\cal X}_{\sL} , {\cal P}_{\sL})_\pm = (-\sT''_\pm \pm  i \pi +\log\fl_\pm \mp  \frac{\hbar}2, \half (\sT_\pm  +\sT''_\pm   -\log \fl_\pm   ) \big{)}\;.
 \end{align}
$\Pi_{\sL}$ is related to the $\Pi_{\SR}$ in \eqref{SR polarization} in the following way
 \begin{align}
 \left(\begin{array}{c}  {\cal X}_\sL \\ {\cal P}_\sL\end{array}\right)_\pm  &= \left(\begin{array}{cc}2 & 0 \\ - \half &   \half \end{array}\right)\left(\begin{array}{c}  {\cal X}_{\SR} \\ {\cal P}_{\SR}\end{array}\right)_\pm  \pm \left(\begin{array}{c} -\frac{\hbar}2\\  \frac{ i \pi }{2} \end{array}\right)  \; \label{Pi(L) to Pi(RS)} \, .
 \end{align}
In terms of ${\cal X}_{\sL \pm}, {\cal P}_{\sL \pm}$ (position, momentum) operators in $\Pi_{\sL}$, 솓 operator $\sL$ in  \eqref{L,R operators} can be written as
\begin{align}
 \sL =  \left( \prod_{r=1}^\infty \frac{1-q^r e^{{\cal X}_{\sL+}} \fl^{-1}_+}{1-q^{r-1} e^{{\cal X}_{\sL-} }\fl^{-1}_-} \right) \exp \big{[} -\frac{1}{\hbar}\big{(}({\cal P}_{\sL+} + \half \log \fl_+ )^2-({\cal P}_{\sL-} +\half \log \fl_- )^2 \big{)}\big{]}\;.
 \end{align}
 Let us suppress the subscript $\sL$ of ${\cal X}_{\sL \pm}, {\cal P}_{\sL \pm}$ hereafter. For instance, eq.~\eqref{xp-operation} would be written as
 \begin{equation}
_\sL\langle  m,e | e^{ {\cal X}_{ \pm}} =\; _\sL \langle   m,e \mp 1 | q^{ \frac{m}{2}}, \qquad
_\sL\langle m,e | e^{ {\cal P}_{ \pm}} =\; _\sL \langle m \pm 1 ,e | q^{ \frac{e}{2}}
  \, .
 \end{equation}
Using eq.~\eqref{basis polarization transformation}, one can find a relation between $_\sL \langle(m,e)|$ and $ _{\SR}\langle (m,e)|$ such as
\footnote
{
The $(-1)$ in eq.~\eqref{sr-l} should be regarded as $(e^{ i \pi})$, thus the kets are related as follows,
\begin{equation}
| m, e \rangle_{\SR} =  q^{ \frac{1}{4}(m-e) }(-1)^m | m, e \rangle_{\sL}
\, .
\end{equation}
}
 \begin{align}
_{\SR}\langle m,e| &=  _{\sL}\langle 2m,\half(e-m)| q^{\frac{1}4 (m-e)} (-1)^{- m} \, ,  \label{sr-l} \\
_\sL \langle m, e| &= _{\SR}\langle  \half m, 2e+\half m| q^{\half e} (-1)^{\frac{m}{2}}  \, .
\end{align}
The second equation results in the following inner product of $ _\sL \langle m,e|$ basis
\begin{equation}
_\sL \langle m_2, e_2| m_1,e_1 \rangle_\sL  = q^{e_1} \delta_{m_1,m_2} \delta_{e_1,e_2} \;, \label{l-inner}
 \end{equation}
which follows from the inner product of SR basis given in eq.~\eqref{Inner product on SR basis}.  Using eq.~\eqref{sr-l},
The evaluation of $\sL$ operator in $_{\SR} \langle  m,e |$ basis can be rewritten as follows
\begin{align}
&_{\SR}\langle (m_2, e_2),(m_\eta, e_\eta) |\sL | (m_1,e_1), (m_\eta, 0)\rangle_{\SR}\nn
\\
&= q^{\frac{1}4(m_2-e_2+m_1-e_1)} (-1)^{-m_2+m_1}\; _\sL \langle (2m_2, \frac{e_2-m_2}2),(m_\eta, e_\eta)|\sL| (2m_1, \frac{e_1-m_1}2),(m_\eta, 0)\rangle_\sL \label{l-sr}
\end{align}
To evaluate the right hand side, we first note that the infinite product part of $ \sL$ operator acts on the bra as follows
 \begin{align}
 &_\sL \langle (m_2,e_2),(m_\eta, e_\eta)|\prod_{r=1}^\infty \frac{1-q^r e^{{\cal X}_+} \fl^{-1}_+}{1-q^{r-1} e^{{\cal X}_- } \fl^{-1}_-}  \nn
  \\
 &= \oint \frac{du_2}{2\pi i u_2} \oint \frac{du_\eta}{2\pi i u_\eta} u^{-e_2} u_\eta^{-e_\eta}  \;_\sL \langle (m_2,u_2),(m_\eta, u_\eta)|\prod_{r=1}^\infty \frac{1-q^r e^{{\cal X}_+}\fl^{-1}_+}{1-q^{r-1} e^{{\cal X}_- } \fl^{-1}_-}\nn
 \\
 &=\sum_{e'} \;_\sL \langle (m_2,e_2+e') , (m_\eta, e_\eta -e') |\CI_\Delta (m_\eta - m_2, e')\;, \label{bra}
 \end{align}
 and that the exponential part of $\sL$ operator acts on the ket as follows
 \begin{align}
 &\exp \big{[} -\frac{1}{\hbar}(({\cal P}_+ + \half \log \fl_+)^2-({\cal P}_-+ \half \log \fl_-)^2)\big{]}|(m_1,e_1),(m_\eta, e_\eta)\rangle_\sL  \nn
 \\
 &=(\;_\sL\langle (m_1,e_1),(m_\eta, e_\eta) | \exp \big{[} - \frac{1}{\hbar}(({\cal P}_-+ \half \log  \fl_-)^2-({\cal P}_++ \half \log \fl_+)^2 )\big{]} )^\dagger \nn
 \\
 &= |(m_1+2e_1 +m_\eta,e_1),(m_\eta, e_\eta-e_1- \half m_\eta)\rangle_\sL \, .  \label{ket}
 \end{align}
 Here we used the fact that the adjoint of operators are given by
\begin{equation}
( {\cal P}_{\pm, \sL} )^\dagger = {\cal P}_{\mp, \sL}, \quad  (\fl_{ \pm})^\dagger =\fl_{\mp}\; ,
\end{equation}
 and that they can be written as ${\cal P}_{ \pm} = \pm \partial_m+ \frac{ \hbar}{2}e, \ln \fl_{ \pm} = \mp \partial_{e_{ \eta}} + \frac{ \hbar}{2}m_{ \eta} $.
Using eq.~\eqref{bra}, \eqref{ket}, and then \eqref{l-inner}, the evaluation of $\sL$ operator in eq.~\eqref{l-sr} can be rewritten as
\begin{align}
&_{\SR}\langle (m_2, e_2),(m_\eta, e_\eta) |\sL | (m_1,e_1), (m_\eta, 0)\rangle_{\SR} \nn
\\
&=\delta_{\ldots} \delta_{\ldots}  (-1)^{m_1-m_2} q^{\frac{1}4(e_1-e_2-m_1+m_2)}  \CI_{\Delta} (m_{\eta} - 2 m_2, \half (e_1-e_2-m_1+m_2)) \; \nn \\
&
=\delta_{\ldots} \delta_{\ldots}  (-1)^{e_2-e_1} q^{\frac{1}4(e_1-e_2-m_1+m_2)}     \CI_{\Delta} (-e_1-m_1, \frac{1}{2}(e_1-e_2-m_1+m_2))
\nn \\
&= \delta_{ \ldots} \delta_{ \ldots}  (-1)^{e_2-e_1}  q^{\frac{1}4(e_1-e_2-m_1+m_2)} \CI_{\Delta} (\frac{-e_1+m_1+e_2-m_2}{2}, m_1+e_1)
\nn
\end{align}
where $ \delta_{ \ldots} \delta_{ \ldots}$ denotes the following combination of Kronecker delta functions, 
\begin{align}
 \delta_{\ldots}  \delta_{\ldots}
&= \delta (-e_1+2m_2-m_1-m_\eta)\delta(e_2-e_1+2e_{\eta} + m_2 - m_1 )\;.
\label{delta-abb}
\end{align}
In the third line,  we changed the first argument in the tetrahedron index and the power of $(-1)$ using Kronecker delta functions. In the last line,
we used the identity of $\CI_{\Delta}$ in eq.~\eqref{identity for tetrahedron index}.
This completes the derivation of the first part of \eqref{shear index for L,R}.

For $\varphi=\sR$, we basically repeat the previous derivation for $ \varphi = \sL$ using a polarization $\Pi_{\sR}$ instead of the polarization $ \Pi_{\sL}$.
$\Pi_{\sR}$ is defined by
\begin{equation}
\Pi_\sR  =({\mathcal X}_\sR, {\mathcal P}_{\sR})_{ \pm} =( \sT'_{ \pm}  \pm i \pi \pm \frac{ \hbar}{2} ,  \half ( \sT'_{ \pm}+\sT_{ \pm} ) ) \;,
\end{equation}
thus it is related to $ \Pi_{\SR}$ as follows
\begin{equation}
\begin{pmatrix}
{\mathcal X_{\sR}} \\
{\mathcal P_{\sR}}
\end{pmatrix}_{ \pm}
=  \begin{pmatrix} 1&-1 \\ 1& 0 \end{pmatrix}
\cdot
\begin{pmatrix}
{\mathcal X}_{ \SR} \\
{\mathcal P}_{\SR}
\end{pmatrix}_{ \pm}
 \pm
 \begin{pmatrix}  \frac{ \hbar}{2} + i \pi \\  0 \end{pmatrix}
\end{equation}
In terms of momentum and position operators in this polarization, $\sR$ operator in \eqref{L,R operators} can be written as
\begin{equation}
\sR = \left( \prod_{r=1}^{ \infty}   \frac{1-q^r e^{{\mathcal X}_{\sR-}}}{1-q^{r-1} e^{{\mathcal X}_{\sR+}}}  \right)
\exp [ \frac{1}{ \hbar} ( {\mathcal P}_{\sR+}^2-{\mathcal P}_{\sR-}^2 ) ]\;.
\end{equation}
The basis change between SR- and $\sR$-basis can be obtained as
\begin{align}
_{\SR} \langle m,e  | = _\sR \langle  m-e, m  | (-1)^{m} q^{  \frac{m}{2}} \;,\nn
\quad
_\sR \langle m, e |   = _{\SR}\langle e,e-m  | (-1)^{-e} q^{ - \frac{e}{2}} \;,\nn
\end{align}
where the inner product of $\sR$-basis is
\begin{align}
& _\sR \langle m_2, e_2  | m_1, e_1  \rangle_{\sR}   =\delta_{m_1, m_2} \delta_{e_1,e_2} q^{-e_2} \;.\nn
\end{align}
In the $_\sR \langle m,e|$ basis, the expectation value of $\sR$ can be evaluated similarly to eq.~\eqref{bra} and \eqref{ket},
which results in
\begin{equation}
_{\sR}\langle  (m_2, e_2) | \sR | (m_1, e_1) \rangle_{\sR}
= \sum_{ e^{ \prime}}
\CI_{ \Delta} (-m_2, e^{ \prime}) _{\sR }\langle m_2, e_2 - e^{ \prime} |  m_1 - 2 e_1, e_1 \rangle_\sR\;.  \nn
\end{equation}
To evaluate the charge shifts of the ket basis, we used the adjoint relation of ${\mathcal P_{\sR,\pm}}$
\begin{equation}
( {\mathcal P_{\sR, \pm }})^{ \dagger} = {\mathcal P_{\sR, \mp}} \,
\end{equation}
which can be deduced from the adjoint relation of ${\mathcal P}_{\SR, \pm}$.
Thus, we obtain
\begin{align}
_{\SR}\langle  m_2,e_2  |  \sR | m_1, e_1 \rangle_{\SR}
&= (-1)^{m_2-m_1} q^{  \frac{m_1+m_2}{2}} _{\sR}\langle m_2-e_2 , m_2 | \sR |m_1 - e_1, m_1 \rangle_{\sR} \nn
\\
&=(-1)^{m_2+e_1} q^{ \half (m_2+e_1)}  \CI_{ \Delta} (-e_1 -m_2, m_1+e_1) \delta_{m_2+m_1-e_2+e_1,0} \,.
\nn
\end{align}
In the second line, the Kronecker delta function was used to change the first argument of $\CI_{ \Delta}$, then
the triality relation \eqref{Triality} was used in turn. This gives a derivation for the second part in eq.~\eqref{shear index for L,R}.
\\

\section{Basis change between SR  and FN basis} \label{basis from FN to shear}

Recall that $\Pi_{\SR}= ({\cal S},{\cal R}) := (\frac{1}{2}(\sT+\sT'), \frac{1}2 (\sT-\sT'))$. Explicit expressions for $\ss=\exp(\mathcal{S}), \sr = \exp(\mathcal{R})$  in terms of Fenchel-Nielsen operators, $(\hat{\lambda}, \hat{\tau})$ are given by
\begin{align}
&\ss_{\pm} =\exp (\CS_\pm) =q^{\pm 1/4} \frac{1}{(q^{\mp 1/4} \hat{ \lambda}^{-1}_{\pm} \hat{\tau}_\pm^{1/2}-q^{\pm 1/4}  \hat{\tau}_\pm^{-1/2}\hat{\lambda}_\pm) }(\hat{\tau}_{\pm}^{1/2} - \hat{\tau}_{\pm}^{-1/2}) \;. \nonumber
\\
&\sr_{\pm} =  \exp (\CR_\pm) = q^{\pm 1/4} \frac{1}{\hat{\lambda}_\pm -\hat{ \lambda}_\pm^{-1}} (\hat{\tau}^{-1/2}_\pm - \hat{\tau}^{1/2}_\pm)\frac{1}{\hat{\lambda}_\pm - \hat{\lambda}_\pm^{-1}} (q^{\mp 1/4} \hat{\lambda}^{-1}_{\pm} \hat{\tau}_\pm^{1/2}-q^{\pm 1/4}  \hat{\tau}_\pm^{-1/2} \hat{\lambda}_\pm) \;.
 \label{s,r variable }
\end{align}
We will express SR fugacity basis $\langle m,u|:=_{\widetilde{\SR}}\langle m,u|$ in terms of FN fugacity basis $\langle \tilde{m},\tilde{u}|:=_{\FN}\langle \tilde{m},\tilde{u}|$,
\begin{align}
&\langle m,u| = \oint \frac{d\tilde{u}}{2\pi i \tilde{u}}\Delta(\tilde{m},\tilde{u},q) \langle m,u|\tilde{m},\tilde{u}\rangle \langle \tilde{m},\tilde{u}|\;.
\end{align}
By imposing the following conditions, we obtain difference equations for the basis change coefficients $\langle m,u|\tilde{m},\tilde{u}\rangle$.
\begin{align}
\langle m,u|\ss_\pm |\tilde{m},\tilde{u}\rangle &= x_\pm \cdot \langle m,u| \tilde{m},\tilde{u}\rangle \nn
\\
&=\langle m,u| \ss_\pm (\hat{\lambda}_\pm, \hat{\tau}_\pm )|\tilde{m},\tilde{u}\rangle = s_\pm^T(\tilde{x}_\pm ,\tilde{p}_\pm)\cdot \langle m,u|\tilde{m},\tilde{u}\rangle \nn
\\
\langle m,u|\sr_\pm |\tilde{m},\tilde{u}\rangle &= p_\pm \cdot\langle m,u| \tilde{m},\tilde{u}\rangle \nn
\\
&=\langle m,u| \sr_\pm (\hat{\lambda}_\pm, \hat{\tau}_\pm )|\tilde{m},\tilde{u}\rangle = r_\pm^T(\tilde{x}_\pm, \tilde{p}_\pm )\cdot \langle m,u|\tilde{m},\tilde{u}\rangle \;. \label{difference equation for basis change matrix}
\end{align}
Here $\sO^{\rm T}$ denote the transpose of $\sO$, $\sO_\pm^{\rm T} := (\sO^\dagger_\pm)^*$. For SR operators, their transpose operators are
\begin{align}
(\ss_\pm, \sr_\pm)^{\rm T} = (\ss_\pm , \sr_\pm)/.\{ q\rightarrow q^{-1} ,  \tau\rightarrow \tau^{-1}\} \;.
\end{align}
The operators $(x,p)_\pm,(\tilde{x},\tilde{p})_\pm$ are given
\begin{align}
x_\pm = q^{\frac{m}2} u^{\pm 1}\;,\quad p_\pm = e^{\pm \partial_{m} + \frac{1}2\hbar u\partial_{u}}\;, \quad \tilde{x}_\pm = q^{\frac{\tilde{m}}2} \tilde{u}^{\pm 1}\;,\quad \tilde{p}_\pm = e^{\pm \partial_{\tilde{m}} + \frac{1}2\hbar \tilde{u}\partial_{\tilde{u}}} \;.
\end{align}
To make the action of FN, SR operators simple, we introduce new variables $(a,b),(s,t)$ defined as follow
\begin{align}
q^a := q^{\frac{\tilde{m}}2}\tilde{ u} ,\quad q^b = q^{\frac{\tilde{m}}2}\tilde{u}^{-1}, \quad q^s := q^{\frac{m}2} u ,\quad q^t = q^{\frac{m}2}u^{-1}\;.
\end{align}
Let's denote $\langle m,u|\tilde{m},\tilde{u}\rangle$ in terms of these variables as $C(a,b|s,t)$.
\begin{align}
C(a,b|s,t):=\langle m, u|\tilde{m},\tilde{u}\rangle /.\{\tilde{m}\rightarrow   a+b, m\rightarrow s+t, \tilde{u}\rightarrow q^{\frac{a-b}2}, u \rightarrow q^{\frac{s-t}2}\} \;.
\end{align}
One advantage of these variables is that   $+$ ($-$) type operators act only on $a,s$ ($b,t$) variables. Thus the difference equations \eqref{difference equation for basis change matrix} for $C(s,t|a,b)$ factorizes into $\pm$ parts and we can set
\begin{align}
C (s,t|a,b) = C_+ (s|a) C_- (t|b) \;. \label{basis change coefficient}
\end{align}
Then, the difference equations for $C_+$ are
\begin{align}
&(1-q^{-a+s}) C_+(s|a) = (1-q^{s+a+1}) C_+(s|a+1) \;, \nn
\\
&(q^a - q^{-a}) C_+(s+1|a) \nn
\\
&=\frac{q^{-(a+1/2)} C_+(s|a)-q^{a+1/2} C_+(s|a+1)}{q^{a+1/2}-q^{-(a+1/2)}}- \frac{q^{-(a-1/2)} C_+(s|a-1)-q^{a-1/2} C_+(s|a)}{q^{a-1/2}-q^{-(a-1/2)}} \;. \nn
\end{align}
For $C_-$, the difference equations are
\begin{align}
&(1-q^{b+t}) C_-(t|b) = (1-q^{t-b-1}) C_-(t|b+1) \;, \nonumber
\\
&(q^b -q^{-b})C_-(t-1|b) \nonumber
\\
&=\frac{q^{-(b-1/2)} C_-(t|b)-q^{b-1/2} C_-(t|b-1)}{q^{b-1/2}-q^{-(b-1/2)}}- \frac{q^{-(b+1/2)} C_-(t|b+1)-q^{b+1/2} C_-(t|b)}{q^{b+1/2}-q^{-(b+1/2)}} \;.
\end{align}
We use the fact that
\begin{align}
(x_+,x_-, p_+, p_-) = (q^s, q^t, e^{\partial_s}, e^{-\partial_t})\;, \quad (\tilde{x}_+,\tilde{x}_-, \tilde{p}_+, \tilde{p}_-) = (q^a, q^b, e^{\partial_a}, e^{-\partial_b})\;.
\end{align}
Solving the two difference equations, we find the following solutions
\begin{align}
&C_+ (s|a) =(-1)^{a} q^{-a(a+1)/2-as}\prod_{r=0}^\infty \frac{1-q^{r+1}q^{-a+s}}{1-q^{r}q^{-a-s}}\;, \nn
\\
&C_- (t|b) = (-1)^b q^{b(b-1)/2+bt}\prod_{r=0}^\infty \frac{1-q^{r+1}q^{-b-t}}{1-q^{r}q^{-b+t}} \;. \nn
\end{align}
Therefore,
\begin{align}
&C(s,t|a,b)= (-1)^{a+b}q^{-a(a+1)/2-as+b(b-1)/2+bt}\prod_{r=0}^\infty \frac{1-q^{r+1}q^{-a+s}}{1-q^{r}q^{-a-s}}\prod_{r=0}^\infty \frac{1-q^{r+1}q^{-b-t}}{1-q^{r}q^{-b+t}}\;. \label{sol for C(s,t|a,b)}
\end{align}
In the original fugacity variables$(m,u), (\tilde{m},\tilde{u})$, the basis change matrix is given by
\begin{align}
&\langle m,u| \tilde{m},\tilde{u}\rangle = C(s,t|a,b)|_{a\rightarrow \frac{\tilde{m}}2 +\log_q \tilde{u},b\rightarrow \frac{\tilde{m}}2 -\log_q \tilde{u},s \rightarrow \frac{m}2 +\log_q u, t\rightarrow \frac{m}2 +\log_q u} \nn
\\
&=  (- \tilde{u}q^{1/2})^{-\tilde{m}}  \frac{\CI_{\Delta}(\tilde{m}-m, \tilde{u}/u )}{\CI_{\Delta}(\tilde{m}+m,  u \tilde{u}q^{-1})} u^{-\tilde{m}} \tilde{u}^{-m} \;, \nn
\\
&= (-q^{\half} u)^m \CI_{\Delta} (-m-\tilde{m}, u^{-1}\tilde{u}^{-1})\CI_{\Delta}(\tilde{m}-m, \tilde{u}/u)\;.
\label{eq:D}
\end{align}
In the SR charge basis, we have
\begin{align}
\langle m, e| \tilde{m}, \tilde{u}\rangle = \sum_{e_1 \in \mathbb{Z}} (- q^{\half})^m \tilde{u}^{-2e_1 -e+m} \CI_{\Delta}(-m-\tilde{m} ,e_1)\CI_{\Delta}(-m+\tilde{m}, -e+m-e_1) \;. \label{eq:D-charge basis}
\end{align}
This basis change matrix element can be thought of as $SL(2,\mathbb{C})$ CS partition function on a mapping cylinder $\Sigma_{1,1}\times_{\varphi} I$ with $\varphi= \textrm{identity}$ in the polarization where positions are $(\ss_{\textrm{bot}}, \lambda_{\rm top}, \fl)$ and momenta are $(\sr_{\rm bot}, \tau_{\rm top}, \sm)$. Identifying FN operators $(\lambda, \tau)$ as UV  operators and SR operator $(\ss, \sr)$ as IR operators, the mapping cylinder is called a RG manifold in  \cite{Dimofte:2013lba}.

Note that the SR  basis is Weyl-reflection invariant and thus the states are in $\mathcal{H}_{SL(2,\mathbb{C})} \subset \widetilde{\CH}_{SL(2,\mathbb{C})}$. For the charge basis $| m,e\rangle$ to be an (non-zero) element in $\mathcal{H}_{SL(2,\mathbb{C})}$, we need to impose following conditions
\begin{align}
&\langle m,e|\tilde{m},\tilde{u}\rangle \neq 0\;,  \quad \textrm{for some $\tilde{m} \in \mathbb{Z}/2$}\;.
\end{align}
This condition implies that
\begin{align}
m,e\in \mathbb{Z}/2, \quad m+e \in \mathbb{Z}\;. \label{Range for SR charge}
\end{align}
Furthermore, we claim that SR charge basis are complete basis for $\mathcal{H}_{SL(2,\mathbb{C})}$. How can we prove the completeness?
One simple answer uses the fact that  a operator $\ss_+\ss_-$ is self-adjoint. Since $\langle m,e|$ are eigenstates for the operator with eigenvalues $q^{m}$, they form a complete basis. Using the property  $(\ss_\pm, \sr_\pm)^\dagger = (\ss_\mp, \sr_\mp)$ and
\begin{align}
\langle m,e|\ss_+ = \langle  m,e\mp 1| q^{\frac{m}2}\;, \quad \langle m,e|\sr_\pm = \langle m\pm 1, e| q^{\frac{e}2}\;,
\end{align}
One can see that
\begin{align}
\langle m,e|m',e'\rangle = \kappa \delta_{m,m'}\delta_{e,e'}\;.
\end{align}
Here $\kappa$ is $(m,e)$-independent constant and it is 1 in \eqref{eq:D}. From this orthonormality and the fact the basis $\langle m,e|$ are complete basis in $\mathcal{H}_{SL(2,\mathbb{C})}$, one obtain following completeness relation
\begin{align}
\mathds{1}_{\mathcal{H}_{SL(2,\mathbb{C})}} = \sum_{(m,e)}| m,e\rangle \langle m,e| \;.
\end{align}
More directly, the completeness relation is equivalent to the following identity, 
\begin{align}
&\sum_{(m,e)}\langle \tilde{m}_b, \tilde{u}_b| m,e\rangle \langle m,e| \tilde{m}_t, \tilde{m}_t\rangle  = \delta(\tilde{m}_b - \tilde{m}_t) \frac{\delta(\tilde{u}_b - \tilde{u}_t)}{\Delta(\tilde{m}_t,\tilde{u}_t)}\;.\nn
\end{align}
using the explicit expression in \eqref{eq:D-charge basis}. While we do not have an exact proof, we have confirmed it by series expansion in $q$. More precisely, we checked  that
\begin{align}
&\sum_{\tilde{m}\in \mathbb{Z}/2}\oint \frac{d\tilde{u}_t}{2\pi i \tilde{u}_t}\Delta(\tilde{m}_t, \tilde{m}_t)\big{(}\sum_{(m,e)}\langle \tilde{m}_b, \tilde{u}_b| m,e\rangle \langle m,e| \tilde{m}_t, \tilde{m}_t\rangle \big{)} f(\tilde{m}_t,\tilde{u}_t  ) = f(\tilde{m}_b, \tilde{u}_b)\;, \nn
\end{align}
by expansion in $q$ for various Weyl-reflection invariant trial function $f(m,u)$, i.e., $f(m,u)= f(-m, u^{-1})$.

\end{document}